\documentclass[prd,amsmath,floats,amssymb,authoryear,floatfix,superscriptaddress,nofootinbib,onecolumn]{revtex4} 
\usepackage[plainpages=false, colorlinks=true, anchorcolor=blue, linkcolor=blue, citecolor=blue, bookmarks=false]{hyperref}
\usepackage{graphicx}
\usepackage{hhline} 
\usepackage{natbib}
\usepackage{appendix}
\usepackage{multirow}
\usepackage{makecell}

\renewcommand{\thetable}{\textbf{\arabic{table}}}

\newcommand{\civ}{C\,\textsc{iv}}

\makeatletter
\renewcommand{\fnum@figure}{\textbf{Figure~\thefigure}}
\renewcommand{\fnum@table}{\textbf{Table~\thetable}}
\makeatother

\newcommand{\rthis}[1]{\textcolor{black}{#1}}
\usepackage{float}
\usepackage[caption=false]{subfig}
\usepackage[paperwidth=210mm,paperheight=297mm,centering,hmargin=1cm,vmargin=2.5cm]{geometry}

\begin{document}
\preprint{APS/123-QED}

\title{Constraints on Dark Energy Models Using Late Universe Probes}

\author{Shubham Barua}
 \altaffiliation{Email:ph24resch01006@iith.ac.in}
\author{Shantanu Desai}
 \altaffiliation{Email:shntn05@gmail.com}
\affiliation{
 Department of Physics, IIT Hyderabad Kandi, Telangana 502284,  India}




\begin{abstract}
We use late Universe probes - Type Ia Supernovae from the PantheonPlus compilation, Quasars, and Dark Energy Spectroscopic Instrument (DESI) Data Release 1 (DR1) BAO data - along with Cosmic Chronometers or Megamasers to constrain various dark energy parameterizations. These include the standard $\Lambda$CDM model ($w_0=-1,w_a=0$), as well as the Chevallier-Polarski-Linder (CPL), Barboza-Alcaniz (BA), Jassal-Bagla-Padmanabhan (JBP), Exponential (EXP), and Transitional Dark Energy (TDE) parameterizations. We find that across all parameterizations, the constrained values of $w_0$ and $w_a$ remain within $(1-2)\sigma$ of the standard $\Lambda$CDM model, irrespective of spatial curvature, dataset combinations, or prior choices. We find from Bayesian model comparison that $\Lambda$CDM remains the most favored model for both flat and non-flat cases, with results remaining robust under different priors. Across all dataset combinations, we reaffirm the fact that LRG1 and LRG2 data points from the DESI BAO dataset are responsible for driving the preference for dynamical dark energy.
\end{abstract}


\maketitle
\section{\label{sec:1}Introduction}

The discovery of the accelerated expansion of the universe \cite{riess_1998, perlmutter_1999} represents a pivotal observational milestone in modern cosmology, providing compelling evidence for the existence of Dark Energy (DE) \cite{tegmark_2004, riess_2004,scranton_2003, tonry_2003, knop_2003, feng_2005, astier_2006, eisenstein_2005,eisenstein_2007,tegmark_2006, sahni_2006,wood_vasey_2007, vikhlinin_2009, stern_2010, sherwin_2011, bennett_2013, hinshaw_2013, dawson_2013, dejong_2013, anderson_2014, weinberg_2013, florian_2014, delubac_2015}. The simplest way to model the  late-time acceleration \cite{des_2016, haridasu_2017, huterer_2018, troxel_2018, scolnic_2018, planck_2020a, planck_2020b, gomez_2019, yang_2020} is through the inclusion of a positive cosmological constant term in Einstein’s field equations, leading to the formulation of the standard $\Lambda$CDM model. To date, this model has demonstrated remarkable success in explaining a wide range of observed cosmological phenomena \cite{planck_2020b, scolnic_2022, brout_2022, holicow_2017, holicow_2020}.

However, in recent decades, a  large number of cosmological tensions  have emerged within the $\Lambda$CDM framework~\cite{valentino_2021, verde_2019,  abdalla_2022, scherer_2025, shubham_2025}. Among these tensions, the most notable is the Hubble tension—a discrepancy exceeding the $5\sigma$ level between the value of the Hubble constant $H_0$ inferred from Cosmic Microwave Background (CMB) observations \cite{planck_2020b} and the value derived from local measurements using the distance ladder method \cite{riess_2022}. This discrepancy has prompted extensive investigations into the validity of the $\Lambda$CDM model and the potential existence of new physics beyond the Standard Model of cosmology \cite{abdalla_2022, knox_2020, poulin_2023, tanvi_2022, poulin_2019, berghaus_2020, kamionkowski_2023, agrawal_2023, vagnozzi_2023, montani_2024, raveri_2020, keeley_2019, shouvik_2024}.

A general approach to parameterizing cosmic acceleration is through an effective equation of state parameter $w$, where $w < -\frac{1}{3}$ is required for accelerated expansion \cite{sahni_2000, carroll_2001, peebles_2003, padmanabhan_2003, copeland_2006, caldwell_2009, li_2011, martin_2012, krauss_1995, Bamba_2012}. While the simplest realization of DE is the cosmological constant ($\Lambda$), where $w=-1$, a more general approach involves parameterizing DE with a free equation of state parameter. This allows for the possibility of alternative explanations for cosmic acceleration. 
Along these lines, numerous studies have extended the idea of vanilla dark energy to  the concept of Dynamical Dark Energy (DDE), where the equation of state is no longer constant but instead varies as a function of the scale factor, denoted as $w(a)$ \cite{yang_2019a, yang_2019b, yang_2021, colgain_2021, mainini_2003, alam_2004, sola_2006, sola_2018, antoniadis_2007, zhao_2012, zhao_2020, pan_2019, celia_2022, Li_2024, shouvik_2025, ParkCG_2024}. Since Cosmic Microwave Background (CMB) data alone is insufficient to place stringent constraints on DDE models \cite{valentino_2017, valentino_2022, valentino_2020}, observations from the local universe become crucial for comprehensive analysis.

In 2024, the Dark Energy Spectroscopic Instrument (DESI) Baryon Acoustic Oscillation (BAO) data \cite{desi_2024} provided evidence supporting DDE under the Chevallier-Polarski-Linder (CPL) parameterization of the dark energy equation of state:$w(a) = w_0 + w_a(1-a)$, where $w_0$ represents the present-day equation of state, and $w_a$ captures its evolution with the scale factor. This evidence emerges from the combined analysis of CMB, BAO, and Supernovae Type Ia (SNe Ia) data, with statistical significance ranging from $(2.5 - 3.9)\sigma$ depending on the specific SNe Ia sample utilized. 

In this work, we investigate six distinct parameterizations of dark energy: the standard $\Lambda$CDM model($w_0 = -1, w_a = 0$), Chevallier-Polarski-Linder (CPL) \cite{cpl_2001, cpl_2003}, Barboza-Alcaniz (BA) \cite{ba_2008} Jassal-Bagla-Padmanabhan (JBP) \cite{jbp_2005}, exponential (EXP) \cite{najafi_2024} and Transitional Dark Energy (TDE) parameterization \cite{keeley_2019}. We employ late-universe observational probes, specifically the PantheonPlus compilation of Type Ia supernovae (SNe Ia) \cite{scolnic_2022}, the DESI 2024 BAO dataset \cite{desi_2024}, quasars (QSO) \cite{cao_2022}, and either the Cosmic Chronometer (CC) \cite{moresco_2023} or Megamaser (MM) \cite{Pesce_2020} dataset, to constrain these parameterizations. Our objectives are to assess the role of the BAO data points LRG1 and LRG2 (from DESI DR1) in driving deviations from the standard $(w_0, w_a) = (-1,0)$ scenario, explore the impact of different prior choices (Uniform or Gaussian), and evaluate the preference for specific model-dataset combinations using Bayesian evidence. We use only late-universe probes \cite{zheng_2024, wu_2025} in this work due to the existence of the $5\sigma$ tension which exists between CMB inferred $H_0$ value and the one constrained using the cosmic distance ladder.

This manuscript is organized as follows. In section \ref{sec:2}, we propose the myriad DE parameterizations we wish to constrain. Section \ref{sec:3} contains a description of all data sets used in this work. Section \ref{sec:4} contains the relations that we used to constrain the parameters. The methodology of our analysis is described in section \ref{sec:5}. In section \ref{sec:results}, we list the results obtained in our analysis \rthis{and in section \ref{sec:behaviour} we discuss the behavior of $w(z)$ obtained from our constraints}. Finally, in section \ref{sec:conclusion} we state our conclusion.

\section{Cosmological Models}
\label{sec:2}
In this section, we describe the cosmological models used in our study. We consider the Friedmann-Lemaitre-Robertson-Walker (FLRW) metric~\cite{zyla_2020} throughout this work. It is given by 
\begin{equation}
\label{eq1}
    ds^2 = -dt^2 + a^2(t)\left[\frac{dr^2}{1-kr^2}+ r^2\left(d\theta^2 +  \text{sin}^2\theta d\phi^2\right)\right]
\end{equation}
and assumes a homogeneous and isotropic universe on large scales. The late universe is dominated by dark energy, which  has been observationally determined and the amount of radiation is negligible. On solving the Friedmann equations using the FLRW metric, we get the following general form of the expansion history of the universe \cite{zheng_2024}
\begin{equation}
\label{eq2}
    H^2(z) = H^2(0)\left[\Omega_m(1+z)^3 + f_{DE}(z) + \Omega_k(1+z)^2\right],
\end{equation}
where $\Omega_m$ is the present dimensionless matter density of the universe, $\Omega_k$ expresses the curvature of the universe and $f(z)_{DE}$ is the term representing dark energy dynamics
\begin{equation}
\label{eq3}
    f_{DE}(z) = \Omega_{DE}e^{\left(3 \int_0^z \frac{1+w(z')}{1+z'}dz'\right)},
\end{equation}
where $\Omega_{DE} = 1 - \Omega_m - \Omega_k$.
Over the years, various dark energy parameterizations represented by their equation of state have been proposed \cite{wolf_2023}. 
In this work, we consider the following dynamical dark energy models:
\begin{itemize}
    \item $\Lambda$CDM model - In this case we consider $w_{DE} = -1$. This reduces Eqn.~\ref{eq2} to the following well-known form in the presence of non-zero curvature.
    \begin{equation}
    \label{eq4}
        H^2(z) = H^2_0\left[\Omega_m(1+z)^3 + \Omega_k(1+z)^2 + \Omega_{DE}\right].
    \end{equation}
    \item CPL parameterization - This is one of the most popular parameterizations, also known as the Chevallier-Polarski-Linder (CPL) parameterization \cite{cpl_2001, cpl_2003}. The equation of state of this dynamical dark energy model is given by $w(z) = w_0 + \frac{w_az}{1+z}$. After  evaluating the expression for $f_{DE}(z)$ from  Eq.~\ref{eq3}  and plugging  in Eq.~\ref{eq3}, we have the Hubble parameter given by:
    \begin{equation}
    \label{eq5}
            H^2(z) = H^2_0\left[\Omega_m(1+z)^3 + \Omega_k(1+z)^2 + \Omega_{DE}(1+z)^{3(1+w_0+w_a)}e^{(\frac{-3w_az}{1+z})}\right].
    \end{equation}
    \item BA parameterization - The equation of state of dark energy for the Barboza-Alcaniz parameterization \cite{ba_2008} is given by $w(z) = w_0 + \frac{w_az(1+z)}{(1+z^2)}$. The Friedmann equation reduces to: 
    \begin{equation}
    \label{eq6}
        H^2(z) = H^2_0\left[\Omega_m(1+z)^3 + \Omega_k(1+z)^2 + \Omega_{DE}(1+z)^{3(1+w_0)}(1+z^2)^\frac{3w_a}{2}\right].
    \end{equation}
    \item JBP parameterization - Also known as the Jassal-Bagla-Padmanabhan parameterization \cite{jbp_2005}, the equation of state for this case is given by $w(z) = w_0 + \frac{w_az}{(1+z)^2}$. On solving the Friedmann equation, we get:
    \begin{equation}
    \label{eq7}
        H^2(z) = H^2_0\left[\Omega_m(1+z)^3 + \Omega_k(1+z)^2 + \Omega_{DE}(1+z)^{3(1+w_0)}e^\frac{3w_az^2}{2(1+z)^2}\right].
    \end{equation}
    \item EXP parameterization - Here we consider the equation of state of DE to assume the form $w(z) = w_0-w_a+w_ae^{\left(\frac{z}{1+z}\right)}$ \cite{pan_2020, dimakis_2016, najafi_2024, wolf_2025}. The first order Taylor expansion of the exponential gives us the CPL parameterization while higher order terms may play an important role. We will henceforth refer to this as the EXP parameterization. Following \cite{najafi_2024}, we consider upto the second order terms such that the equation of state becomes $w(z) = w_0 + w_a\left[\frac{z}{1+z}+\frac{1}{2!}\left(\frac{z}{1+z}\right)^2\right]$. The Hubble parameter then  assumes the following form: 
    \begin{equation}
        \label{eq8}
        H^2(z) = H^2_0\left[\Omega_k (1+z)^2 + \Omega_m (1+z)^3 + \Omega_{DE}(1+z)^{3(1+w_0+w_a)} e^{\left(-\frac{3 w_a z}{1+z} \right)} e^{\left(3 w_a \left(\frac{1}{4(1+z)^2} + \frac{1}{2(1+z)} - \frac{3}{4} \right) \right)} (1+z)^{\frac{3 w_a}{2}} \right].
    \end{equation}
    \item TDE parameterization - This is a more flexible parameterization first introduced in \cite{keeley_2019}, and recently used to study the DESI 2024 BAO data in \cite{keeley_2025}. The parametric form for the equation of state is given by:
    \begin{equation}
        \label{eq9}
        w(z) = \frac{1}{2}\left[(w_0+w_a)+(w_a-w_0)\text{tanh}\left(\frac{z-z_T}{\Delta_z}\right)\right],
    \end{equation}
    where $w_0$ is the value of the equation of state below the $z_T$ (the transition redshift),  while $w_a$ denotes the value of the equation of state above $z_T$. $\Delta_z$ is the width of the transition which controls the rate of change of $w(z)$.
\end{itemize}

\section{Description of Data}
\label{sec:3}

\subsection{Baryon Acoustic Oscillations}
\label{sec:3.1}

Baryon Acoustic Oscillations (BAO) are considered the archetype of statistical standard rulers \cite{ruchika_2024}. They are based on the principle that the large-scale clustering of galaxies exhibits a preferred scale, which, when measured at different redshifts, can be used to constrain the angular diameter distance and, consequently, the expansion history of the universe. In the early universe, the hot plasma of photons and baryons was tightly coupled through Thompson scattering. Oscillations were established in this photon-baryon fluid due to the competing effects of gravitational attraction and radiation pressure. As the universe cooled and recombination occurred, photons decoupled from baryons and began to propagate freely, forming the Cosmic Microwave Background (CMB), while the baryonic oscillations were effectively ``frozen'' in place. This imprint on the baryon density manifests as an excess in clustering at a characteristic scale, known as the sound horizon \cite{eisenstein_1998}.

The sound horizon at recombination represents the maximum distance that acoustic waves could have traveled in the photon-baryon fluid by the time of recombination \cite{sunyaev_1972}. This scale is not only imprinted in the large-scale structure of the universe, observed as a peak in the two-point correlation function and as oscillatory features (``wiggles'') in the galaxy power spectrum, but also appears as the characteristic angular scale of temperature fluctuations in the CMB power spectrum \cite{peebles_1970}. Thus, BAO serves as a robust cosmological standard ruler for probing the universe's expansion history across cosmic time \cite{sutherland_2012, eisenstein_2007}.

In this paper, we use the Dark Energy Spectroscopic Instrument (DESI) survey dataset obtained from its first data release \cite{desi_2024}. The data consists of anisotropic BAO measurements obtained from Luminous Red galaxies (LRGs), Emission Line galaxies (ELGs) and Lyman-$\alpha$ forest quasars and isotropic measurements from Bright galaxy samples (BGS) and quasars QSO samples. The various tracers, their effective redshifts and corresponding BAO measurements are given in Table~\ref{table1}.
\begin{table}[htbp!]
\caption{DESI BAO Data Release 1 \cite{desi_2024}}
\label{table1}
\centering
    \begin{tabular}{|c|c|c|c|c|c|}
        \hline
        \thead{Tracer}       & \thead{Redshift}  & \thead{\boldmath $z_\text{eff}$} & \thead{\boldmath $D_M/r_d$}        & \thead{\boldmath $D_H/r_d$}         & \thead{\boldmath $r$ or $D_V/r_d$} \\
        \hline
        BGS          & 0.1-0.4   & 0.295     &  $-$             & $-$                & $7.93\pm0.15$ \\
        LRG1         & 0.4-0.6   & 0.510     &  $13.62\pm0.25$  & $20.98\pm0.61$     & $-0.445$ \\
        LRG2         & 0.6-0.8   & 0.706     &  $16.85\pm0.32$  & $20.08\pm0.60$     & $-0.420$ \\
        LRG3$+$ELG1  & 0.8-1.1   & 0.903     &  $21.71\pm0.28$  & $17.88\pm0.35$     & $-0.389$ \\
        ELG2         & 1.1-1.6   & 1.317     &  $27.79\pm0.69$  & $13.82\pm0.42$     & $-0.444$ \\
        QSO          & 0.8-2.1   & 1.491     &  $-$             & $-$               & $26.7\pm0.67$ \\
        Lya QSO      & 1.77-4.16 & 2.330  &  $39.71\pm0.94$     & $8.52\pm0.17$      & $-0.477$ \\
        \hline
    \end{tabular}
\end{table}
Numerous discussions have focused on the inferred deviation from the $\Lambda$CDM model, with numerous questions surrounding this discrepancy thoroughly investigated. In \cite{colgain_2024}, it was determined that the data point at $z_\text{eff} = 0.51$ is an outlier. \cite{Wang_2024,Liu_2024} highlighted that the LRG1 ($z_\text{eff} = 0.51$) and LRG2 ($z_\text{eff} = 0.71$) data points are the primary drivers of the observed deviation from the $\Lambda$CDM model. Furthermore, inconsistencies have been confirmed between BAO measurements from the Sloan Digital Sky Survey (SDSS) and the DESI collaboration \cite{ghosh_2024}.

\subsection{Type Ia Supernovae}
\label{sec:3.2}

Type Ia supernovae (SNe) are thought to be the result of the explosion of a carbon-oxygen white dwarf in a binary system as it goes over the Chandrashekhar limit, either due to accretion from a donor or mergers. They act as standard candles in determining cosmic distances through the use of the cosmic distance ladder. They are invaluable as they are bright enough to be seen at large cosmic distances, common enough to be found and can be standardized. These objects were also responsible for providing the first evidence of an accelerating expansion phase of the universe \cite{weinberg_2013, riess_1998, huterer_2018, perlmutter_1999}.

The PantheonPlus sample is a comprehensive collection of 18 supernovae surveys namely, DES \cite{des_sn}, Foundation \cite{foundation_sn}, PS1 \cite{ps1_sn}, SNLS \cite{snls_sn}, SDSS \cite{sdss_sn}, HST \cite{hst_sn1, hst_sn2, hst_sn3, hst_sn4, hst_sn5, hst_sn6}, LOSS-1 \cite{loss1_sn}, LOSS-2 \cite{loss2_sn}, SOUSA \cite{sousa_sn}, CSP \cite{csp_sn}, CfA1 \cite{cfa1_sn}, CfA2 \cite{cfa2_sn}, CfA3 \cite{cfa3_sn}, CfA4 \cite{cfa4_sn}, Low-z, CNIa0.02 \cite{cn1a_sn} and other low redshift samples \cite{small1_sn, small2_sn, small3_sn, small4_sn, small5_sn, small6_sn, small7_sn, small8_sn, small9_sn}. The data and the full covariance matrix can be found at \url{https://github.com/PantheonPlusSH0ES/DataRelease}. It also consists of SH0ES Cepheid host distance, which facilitates the breaking of the degeneracy between $M$ and $H_0$. However, we avoid the usage of the SH0ES distances.

PantheonPlus consists of 1701 light curves for 1550, spectroscopically confirmed SNe Ia in the redshift range $0.001 \le z \le 2.26$ \cite{scolnic_2022}. In our analysis, to avoid strong peculiar velocity dependence of the SNe Ia sample, we exclude low redshift ($z\le0.01$) samples and work with 1590 light curves \cite{brout_2022}. 

\subsection{Cosmic Chronometers}
\label{sec:3.3}

The idea behind cosmic chronometers \cite{jimenez_2002, Loubser_2025} is that under minimal assumptions (FLRW metric) and the relation $a(t) = \frac{1}{1+z}$, the Hubble parameter can be derived as 
\begin{equation}
\label{eq10}
    H(z) = -\frac{1}{1+z}\frac{dz}{dt}.
\end{equation}
This is based on the idea that measuring the differential age of the Universe makes it possible to obtain a direct and model-independent determination of $H(z)$. This approach utilizes the age of the oldest galaxies at different redshifts to estimate $H(z)$. The assumption is that `red and dead' galaxies form their stellar population early and further evolution proceeds with minimal subsequent star formation. The age of these galaxies can be associated with the age of the Universe at their formation redshifts. 

In this paper, we use measurements of 32 cosmic chronometers~\cite{moresco_2012, moresco_2015, moresco_2016, moresco_2020, moresco_2023, stern_2010, ratsimbazafy_2017, zhang_2014, joan_2005} in the redshift range $0.07\le z\le1.965$. We use the covariance matrix for computations as described in \cite{moresco_2020}. The $H(z)$ values are tabulated in Table~\ref{table2} and can also be found in \cite{moresco_2023}. 
The last 15 $H(z)$ measurements are correlated and we use the covariance matrix provided in ~\cite{moresco_2023}.
\begin{table}[htbp!]
\caption{32 $H(z)$ Data \cite{moresco_2023}}
\label{table2}
\centering
    \begin{tabular}{|c|c|c|}
        \hline
        \thead{\boldmath $z$} & \thead{\boldmath $H_0$ (km/s/Mpc)} & \thead{Reference}\\
        \hline
        $0.07$   & $69.0\pm19.6$  & \cite{zhang_2014} \\
        $0.09$   & $69.0\pm12.0$  & \cite{joan_2005} \\
        $0.12$   & $68.6\pm26.2$  & \cite{zhang_2014} \\
        $0.17$   & $83.0\pm8.0$   & \cite{joan_2005} \\
        $0.2$    & $72.9\pm29.6$  & \cite{zhang_2014} \\
        $0.27$   & $77.0\pm14.0$  & \cite{joan_2005} \\
        $0.28$   & $88.8\pm36.6$  & \cite{zhang_2014} \\
        $0.4$    & $95.0\pm17.0$  & \cite{joan_2005} \\
        $0.47$   & $89.0\pm50.0$  & \cite{ratsimbazafy_2017} \\
        $0.48$   & $97.0\pm62.0$  & \cite{stern_2010} \\
        $0.75$   & $98.8\pm33.6$  & \cite{borghi_2022} \\
        $0.88$   & $90.0\pm40.0$  & \cite{stern_2010} \\
        $0.9$    & $117.0\pm23.0$ & \cite{joan_2005} \\
        $1.3$    & $168.0\pm17.0$ & \cite{joan_2005} \\
        $1.43$   & $177.0\pm18.0$ & \cite{joan_2005} \\
        $1.53$   & $140.0\pm14.0$ & \cite{joan_2005} \\
        $1.75$   & $202.0\pm40.0$ & \cite{joan_2005} \\
        $0.1791$ & $74.91$        & \cite{moresco_2020} \\
        $0.1993$ & $74.96$        & \cite{moresco_2020} \\
        $0.3519$ & $82.78$        & \cite{moresco_2020} \\
        $0.3802$ & $83.0$         & \cite{moresco_2020} \\
        $0.4004$ & $76.97$        & \cite{moresco_2020} \\
        $0.4247$ & $87.08$        & \cite{moresco_2020} \\
        $0.4497$ & $92.78$        & \cite{moresco_2020} \\
        $0.4783$ & $80.91$        & \cite{moresco_2020} \\
        $0.5929$ & $103.8$        & \cite{moresco_2020} \\
        $0.6797$ & $91.6$         & \cite{moresco_2020} \\
        $0.7812$ & $104.5$        & \cite{moresco_2020} \\
        $0.8754$ & $125.1$        & \cite{moresco_2020} \\
        $1.037$  & $153.7$        & \cite{moresco_2020} \\
        $1.363$  & $160.0$        & \cite{moresco_2020} \\
        $1.965$  & $186.5$        & \cite{moresco_2020} \\
        \hline
    \end{tabular}
\end{table}

\subsection{Quasars}
\label{sec:3.4}

\citet{cao_2022} established that \civ~quasar (QSO) data are standardizable through the \civ R-L relation since the parameters of this relation are independent of any cosmological model. It was also shown that \civ~QSO data are consistent with $H(z) +$BAO data. We use 38 high-quality \civ~ reverberation mapped (RM) QSOs. These 38 \civ RM sources were compiled by \cite{kaspi_2021} with significantly measured time-delays. It covers a redshift range of $0.001064 \le z \le 3.368$.  The sources, their redshifts, flux density at 1350 \AA, monochromatic luminosity at 1350 \AA and the rest-frame \civ~time-lag $\tau$ have been listed in Table~\ref{table3}. 

\subsection{Megamasers}
\label{sec:3.5}

Water megamasers in accretion disks around supermassive black holes (SMBH) are a unique way of determining geometric distances to their host galaxies without referring to the cosmic distance ladder. The Megamaser Cosmology Project \cite{Pesce_2020, Pesce_H0_2020, reid_2019} is a multi-year campaign to survey active galactic nuclei (AGN) for the presence of water megamasers, monitor their spectral evolution and map their structures using very long baseline interferometry (VLBI). 

In this work, as an alternative to cosmic chronometers (as described in Section \ref{sec:5}), we use Megamasers (hereafter referred to as MM) to obtain constraints on $H_0$. Following the method of \cite{Pesce_H0_2020,Shubham2025}, we take the peculiar velocities of the galaxies into account by considering $\sigma_\text{pec} = 250$km s$^{-1}$ into the velocity uncertainties. The expected velocities($v_i$) of the six Megamaser host galaxies are taken as nuisance parameters. The total likelihood is given by
\begin{equation}
\label{eq11}
    \mathcal{L}_\text{tot} = \mathcal{L}_\text{v}\mathcal{L}_\text{D},
\end{equation}
where 
\begin{equation}
\label{eq12}
    \mathcal{L}_\text{v} = \Pi_i \frac{1}{\sqrt{2\pi(\sigma_\text{v,i}^2 + \sigma_\text{pec}^2)}}\text{exp}\left[-\frac{1}{2}\frac{(v_i - \hat{v_i})^2}{\sigma_\text{v,i}^2 + \sigma_\text{pec}^2}\right]  \text{and}
\end{equation}
\begin{equation}
\label{eq13}
    \mathcal{L}_\text{v} = \Pi_i \frac{1}{\sqrt{2\pi\sigma_\text{D,i}^2}}\text{exp}\left[-\frac{1}{2}\frac{(D_{A,i} - \hat{D}_{A,i})^2}{\sigma_\text{$D_{A,i}$}^2}\right].
\end{equation}
Here, $D_A$ is the angular diameter distance to each galaxy, $\hat{v}$ and $\hat{D}_A$ represent the measured velocities and angular distances, respectively. The quantities without a hat are the theoretically measured quantities.

\begin{table}[htbp!]
\caption{\civ  QSO data \cite{cao_2022}}
\label{table3}
\centering
    \begin{tabular}{|c|c|c|c|c|c|}
        \hline
        \thead{Object} & \thead{\boldmath$z$} & \thead{\boldmath log(F$_{1350}$/erg s$^{-1}$ cm$^{-2}$)} & \thead{\boldmath log(L$_{1350}$/erg s$^{-1}$)} & \thead{\boldmath $\tau$(days)} & \thead{Reference}\\
        \hline
        NGC 4395     & 0.001064 & $-11.4848\pm0.0272$ & $39.9112\pm0.0272$ & $0.040^{+0.024}_{-0.018}$ & \cite{peterson_2005,peterson_2006} \\
        NGC 3783     & 0.00973  & $-9.7341\pm0.0918$  & $43.5899\pm0.0918$ & $3.80^{+1}_{-0.9}$        
        & \cite{peterson_2005,peterson_2006} \\
        NGC 7469     & 0.01632  & $-9.9973\pm0.0712$  & $43.7803\pm0.0712$ & $2.5^{+0.3}_{-0.2}$
        & \cite{peterson_2005,peterson_2006} \\
        3C 390.3     & 0.0561   & $-10.8036\pm0.2386$ & $44.0719\pm0.2386$ & $35.7^{+11.4}_{-14.6}$
        & \cite{peterson_2005,peterson_2006} \\
        NGC 4151     & 0.00332  & $-9.7544\pm0.1329$  & $42.6314\pm0.1329$ & $3.34^{+0.82}_{-0.77}$
        & \cite{metzroth_2006} \\
        NGC 5548     & 0.01676  & $-10.2111\pm0.0894$ & $43.5899\pm0.0894$ & $4.53^{+0.35}_{-0.34}$
        & \cite{deRosa_2015} \\
        CTS 286      & 2.551    & $-11.6705\pm0.0719$ & $47.0477\pm0.0719$ & $459^{+71}_{-92}$
        & \cite{lira_2018} \\
        CTS 406      & 3.178    & $-12.0382\pm0.0402$ & $46.9101\pm0.0402$ & $98^{+55}_{-74}$ 
        & \cite{lira_2018} \\
        CTS 564      & 2.653    & $-11.7615\pm0.0664$ & $46.9978\pm0.0664$ & $115^{+184}_{-29}$
        & \cite{lira_2018} \\
        CTS 650      & 2.659    & $-11.8815\pm0.1068$ & $46.8802\pm0.1068$ & $162^{+33}_{-10}$
        & \cite{lira_2018} \\
        CTS 953      & 2.526    & $-11.7082\pm0.0868$ & $46.9996\pm0.0868$ & $73^{+115}_{-58}$
        & \cite{lira_2018} \\
        CTS 1061     & 3.368    & $-11.4788\pm0.0405$ & $47.5299\pm0.0405$ & $91^{+111}_{-24}$
        & \cite{lira_2018} \\
        J 214355     & 2.607    & $-11.7786\pm0.0485$ & $46.9624\pm0.04485$& $136^{+100}_{-90}$
        & \cite{lira_2018} \\
        J 221516     & 2.709    & $-11.6263\pm0.0569$ & $47.1550\pm0.0569$ & $153^{+91}_{-12}$
        & \cite{lira_2018} \\
        DES J0228-04 & 1.905    & $-11.9791\pm0.0405$ & $46.4298\pm0.0405$ & $123^{+43}_{-42}$
        & \cite{Hoormann_2019} \\
        DES J0033-42 & 2.593    & $-12.2248\pm0.0201$ & $46.5105\pm0.0201$ & $95^{+16}_{-23}$
        & \cite{Hoormann_2019} \\
        RMID 032     & 1.715    & $-13.8040\pm0.0210$ & $44.4928\pm0.0210$ & $21.1^{+22.7}_{-8.3}$
        & \cite{grier_2019} \\
        RMID 052     & 2.305    & $-13.1121\pm0.0021$ & $45.4990\pm0.0021$ & $32.6^{+6.9}_{-2.1}$
        & \cite{grier_2019} \\
        RMID 181     & 1.675    & $-13.7265\pm0.0149$ & $44.5451\pm0.0149$ & $102.1^{+26.8}_{-10}$
        & \cite{grier_2019} \\
        RMID 249     & 1.717    & $-13.3140\pm0.0099$ & $44.9841\pm0.0099$ & $22.8^{+31.3}_{-11.5}$
        & \cite{grier_2019} \\
        RMID 256     & 2.244    & $-13.4939\pm0.0030$ & $45.0888\pm0.0030$ & $43.1^{+49}_{-15.1}$
        & \cite{grier_2019} \\
        RMID 275     & 1.577    & $-12.5961\pm0.0010$ & $45.6110\pm0.0010$ & $76.7^{+10}_{-3.9}$
        & \cite{grier_2019} \\
        RMID 298     & 1.635    & $-12.6497\pm0.0010$ & $45.5960\pm0.0010$ & $82.3^{+64.5}_{-24.5}$
        & \cite{grier_2019} \\
        RMID 312     & 1.924    & $-13.3424\pm0.0040$ & $45.0770\pm0.0040$ & $70.9^{+9.6}_{-3.3}$
        & \cite{grier_2019} \\
        RMID 332     & 2.581    & $-13.1795\pm0.0020$ & $45.5510\pm0.0020$ & $83.8^{+23.3}_{-6.5}$
        & \cite{grier_2019} \\
        RMID 387     & 2.426    & $-12.9782\pm0.0010$ & $45.6870\pm0.0010$ & $48.4^{+34.7}_{-10.1}$
        & \cite{grier_2019} \\
        RMID 401     & 1.822    & $-12.8714\pm0.0030$ & $45.4900\pm0.0030$ & $60.6^{+36.7}_{-13.0}$
        & \cite{grier_2019} \\
        RMID 418     & 1.418    & $-13.0533\pm0.0030$ & $45.0398\pm0.0030$ & $58.6^{+51.6}_{-21.3}$
        & \cite{grier_2019} \\
        RMID 470     & 1.879    & $-13.5732\pm0.0060$ & $44.8210\pm0.0060$ & $27.4^{+63.5}_{-22.0}$
        & \cite{grier_2019} \\
        RMID 527     & 1.647    & $-13.4655\pm0.0030$ & $44.7880\pm0.0030$ & $47.3^{13.3}_{-5}$
        & \cite{grier_2019} \\
        RMID 549     & 2.275    & $-13.2283\pm0.0020$ & $45.3690\pm0.0020$ & $68.9^{+31.6}_{-9.6}$
        & \cite{grier_2019} \\
        RMID 734     & 2.332    & $-13.0935\pm0.0010$ & $45.5299\pm0.0010$ & $68^{+38.2}_{-11.5}$
        & \cite{grier_2019} \\
        RMID 363     & 2.635    & $-12.2525\pm0.0206$ & $46.4997\pm0.0206$ & $300.4^{+17.1}_{-4.7}$
        & \cite{shen_2019} \\
        RMID 372     & 1.745    & $-12.6952\pm0.0198$ & $45.6201\pm0.0198$ & $67^{+20.4}_{-7.4}$
        & \cite{shen_2019} \\
        RMID 651     & 1.486    & $-12.7234\pm0.0198$ & $45.4200\pm0.0198$ & $91.7^{+56.3}_{-22.7}$
        & \cite{shen_2019} \\
        S5 0836+71   & 2.172    & $-11.5354\pm0.0680$ & $47.0128\pm0.0680$ & $230^{+91}_{-59}$
        & \cite{kaspi_2021} \\
        SBS 1116+603 & 2.646    & $-11.5013\pm0.0485$ & $47.2553\pm0.0485$ & $65^{+17}_{-37}$
        & \cite{kaspi_2021} \\
        SBS 1425+606 & 3.192    & $-11.2978\pm0.0356$ & $47.6551\pm0.0356$ & $285^{+30}_{-53}$
        & \cite{kaspi_2021} \\
        \hline
    \end{tabular}
\end{table}

\section{Cosmological Relations}
\label{sec:4}

In this section, we state the relations used in this work. Constraints using SNe Ia are based on measurements of the luminosity distance ($D_L$), which is given by: 
\begin{equation}
\label{eq14}
    D_L(z)= 
\begin{cases}
    \frac{c(1+z)}{H_0\sqrt{|\Omega_k|}}\text{sinh}\left[\frac{H_0\sqrt{|\Omega_k|}}{c}D_C(z)\right],  & \text{if } \Omega_k > 0\\
    (1+z)D_C(z),    & \text{if } \Omega_k = 0\\
    \frac{c(1+z)}{H_0\sqrt{|\Omega_k|}}\text{sin}\left[\frac{H_0\sqrt{|\Omega_k|}}{c}D_C(z)\right],  & \text{if } \Omega_k < 0,\\
\end{cases}
\end{equation}
where 
\begin{equation}
\label{eq15}
    D_C(z) = c\int_0^z\frac{dz'}{H(z')}
\end{equation}
is the comoving distance.
The theoretical distance modulus is given by
\begin{equation}
\label{eq16}
    \mu_\text{th}(z, \boldsymbol{\theta}) = 5\text{log}\left[\frac{D_L(z)}{\text{Mpc}}\right] + 25,
\end{equation}
while $\mu_{\text{obs}}$ is given by
\begin{equation}
\label{eq17}
    \mu_\text{obs} = m_{B}^\text{corr} - M,
\end{equation}
where $m_{B}^\text{corr}$ is the corrected apparent magnitude, $M$ is the peak absolute $B$-band magnitude and $\theta$ represents the model parameters. The residual vector can be constructed as $\Delta\mu_i = \mu_\text{obs,i} - \mu_{th}(z_i;\boldsymbol{\theta})$. Using this, the likelihood is given by 
\begin{equation}
\label{eq18}
    \mathcal{L}(\boldsymbol{\theta}) = \frac{1}{\sqrt{(2\pi)^N|C_\text{SNe}|}}e^{\left(-\frac{1}{2}\chi^2\right)},
\end{equation}
where $|C_\text{SNe}|$ denotes the determinant of the covariance matrix, N is the no. of data points and $\chi^2 = \Delta\mu^TC^{-1}_\text{SNe}\Delta\mu$.

The DESI collaboration has provided  the values of $D_M/r_d$, $D_H/r_d$ and $D_V/r_d$. By measuring the redshift interval $\Delta z$ along the line-of-sight, we get an estimate of the Hubble distance at redshift $z$:
\begin{equation}
\label{eq19}
    D_H(z) = \frac{c}{H(z)}.
\end{equation}
The comoving angular diameter distance \cite{augbourg_2015} can be found by measuring the angle $\Delta \theta$ subtended by the BAO feature at redshift $z$, along the transverse direction and is given by 
\begin{equation}
\label{eq20}
    D_M(z) = (1+z)D_A(z).
\end{equation}
Here, $D_A(z)$ is the angular diameter distance \cite{seo_2003} and is related to $D_L(z)$ using the famous Etherington relation \cite{bora_2021}
\begin{equation}
\label{eq21}
    D_L(z) = (1+z)^2D_A(z).
\end{equation}
This assumes photon conservation, propagation along null geodesics and a metric theory of gravity. The measurements of the spherically averaged distance ($D_V$) \cite{eisenstein_2005} is related to $D_M$ and $D_H$ by
\begin{equation}
\label{eq22}
    D_V(z) = \left[zD_M(z)^2D_H(z)\right]^{1/3}.
\end{equation}
We constrain the three distance measurements and apply the proper normalization factor: 
\begin{itemize}
    \item Isotropic measurements ($D_V(z)$) - $\sum_{i}\text{ln}(2\pi\sigma_{D_V}^2)$.
    \item Anisotropic measurements ($D_H(z)$ and $D_M(z)$) - $\frac{1}{\sqrt{(2\pi)^N|C_\text{BAO}|}}$.
\end{itemize}
In the above expression, $N$ is the number of points in the anisotropic measurements and $C_\text{BAO}$ is the covariance matrix which can be constructed from the uncertainties and the correlation factor.

For cosmic chronometers, the normalization factor is given by $\frac{1}{\sqrt{(2\pi)^N|C_\text{CC}|}}$ where again $N$ is the total number of data points (32 in this case) and $C_\text{CC}$ is the total covariance matrix. For a full discussion on the likelihood construction please refer to \cite{moresco_2020, moresco_2023}.

The R-L relation parameterization utilized here can be found in \cite{cao_2022} where, as in \cite{bentz_2013} the monochromatic luminosity and the rest frame time-delay have been replaced taking into consideration the \civ-region emission properties.
\begin{equation}
\label{eq23}
    \text{log}\frac{\tau}{\text{days}} = \beta_c + \gamma_c\text{log}\frac{L_{1350}}{10^{44}\text{erg s}^{-1}},
\end{equation}
where $\tau, \beta_c$, and $\gamma_c$ are the \civ~time-lag, intercept, and the slope parameters, respectively. The monochromatic luminosity at 1350 \AA~is given by: 
\begin{equation}
\label{eq24}
    L_{1350} = 4\pi D_L^2F_{1350}
\end{equation}
where $F_{1350}$ denotes the measured quasar flux at 1350 \AA in erg s$^{-1}$ cm$^{-2}$ units. 

The likelihood for this data is given by: 
\begin{equation}
\label{eq25}
    \text{ln}\mathcal{L} = -\frac{1}{2}\left[\chi^2 + \sum_{i=1}^{N}\text{ln}(2\pi \sigma_\text{tot,i}^2)\right],
\end{equation}
where 
\begin{equation}
\label{eq26}
    \chi^2 = \sum_{i=1}^N\left[\frac{(\text{log}(\tau_\text{obs,i})-\beta_c-\gamma_c\text{log}(L_\text{1350,i}))^2}{\sigma_\text{tot,i}^2}\right]
\end{equation}
with the uncertainty given by: 
\begin{equation}
\label{eq27}
    \sigma_\text{tot,i}^2 = \sigma_\text{int}^2 + \sigma_{\text{log}\tau_{obs,i}}^2 + \gamma_c^2\sigma_{\text{log}F_{1350,i}}^2,
\end{equation}
where $\sigma_\text{int}$ is the \civ~QSO intrinsic scatter parameter containing the unknown systematic uncertainty and $N$ denotes the number of data points. 

Finally, the total likelihood is given by 
\begin{equation}
\mathcal{L}_\text{tot} = \prod_i \mathcal{L}_i
\label{eq:totalL}
\end{equation}
where $i$ represents the Type Ia SNe, DESI BAO, QSO and CC or MM datasets.

\section{Methodology}
\label{sec:5}
We use Bayesian inference for parameter estimation and model comparison.  We provide a very brief primer on Bayesian inference, and more details can be found in ~\cite{trotta_2017,Kerscher,Krishak} and references therein.

The Bayesian evidence for a particular model is given by $\int \mathcal{L}(D|M,\theta)P(\theta|M)d\theta$ where $\mathcal{L}(D|M,\theta)$ is the likelihood for model $M$ given data $D$ (cf. Eq~\ref{eq:totalL}), and parameter vector ($\theta$), while $P(\theta|M)$ is the prior on the parameter vector $\theta$ given a model $M$. 
We calculate Bayes' Factor  by calculating the ratio of  the Bayesian evidence value for the two models: ($B_{21} = \frac{B_2}{B_1}$ where $B_{21}$ is the Bayes' factor, $B_i$ is the Bayesian evidence for model $i$). To evaluate the significance, we  use \rthis{Kaas and Raftery's} scale~\cite{trotta_2017} to determine the preference for flat or non-flat scenarios for each parameterization.
For the aformentioned calculations of Bayesian evidence and parameter estimations,  we use the {\tt Nautilus}~\cite{nautilus} software, which is based on the Importance nested sampling algorithm.

For each model, our analysis is divided into two parts. First we consider flat cosmologies, wherein $\Omega_k = 0$ in Eqns.~\ref{eq4}, \ref{eq5}, \ref{eq6}, \ref{eq7}, \ref{eq8} and~\ref{eq9} (when we use $w(z)$ to evaluate Eqn.~\ref{eq3} and subsequently, Eqn.~\ref{eq2}). Secondly, we allow $\Omega_k$ to be a free parameter \cite{valentino_2020_omegak, park_2019}. Next, we analyze both flat and non-flat cases by either including or excluding LRG1 and LRG2 observations from the DESI measurements.

Each of these divisions is further categorized based on prior choices:
\begin{itemize}
    \item Uniform priors on $M\in\mathcal{U}(-21, -18)$ and $r_d\in\mathcal{U}(0, 200)$.
    \item Gaussian prior on $M\in\mathcal{N}(-19.253, 0.027)$ and uniform prior on $r_d\in\mathcal{U}(0, 200)$. This specific prior on $M$ is chosen to ensure our analysis remains based solely on late Universe data. The value is derived from the Cepheid calibration of Type Ia supernovae, as determined by SH0ES observations \cite{riess_2022}.
\end{itemize}
The priors applied to each parameter can be found in Table~\ref{table4}. The overall analysis is segregated into two parts - one where we use Cosmic Chronometer data to constrain $H_0$ and the other where we replace Cosmic Chronometers with megamasers. Note that in the case of QSO and megamasers, the observations - $\tau$ and $D_A$, respectively contain asymmetrical errors. For the purpose of this work, we symmetrize the error bars using the formula \cite{cao_2022}:
\begin{equation}
\label{eq29}
    \sigma_x = \frac{1}{2}\left[2\frac{\sigma_{x,+}\sigma_{x,-}}{\sigma_{x,+}+\sigma_{x,-}}+\sqrt{\sigma_{x,+}\sigma_{x,-}}\right],
\end{equation}
where $x$ is the observable and $+$ and $-$ denote the upper and lower error bars, respectively. 

\begin{table}[htbp!]
\caption{Priors imposed on the parameters.}
\label{table4}
\centering
    \begin{tabular}{|c|c|c|}
        \hline
        \thead{Parameter} & \thead{Prior Type} & \thead{Priors}\\
        \hline
        $H_0$          & Uniform   & $\mathcal{U}(50, 200)$ \\
        $\Omega_m$     & Uniform   & $\mathcal{U}(0, 1)$ \\
        $\Omega_k$     & Uniform   & $\mathcal{U}(-0.5, 0.5)$ \\
        $w_0$          & Uniform   & $\mathcal{U}(-3, 2)$ \\
        $w_a$          & Uniform   & $\mathcal{U}(-3, 2)$ \\
        $M$            & Gaussian  & $\mathcal{N}(-19.253, 0.027)$ \\
        $M$            & Uniform   & $\mathcal{U}(-21, -18)$ \\
        $r_d$          & Uniform   & $\mathcal{U}(0, 200)$ \\
        $\beta_c$      & Uniform   & $\mathcal{U}(0, 5)$ \\
        $\gamma_c$     & Uniform   & $\mathcal{U}(0, 10)$ \\
        $\sigma_{int}$ & Uniform   & $\mathcal{U}(0, 5)$ \\
        $z_T$          & Uniform   & $\mathcal{U}(0, 10)$ \\
        $\Delta_z$     & Uniform   & $\mathcal{U}(0, 10)$ \\
        \hline
    \end{tabular}
\end{table}



 The results for comparisons using Bayes' factor can be found in Tables~\ref{table21}, \ref{table22}, \ref{table23} and \ref{table24}.
Henceforth, we shall refer to the BAO$+$SNe Ia$+$QSO dataset combination as ``Base''. When incorporating the CC or MM dataset, the total combination shall be denoted as "Base$+$(CC/MM)".

\section{Results}
\label{sec:results}

This section begins with a generic discussion of all our results, followed by an in-depth analysis of model-dependent results. Finally, we compare our results with recent similar studies. The relevant figures are Figs.~\ref{fig1}-\ref{fig12}. The constraints on each parameterization for different dataset and prior combinations are presented in Tables~\ref{table5}-\ref{table20}. For ease of reference, the relevant sections are provided to guide the reader:
\begin{itemize}
    \item The general results are provided in Sect.~\ref{gen_results}.
     \item Results for  $\Lambda$CDM model (Eqn.~\ref{eq4}) are presented in Sect.~\ref{LCDM} (cf. Fig.~\ref{fig1} and \ref{fig2}).
    \item Findings for the CPL parameterization (Eqn.~\ref{eq5}) are discussed in Sect.~\ref{CPL} (cf. Fig.~\ref{fig3} and \ref{fig4}).
    \item Results for the BA parameterization (Eqn.~\ref{eq6}) are mentioned in Sect.~\ref{BA} (cf. Fig.~\ref{fig5} and Sect.~\ref{fig6}).
    \item The analysis of  JBP parameterization (Eqn.~\ref{eq7}) is given in Sect.~\ref{JBP} (cf. Figs.~\ref{fig7} and \ref{fig8}).
    \item Discussion of EXP parameterization (Eqn.~\ref{eq8}) is done in Sect.~\ref{exp} (cf. Fig.~\ref{fig9} and \ref{fig10}).
    \item Results for TDE parametrization (Eqn.~\ref{eq9}) are given in Sect.~\ref{tde} (cf. Fig.~\ref{fig11} and \ref{fig12})
    \item We compare our results to recent works in Sect.~\ref{comparison}.
\end{itemize}
 
\subsection{General Results}
\label{gen_results}
In this subsection, we note down the values of the nuisance parameters and any trends we find for all the dataset combinations and priors across all DE parametrizations considered in this work.
\begin{itemize}
    \item We note that the CC and MM datasets help to decouple $H_0$ from the other model parameters. This can be seen from the non-correlation of $H_0$ with the other parameters (Figs.~\ref{fig1}-\ref{fig12}). This is why the values of $H_0$ are almost the same across different parameterizations and dataset combinations. 
    \item All the values of QSO nuisance parameters - $(\beta_c, \gamma_c, \sigma_\text{int})$ - are similar. This is due to the fact that these nuisance parameters are independent of the model as discussed in \cite{cao_2022}.
    \item Across all dataset and prior combinations, we find no evidence of statistically significant deviations from $\Omega_k = 0$ in any parameterization.
    \item Introduction of an additional parameter ($\Omega_k$) in a particular parameterization does not have a significant effect on the $w_0$ and $w_a$ values when compared to the flat parameterizations, irrespective of the dataset and prior combinations. This is due to very weak correlation between $\Omega_k$ and $w_0$ and $w_a$.
\end{itemize}
Next, we present  the model-specific results.

\subsection{$\Lambda$CDM Model}
\label{LCDM}

\begin{figure}
    \centering
    \subfloat[Flat $\Lambda$CDM]{\includegraphics[width=0.5\textwidth]{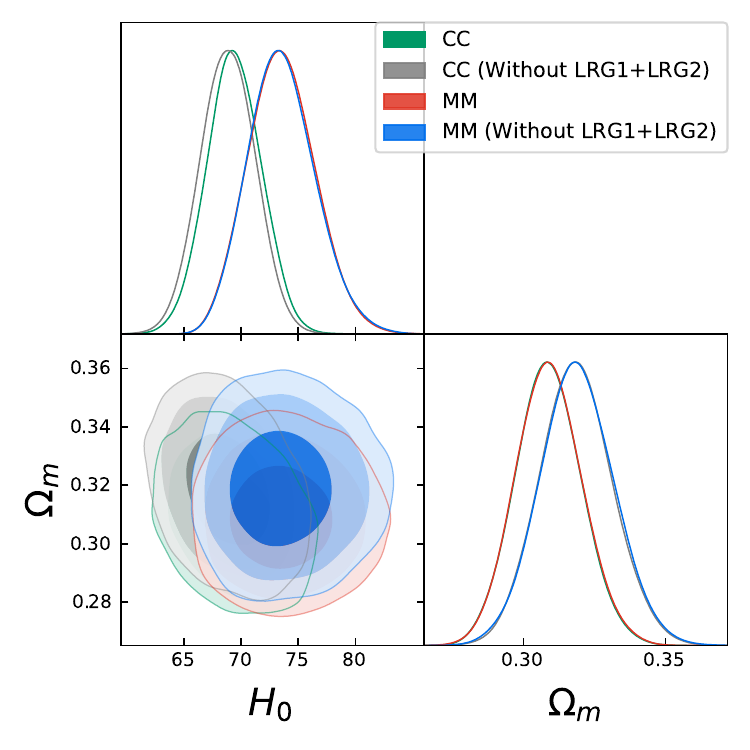}}
    \subfloat[Non-Flat $\Lambda$CDM]{\includegraphics[width=0.5\textwidth]{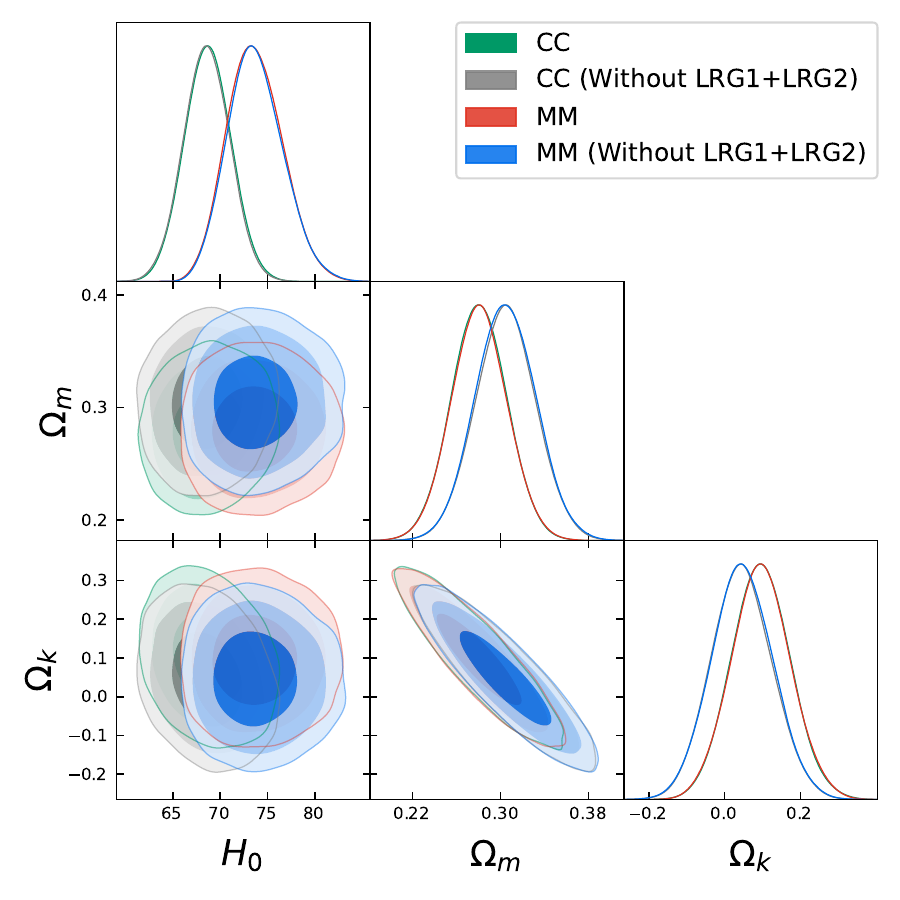}}
    \caption{Constraints on the $\Lambda$CDM model (Eqn.~\ref{eq4}) with uniform priors on $M$ and $r_d$. The left panel corresponds to the flat scenario, while the right panel represents the non-flat scenario. The contours denote the 68\%, 95\%, and 99\% credible regions.}
    \label{fig1}
\end{figure}

\begin{figure}
    \centering
    \subfloat[Flat $\Lambda$CDM]{\includegraphics[width=0.5\textwidth]{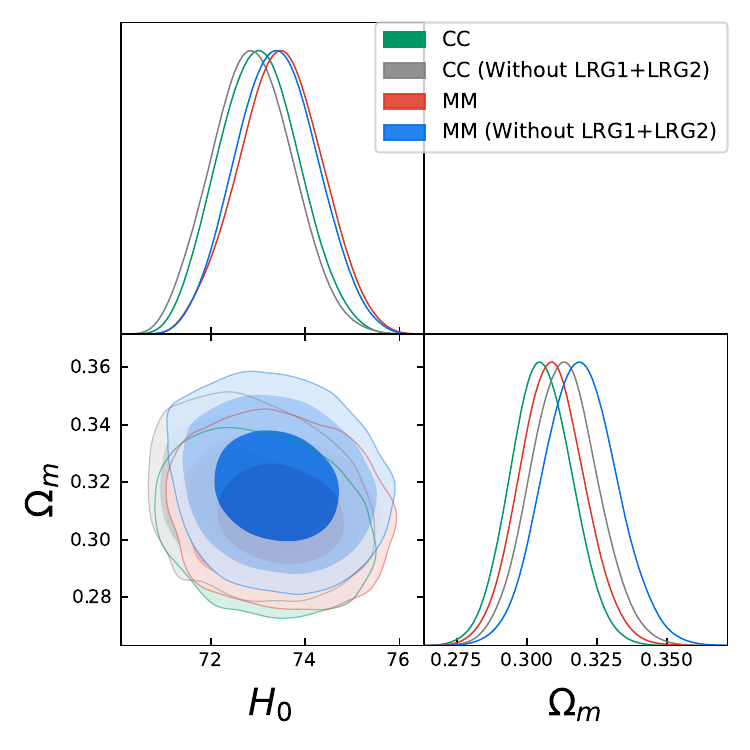}}
    \subfloat[Non-Flat $\Lambda$CDM]{\includegraphics[width=0.5\textwidth]{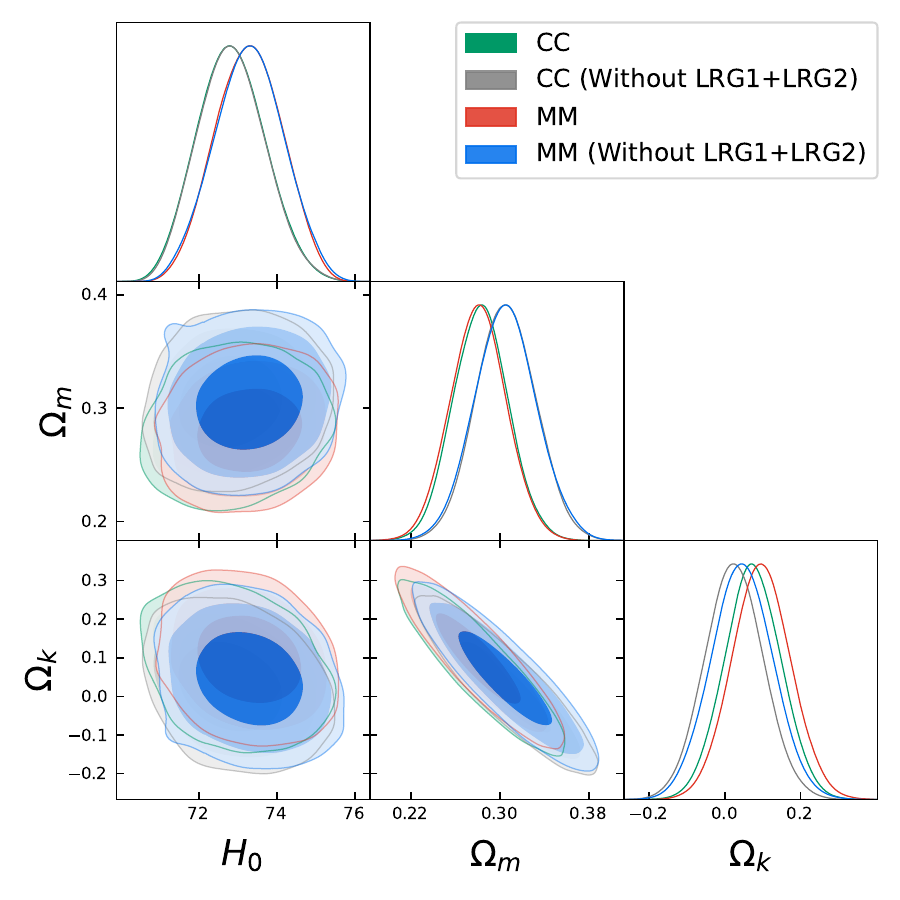}}
    \caption{Constraints on the $\Lambda$CDM model (Eqn.~\ref{eq4}) with Gaussian prior on $M$ and uniform prior on $r_d$. The left panel corresponds to the flat scenario, while the right panel represents the non-flat scenario. The contours denote the 68\%, 95\%, and 99\% credible regions.}
    \label{fig2}
\end{figure}

Refer to Figs.~\ref{fig1} and \ref{fig2}.
\begin{itemize}
    \item Using the Base$+$CC dataset, we notice that $\Lambda$CDM shows a very marginal preference for spatially flat scenario. However, when using Base$+$MM dataset, the preference for spatially flat $\Lambda$CDM model is strong for uniform priors on both $M$ and $r_d$. For Gaussian prior on $M$, the preference again becomes moderate.
    \item When comparing with different DE parameterizations, we consider the $\Lambda$CDM model as the null hypothesis and hence a Bayes’ Factor value of 1 is assigned to it (cf. Tables~\ref{table23} and \ref{table24}). Comparisons with other parameterizations will be considered in the relevant subsections. 
    \item For the  non-flat $\Lambda$CDM model, the curvature constraint remains consistent with $\Omega_k = 0$ within $1.2\sigma$ for all dataset combinations.
\end{itemize}

\subsection{CPL Parametrization}
\label{CPL}

\begin{figure}
    \centering
    \subfloat[Flat CPL]{\includegraphics[width=0.5\textwidth]{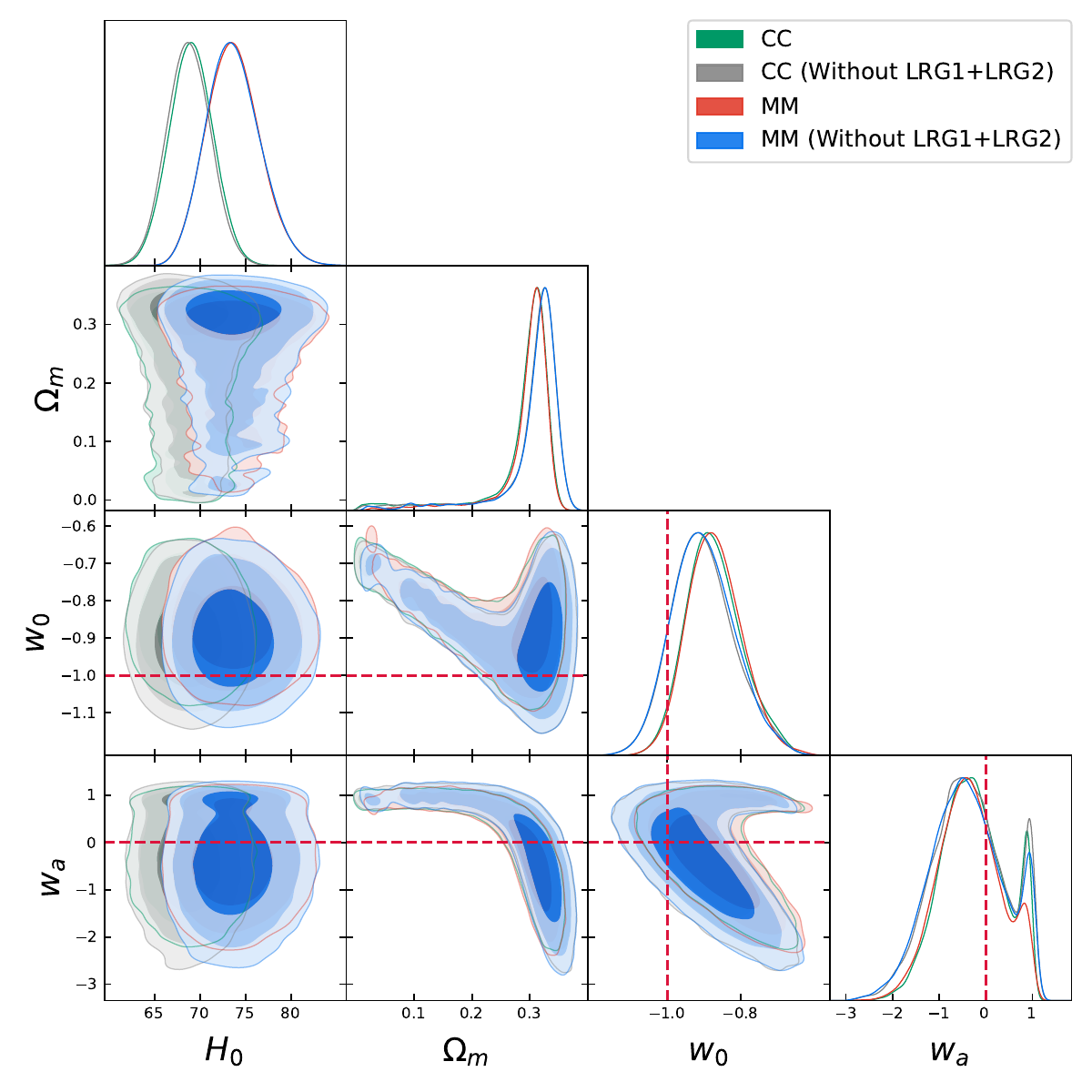}}
    \subfloat[Non-Flat CPL]{\includegraphics[width=0.5\textwidth]{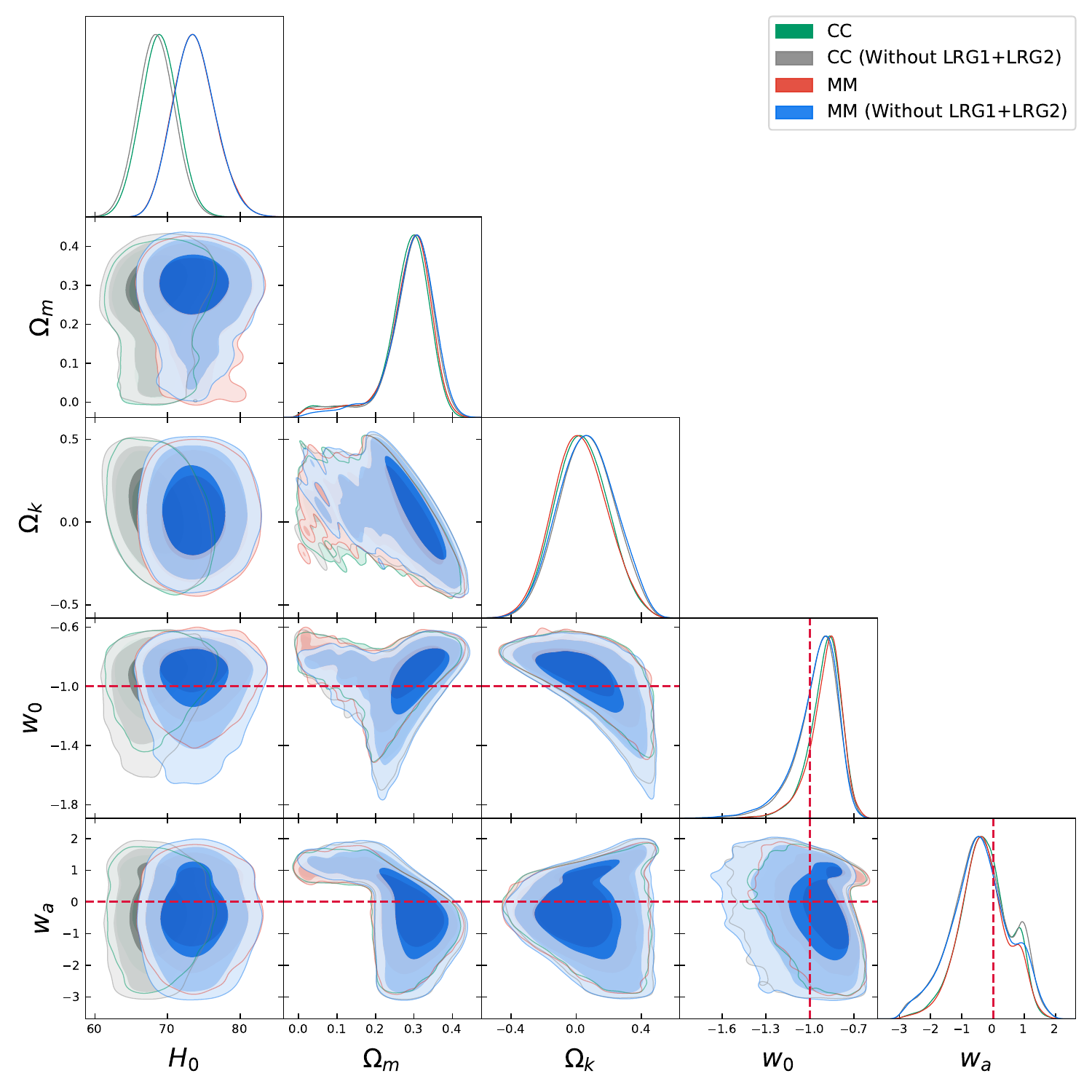}}
    \caption{Constraints on the CPL model (Eqn.~\ref{eq5}) with uniform priors on $M$ and $r_d$. The left panel corresponds to the flat scenario, while the right panel represents the non-flat scenario. The contours denote the 68\%, 95\%, and 99\% credible regions. The red dotted lines indicate the standard $\Lambda$CDM values of $w_0 = -1$ and $w_a = 0$.}
    \label{fig3}
\end{figure}

\begin{figure}
    \centering
    \subfloat[Flat CPL]{\includegraphics[width=0.5\textwidth]{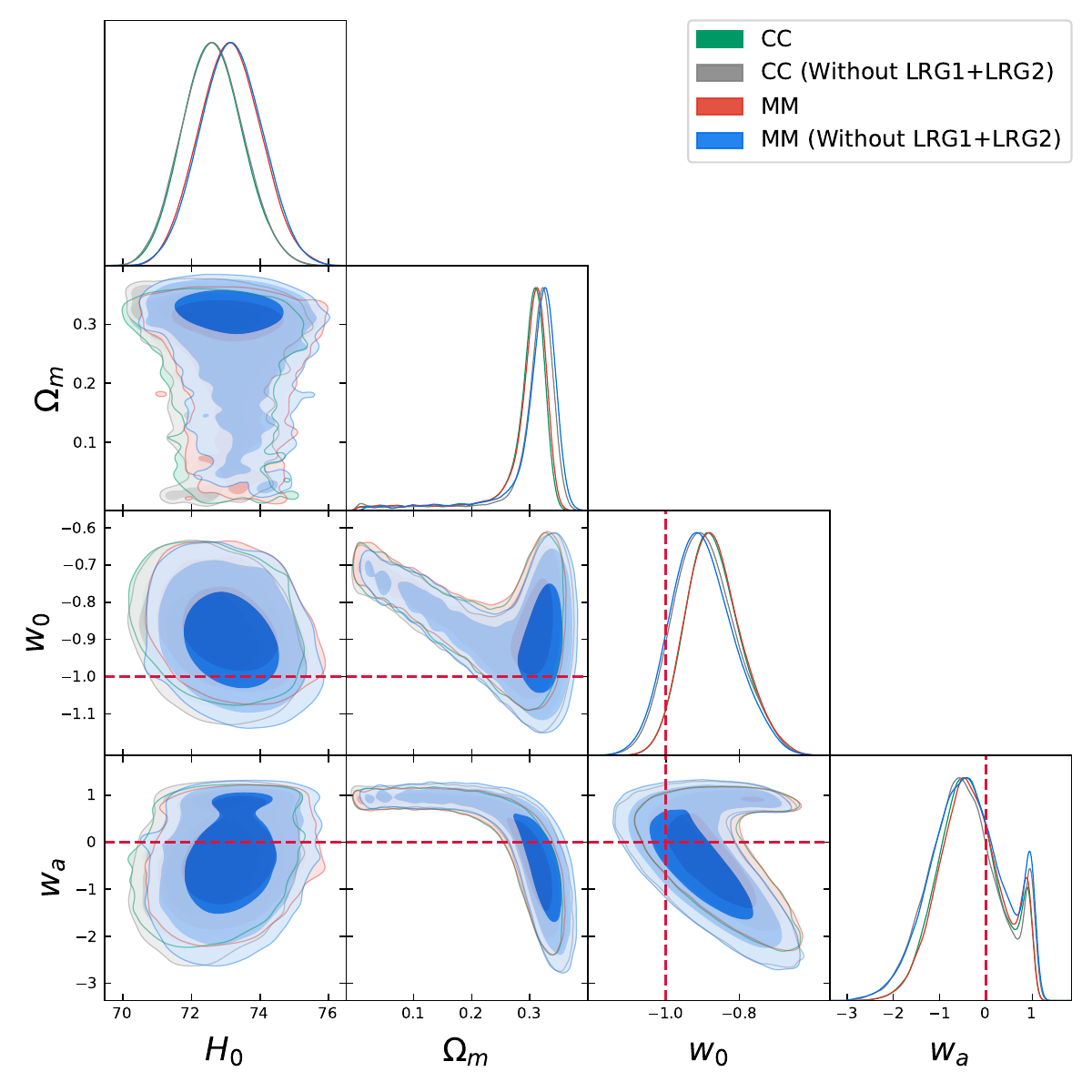}}
    \subfloat[Non-Flat CPL]{\includegraphics[width=0.5\textwidth]{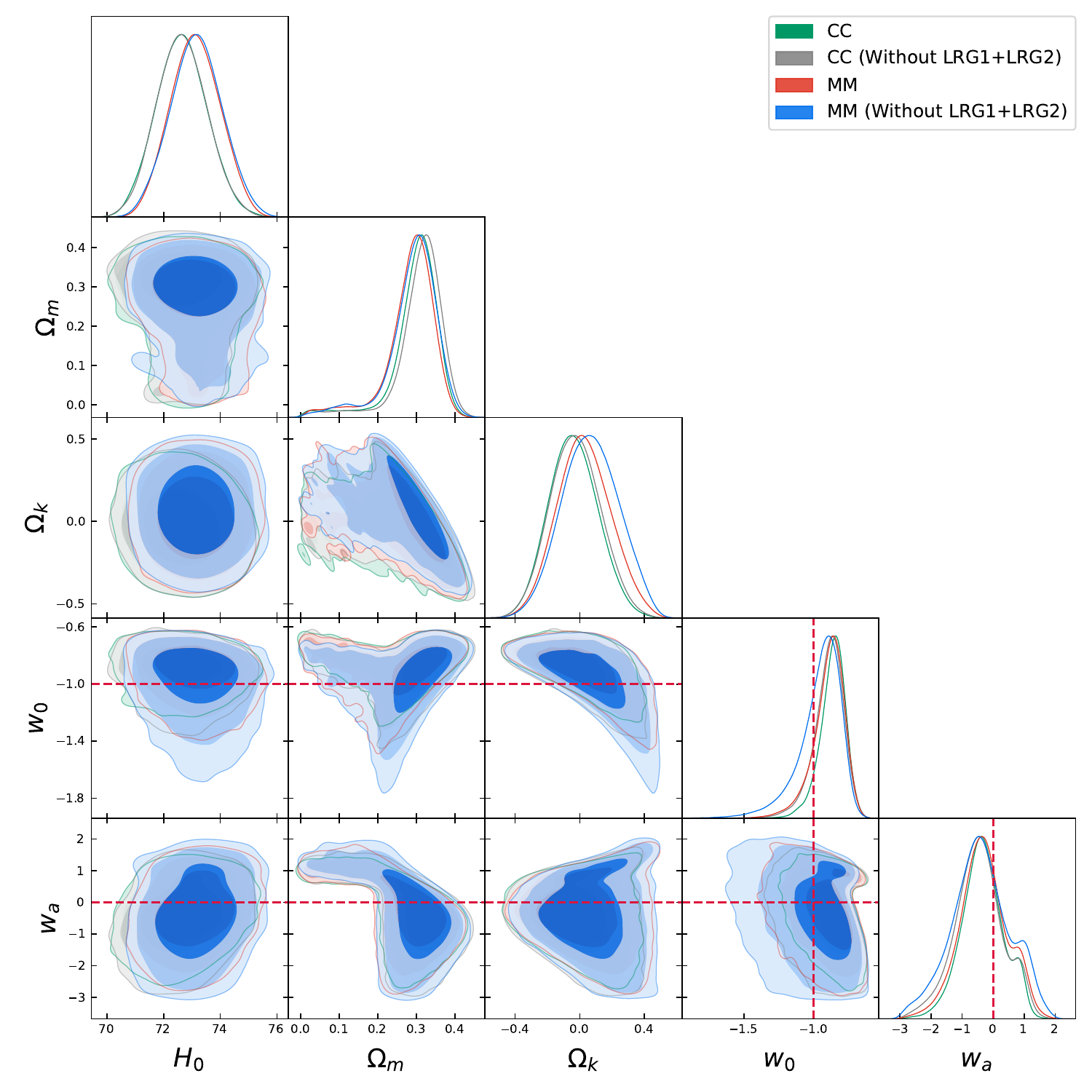}}
    \caption{Constraints on the CPL model (Eqn.~\ref{eq5}) with Gaussian prior on $M$ and uniform prior on $r_d$. The left panel corresponds to the flat scenario, while the right panel represents the non-flat scenario. The contours denote the 68\%, 95\%, and 99\% credible regions. The red dotted lines indicate the standard $\Lambda$CDM values of $w_0 = -1$ and $w_a = 0$.}
    \label{fig4}
\end{figure}

Refer to Figs.~\ref{fig3} and \ref{fig4}.
\begin{itemize}
    \item There is no preference for the flat CPL model over its non-flat counterpart across all data and prior combinations.
    \item For Base$+$CC dataset, a strong preference is shown towards $\Lambda$CDM model (both flat and non-flat) for all priors, irrespective of whether  LRG1 and LRG2 data points are included or not. In the case of Base$+$MM dataset, a very strong preference is shown for flat $\Lambda$CDM parameterization compared to flat CPL when uniform priors are applied on $M$ and $r_d$. This changes to a strong preference in case of a Gaussian prior on $M$ for flat CPL and in all prior and dataset combinations in the case of non-flat CPL.
    \item There is a marginal (1.3 - 1.8)$\sigma$ deviation of $w_0$ from its $\Lambda$CDM value of $-1$ when using Base$+$CC dataset while including and excluding LRG1 and LRG2. This occurs for both flat and non-flat scenarios. When we apply uniform priors on $M$ and $r_d$ (for non-flat CPL), the deviation is within $1\sigma$, irrespective of  whether we include LRG1 and LRG2 or not. For Base$+$MM datasets, $w_0$ values are within $1\sigma$ of -1.0
    \item $w_a$ is consistent (within $0.6\sigma$) with its $\Lambda$CDM value of 0 in all cases.
    \item There is no statistically significant prior effect observed on the $w_0$ and $w_a$ values for both Base$+$CC and Base$+$MM datasets.
\end{itemize}

\subsection{BA Parameterization}
\label{BA}

\begin{figure}
    \centering
    \subfloat[Flat BA]{\includegraphics[width=0.5\textwidth]{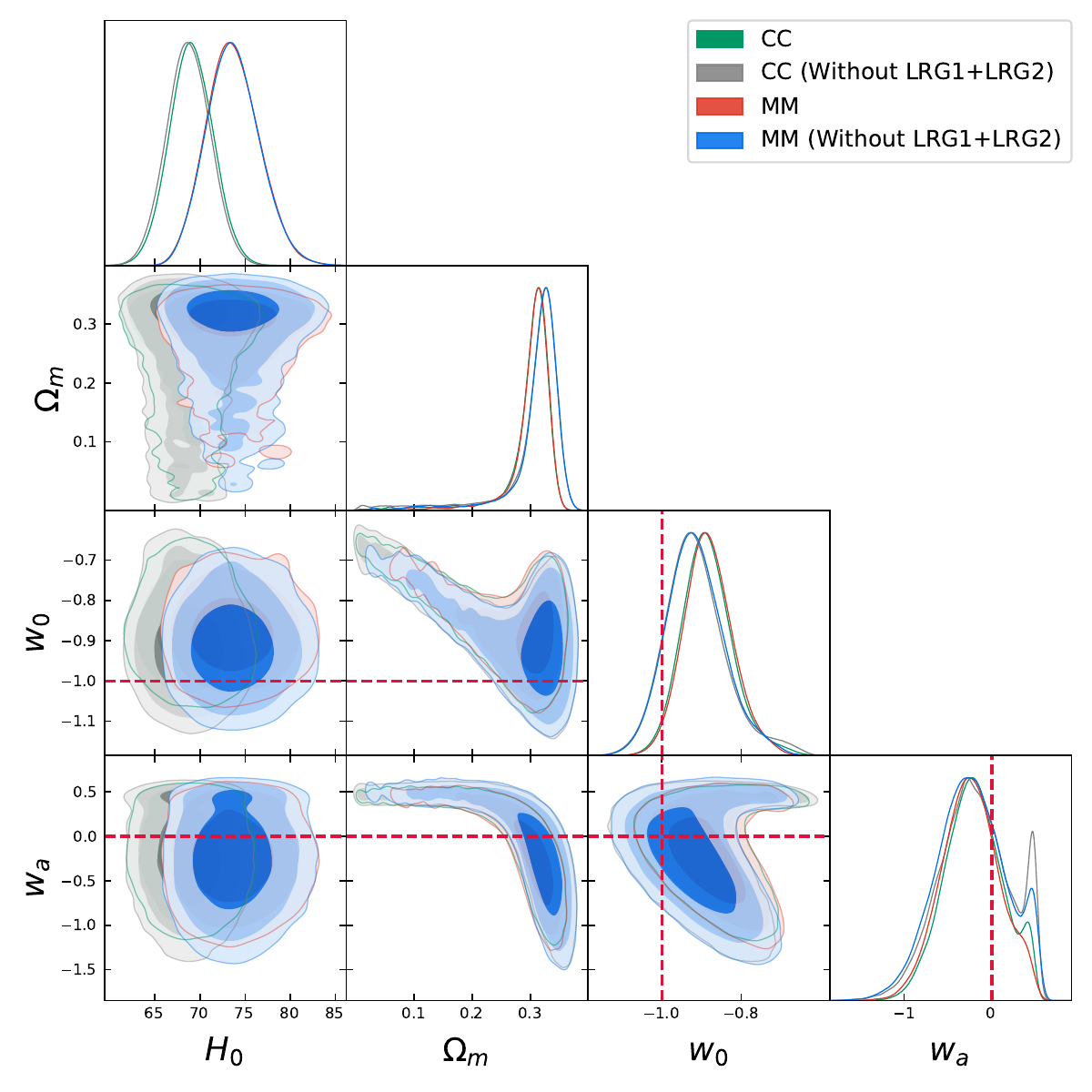}}
    \subfloat[Non-Flat BA]{\includegraphics[width=0.5\textwidth]{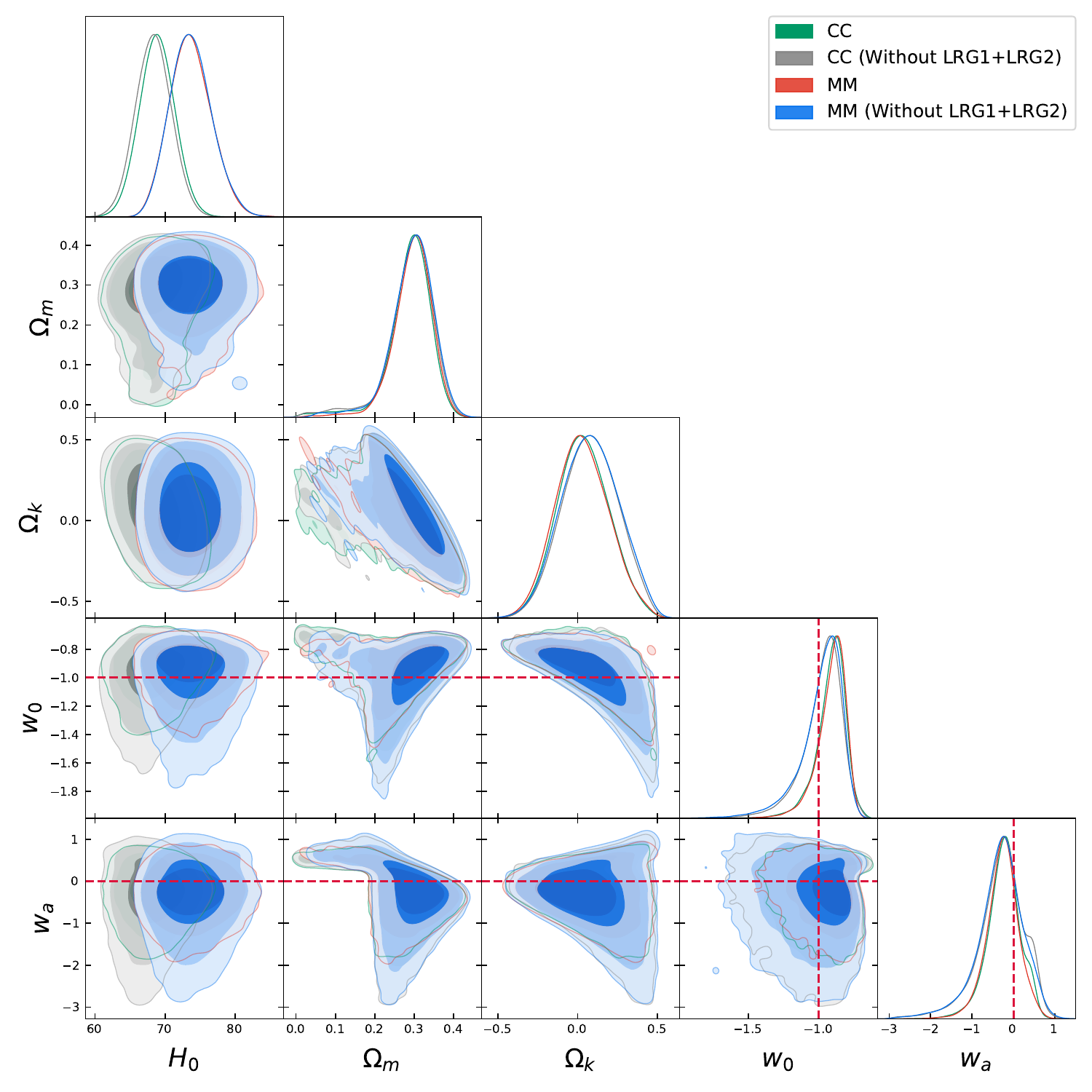}}
    \caption{Constraints on the BA model (Eqn.~\ref{eq6}) with uniform priors on $M$ and $r_d$. The left panel corresponds to the flat scenario, while the right panel represents the non-flat scenario. The contours denote the 68\%, 95\%, and 99\% credible intervals. Red dotted lines indicate the standard $\Lambda$CDM values of $w_0 = -1$ and $w_a = 0$.}
    \label{fig5}
\end{figure}

\begin{figure}
    \centering
    \subfloat[Flat BA]{\includegraphics[width=0.5\textwidth]{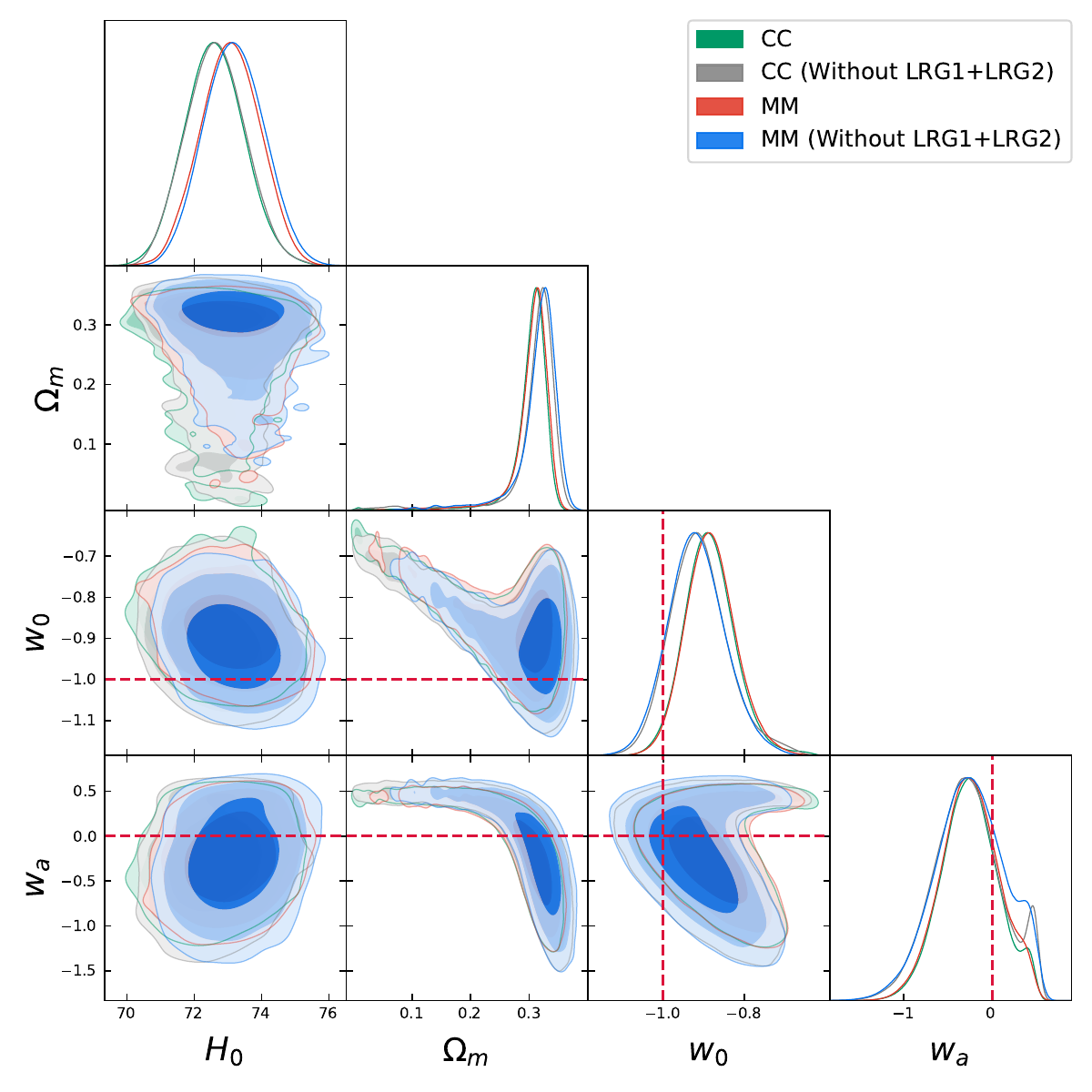}}
    \subfloat[Non-Flat BA]{\includegraphics[width=0.5\textwidth]{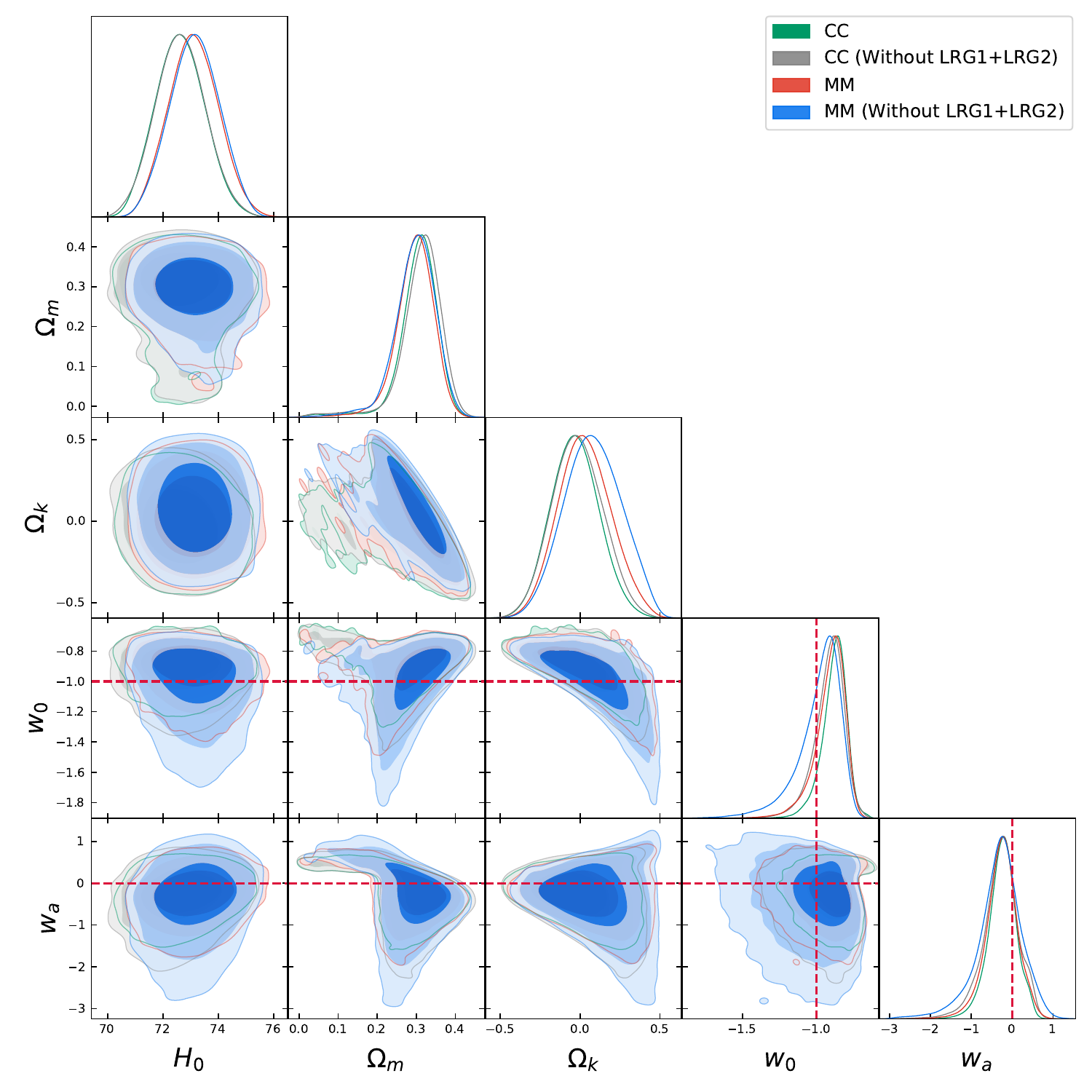}}
    \caption{Constraints on the BA model (Eqn.~\ref{eq6}) with Gaussian prior on $M$ and uniform prior on $r_d$. The left panel corresponds to the flat scenario, while the right panel represents the non-flat scenario. The contours denote the 68\%, 95\%, and 99\% credible intervals. The red dotted lines indicate the standard $\Lambda$CDM values of $w_0 = -1$ and $w_a = 0$.}
    \label{fig6}
\end{figure}
Refer to Figs.~\ref{fig5} and \ref{fig6}.
\begin{itemize}
    \item There is no preference for either the flat or the non-flat BA parameterization for all dataset and prior combinations.
    \item (Flat/Non-flat)$\Lambda$CDM model shows a strong preference compared to (Flat/Non-flat)BA parameterization for all dataset and prior combinations, except when we apply uniform priors on $M$ and $r_d$ while using Base$+$MM datasets. In this case a very strong preference  towards $\Lambda$CDM is obtained. However, this is only the case in the flat BA parameterization. 
    \item For both Base$+$CC and Base$+$MM datasets, we find a $\sim 2\sigma$ deviation of $w_0$ from its corresponding $\Lambda$CDM value of $-1$ when constraining the flat BA parameterization while including LRG1 and LRG2. For the non-flat BA parameterization, this significance reduces to $\sim 1\sigma$. When excluding the LRG1 and LRG2 data points, $w_0$ remains consistent with the value of $-1$ within $1\sigma$ for both flat and non-flat scenarios.
    \item $w_a$ is consistent within $1\sigma$ with its $\Lambda$CDM value of 0 for both flat and non-flat scenarios for all dataset and prior combinations.
    \item We do not find any statistically significant prior effect observed on the $w_0$ and $w_a$ values for both Base$+$CC and Base$+$MM datasets.
\end{itemize}

\subsection{JBP Parameterization}
\label{JBP}

\begin{figure}
    \centering
    \subfloat[Flat JBP]{\includegraphics[width=0.5\textwidth]{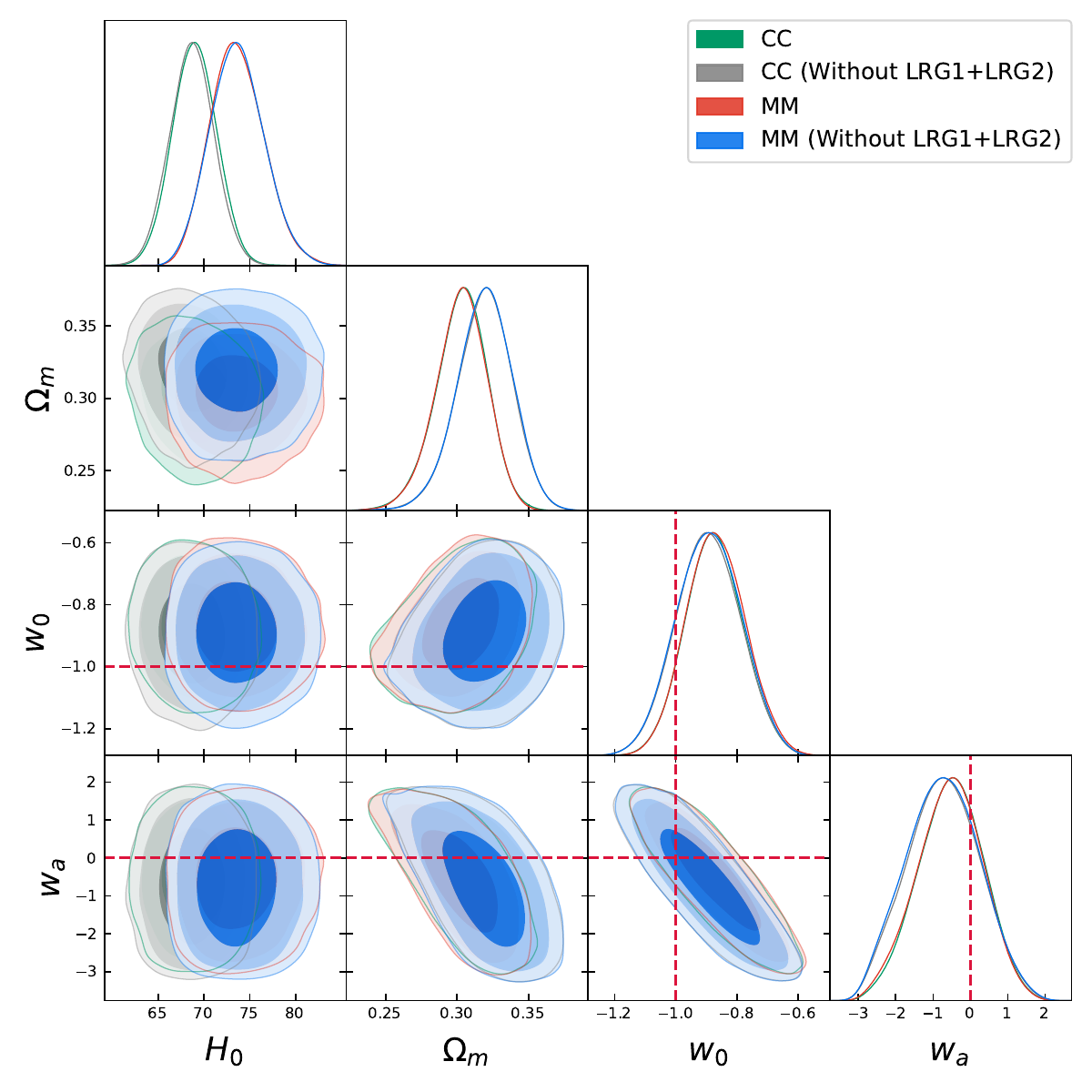}}
    \subfloat[Non-Flat JBP]{\includegraphics[width=0.5\textwidth]{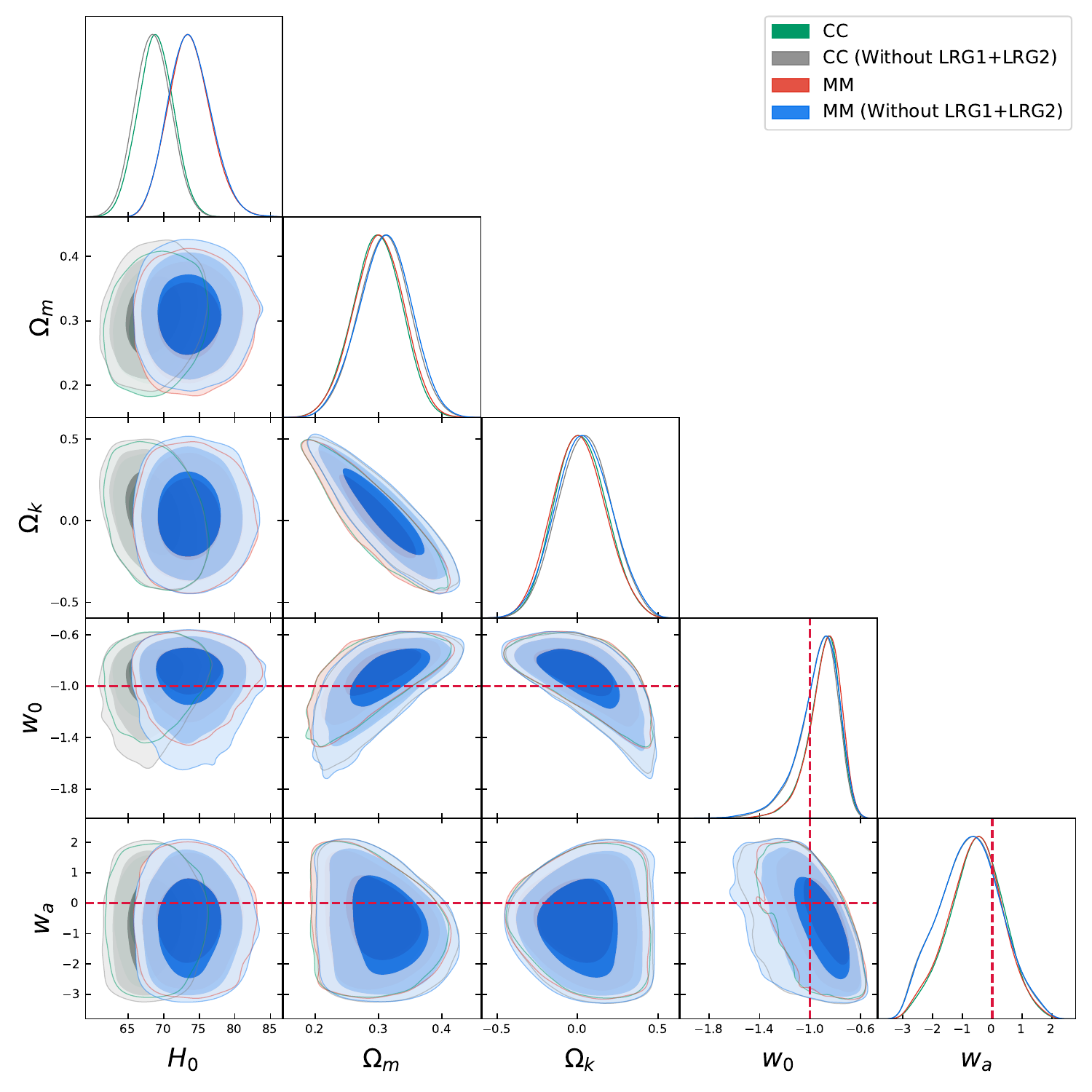}}
    \caption{Constraints on the JBP model (Eqn.~\ref{eq7}) with uniform priors on $M$ and $r_d$. The left panel corresponds to the flat scenario, while the right panel represents the non-flat scenario. The contours denote the 68\%, 95\%, and 99\% credible intervals. Red dotted lines indicate the standard $\Lambda$CDM values of $w_0 = -1$ and $w_a = 0$.}
    \label{fig7}
\end{figure}

\begin{figure}
    \centering
    \subfloat[Flat JBP]{\includegraphics[width=0.5\textwidth]{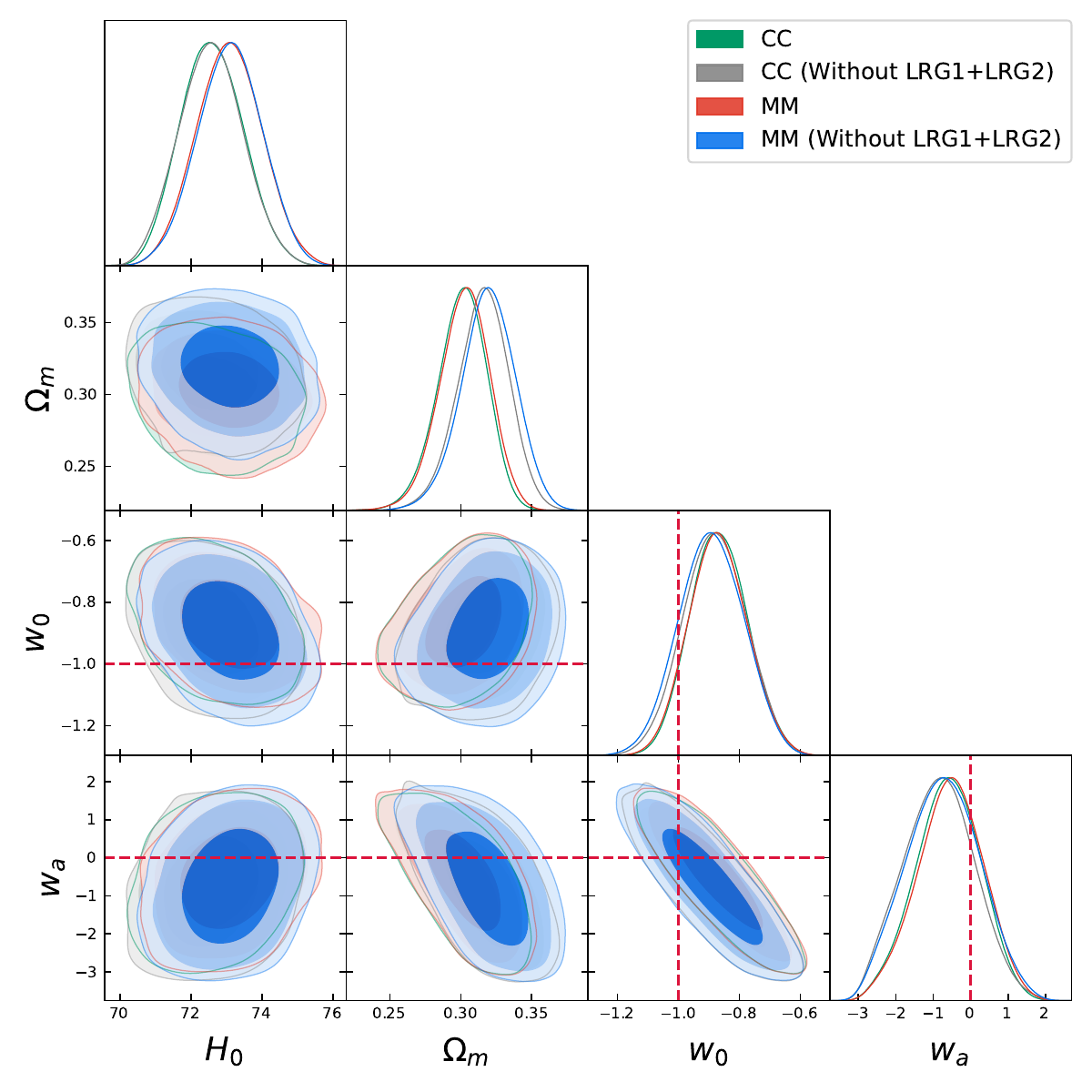}}
    \subfloat[Non-Flat JBP]{\includegraphics[width=0.5\textwidth]{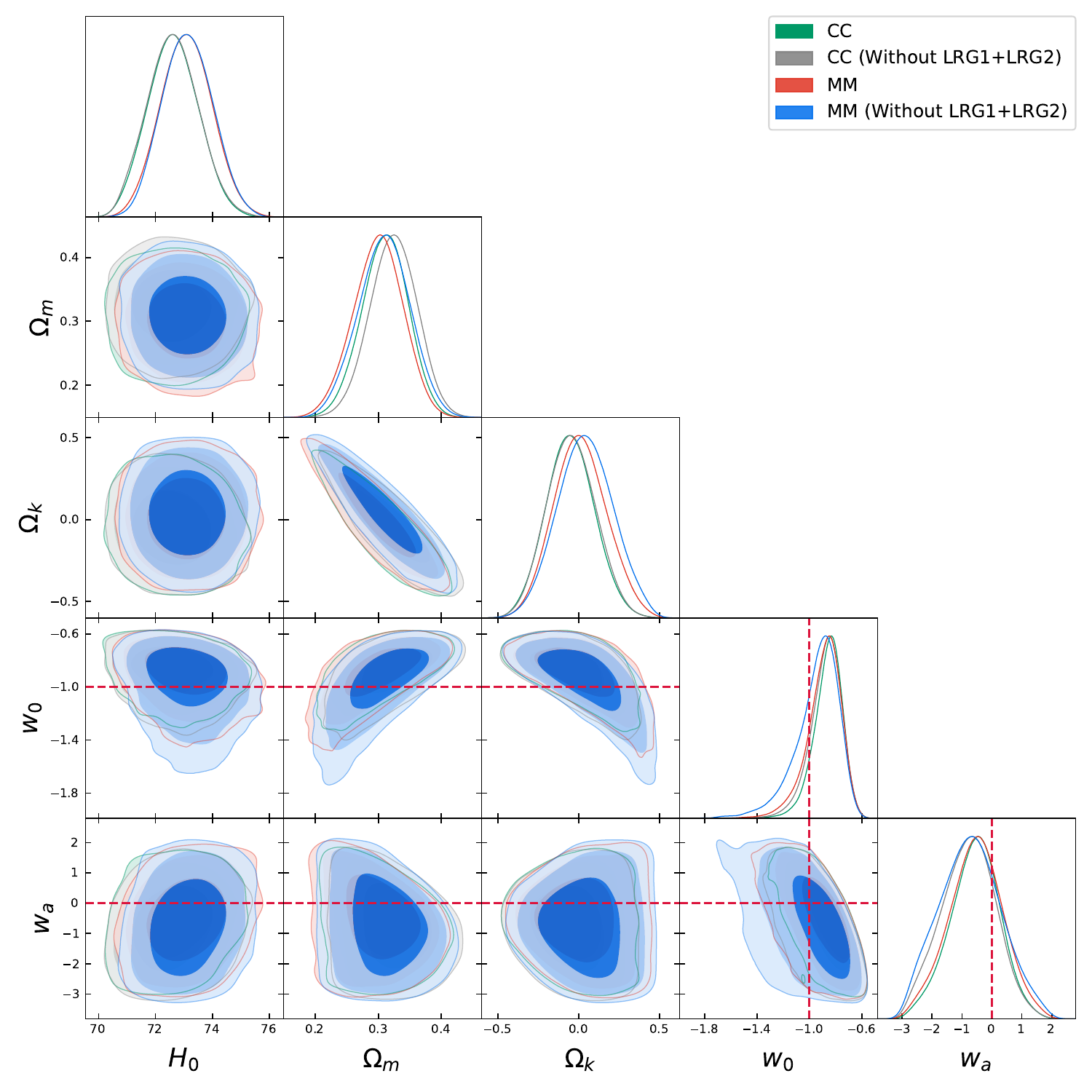}}
    \caption{Constraints on the JBP model (Eqn.~\ref{eq7}) with Gaussian prior on $M$ and uniform prior on $r_d$. The left panel corresponds to the flat scenario, while the right panel represents the non-flat scenario. The contours denote the 68\%, 95\%, and 99\% credible regions. The red dotted lines indicate the standard $\Lambda$CDM values of $w_0 = -1$ and $w_a = 0$.}
    \label{fig8}
\end{figure}
Refer to Figs.~\ref{fig7} and \ref{fig8}.
\begin{itemize}
    \item There is no preference towards either the flat or non-flat scenarios for all dataset and prior combinations.
    \item JBP parameterization shows a similar trend as the CPL and BA parameterizations, wherein using the  Base$+$CC dataset results in a strong preference for $\Lambda$CDM model. Using Base$+$MM dataset for flat JBP and applying uniform prior on $M$ and $r_d$ results in a very strong preference for flat $\Lambda$CDM. However, we notice that when constraining non-flat JBP using Base$+$MM dataset, the preference towards $\Lambda$CDM turns to a moderate one (while including and excluding LRG1 and LRG2).
    \item We notice a $\sim1.5\sigma$ deviation of $w_0$ from its $\Lambda$CDM value of -1, when we include LRG1 and LRG2 in both Base$+$CC and Base$+$MM datasets for flat JBP. In all other cases $w_0$ is within $1\sigma$ of its standard value of -1.
    \item $w_a$ is consistent within $1\sigma$ of its $\Lambda$CDM value of 0. 
    \item No statistically significant prior dependence is found across all dataset combinations for both flat and non-flat scenarios.
\end{itemize}

\subsection{Exponential Parameterization}
\label{exp}

\begin{figure}
    \centering
    \subfloat[Flat Exponential]{\includegraphics[width=0.5\textwidth]{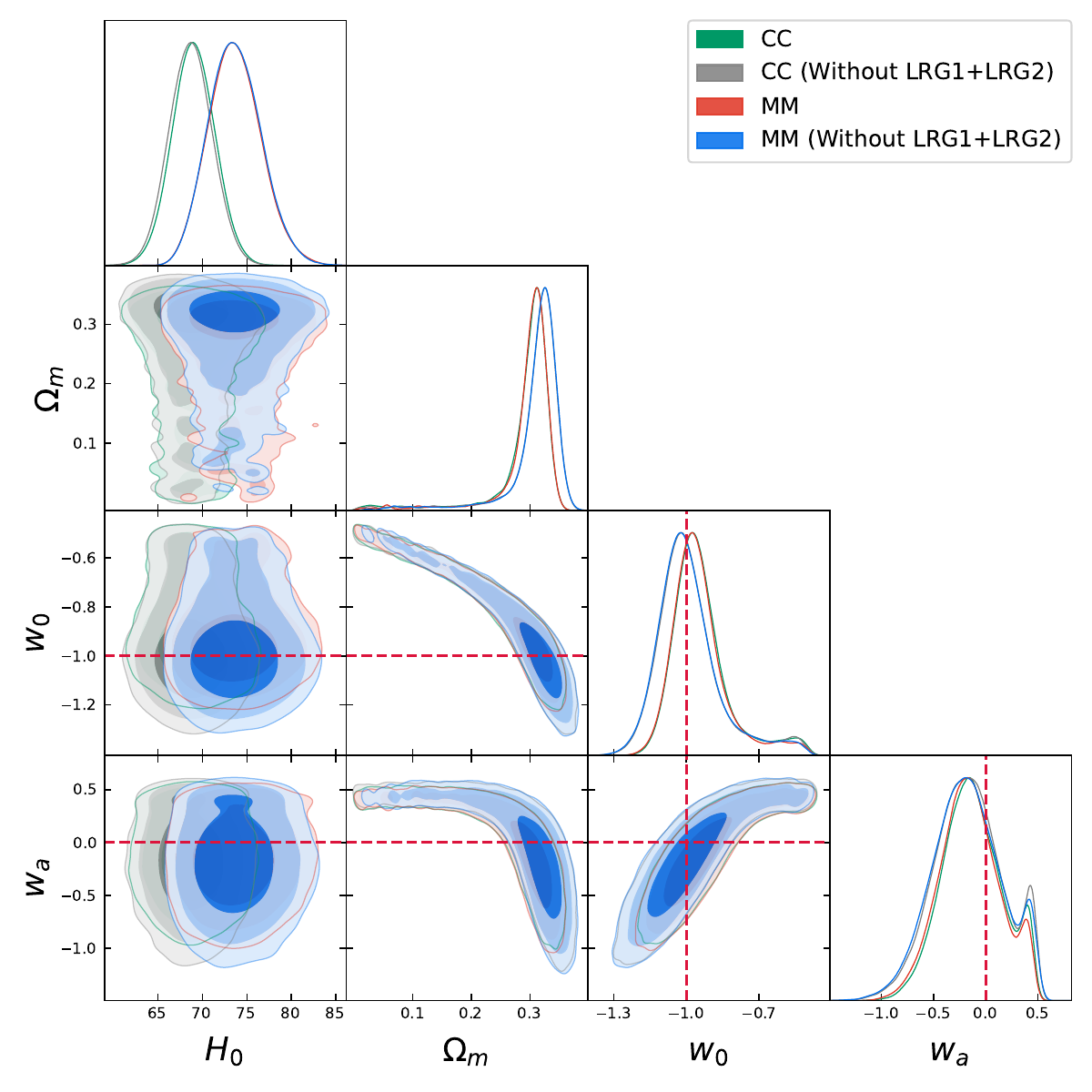}}
    \subfloat[Non-Flat Exponential]{\includegraphics[width=0.5\textwidth]{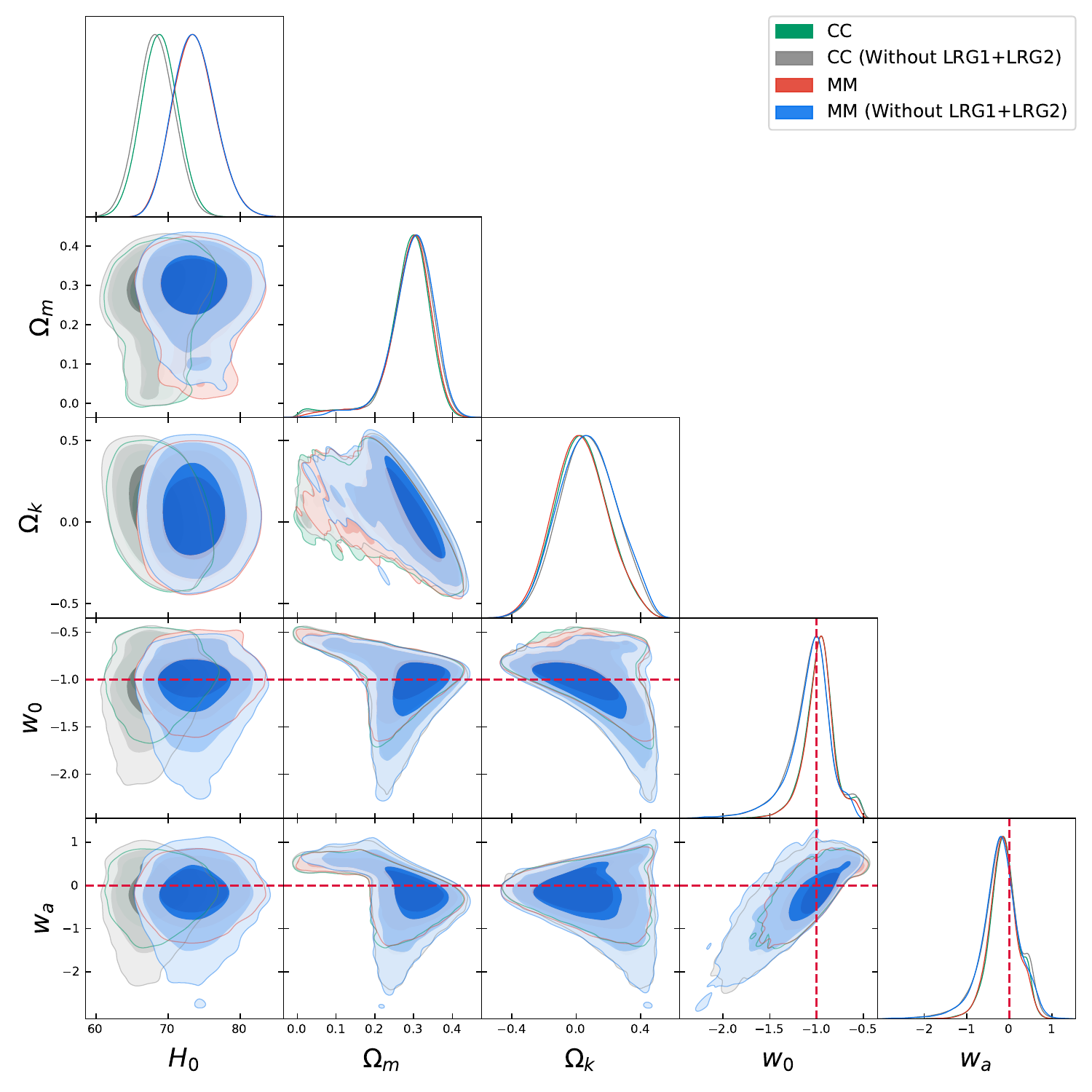}}
    \caption{Constraints on the EXP model (Eqn.~\ref{eq8}) with uniform priors on $M$ and $r_d$. The left panel corresponds to the flat scenario, while the right panel represents the non-flat scenario. The contours denote the 68\%, 95\%, and 99\% credible intervals. Red dotted lines indicate the standard $\Lambda$CDM values of $w_0 = -1$ and $w_a = 0$.}
    \label{fig9}
\end{figure}

\begin{figure}
    \centering
    \subfloat[Flat Exponential]{\includegraphics[width=0.5\textwidth]{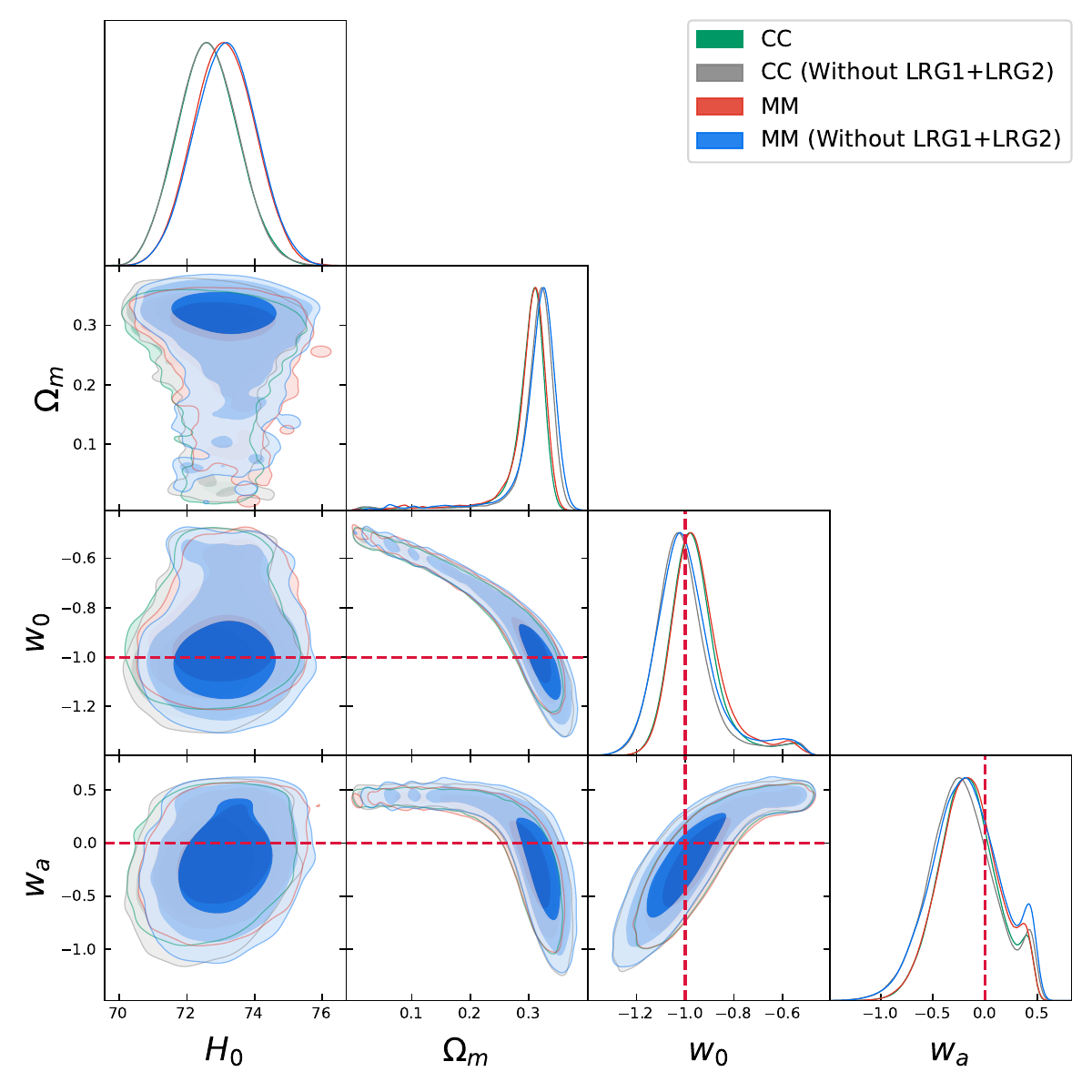}}
    \subfloat[Non-Flat Exponential]{\includegraphics[width=0.5\textwidth]{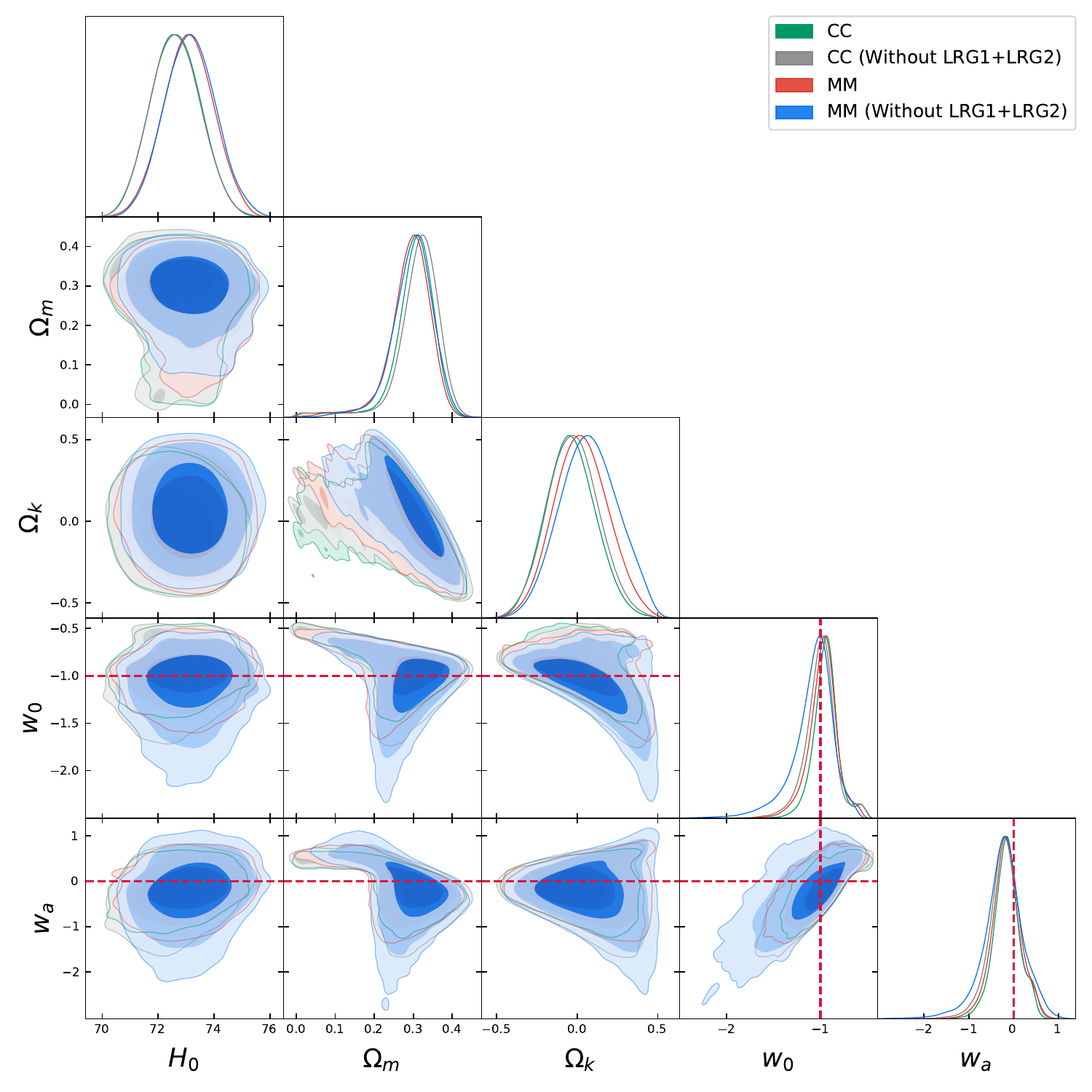}}
    \caption{Constraints on the EXP model (Eqn.~\ref{eq8}) with Gaussian prior on $M$ and uniform prior on $r_d$. The left panel corresponds to the flat scenario, while the right panel represents the non-flat scenario. The contours denote the 68\%, 95\%, and 99\% credible intervals. Red dotted lines indicate the standard $\Lambda$CDM values of $w_0 = -1$ and $w_a = 0$.}
    \label{fig10}
\end{figure}
Refer to Figs.~\ref{fig9} and \ref{fig10}.
\begin{itemize}
    \item Similar to CPL, BA and JBP, we find no preference towards flat or non-flat scenarios.
    \item The trend when comparing to (flat/non-flat) $\Lambda$CDM is similar. It shows a very strong preference towards $\Lambda$CDM when being constrained by Base$+$MM dataset and application of uniform priors on $M$ and $r_d$ for the flat exponential parameterization. In all other cases, the preference towards $\Lambda$CDM is strong.
    \item Contrary to the CPL case where we got some deviation of $w_0$ values from $-1$ (see, Sect.~\ref{CPL}), we find that both $w_0$ and $w_a$ are well within $1\sigma$ $(\text{more precisely}, <0.7\sigma)$ of their $\Lambda$CDM values of $(-1, 0)$.
    \item Prior effects are not pronounced similar to the other parameterizations described above.
\end{itemize}

\subsection{TDE Parameterization}
\label{tde}

\begin{figure}
    \centering
    \subfloat[Flat TDE]{\includegraphics[width=0.5\textwidth]{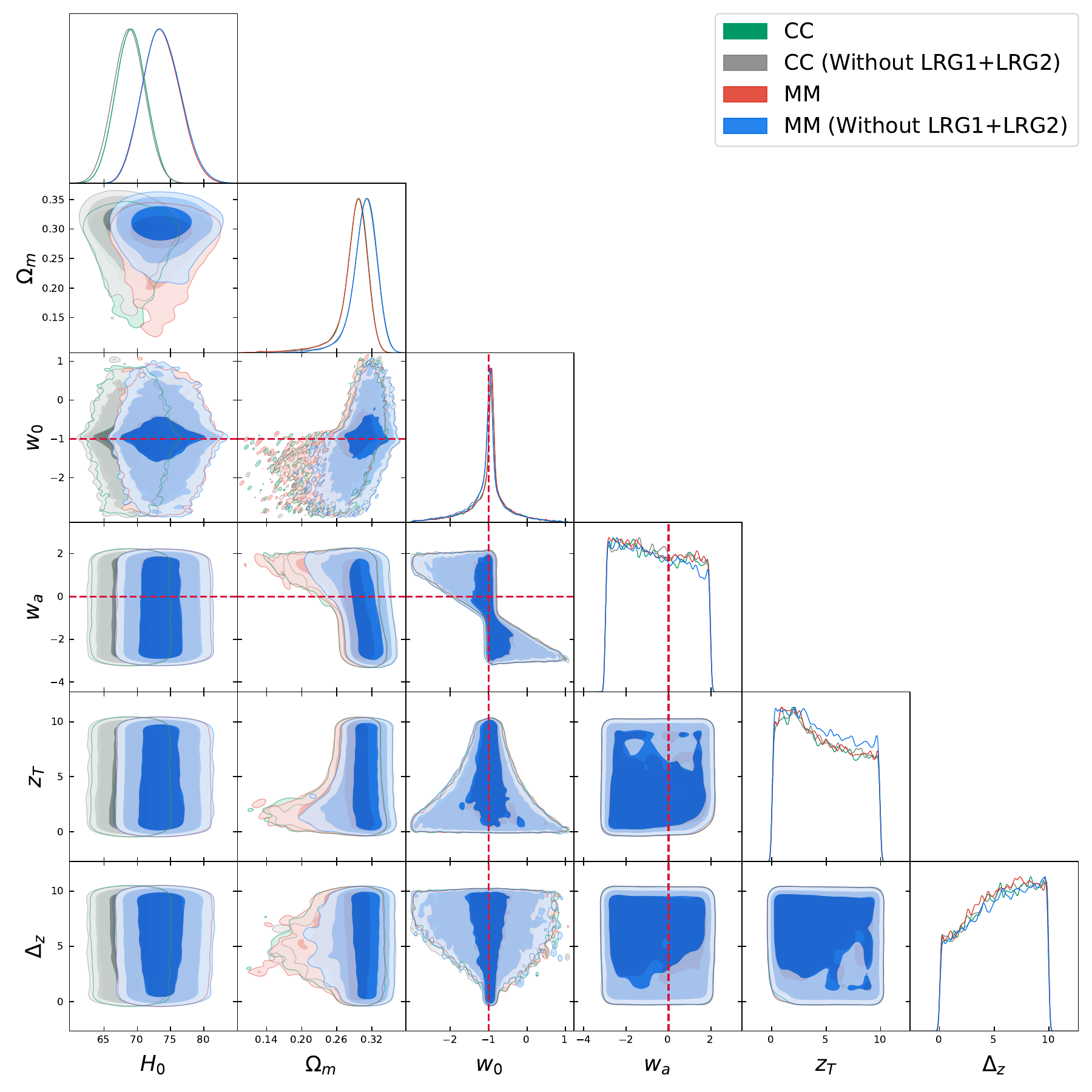}}
    \subfloat[Non-Flat TDE]{\includegraphics[width=0.5\textwidth]{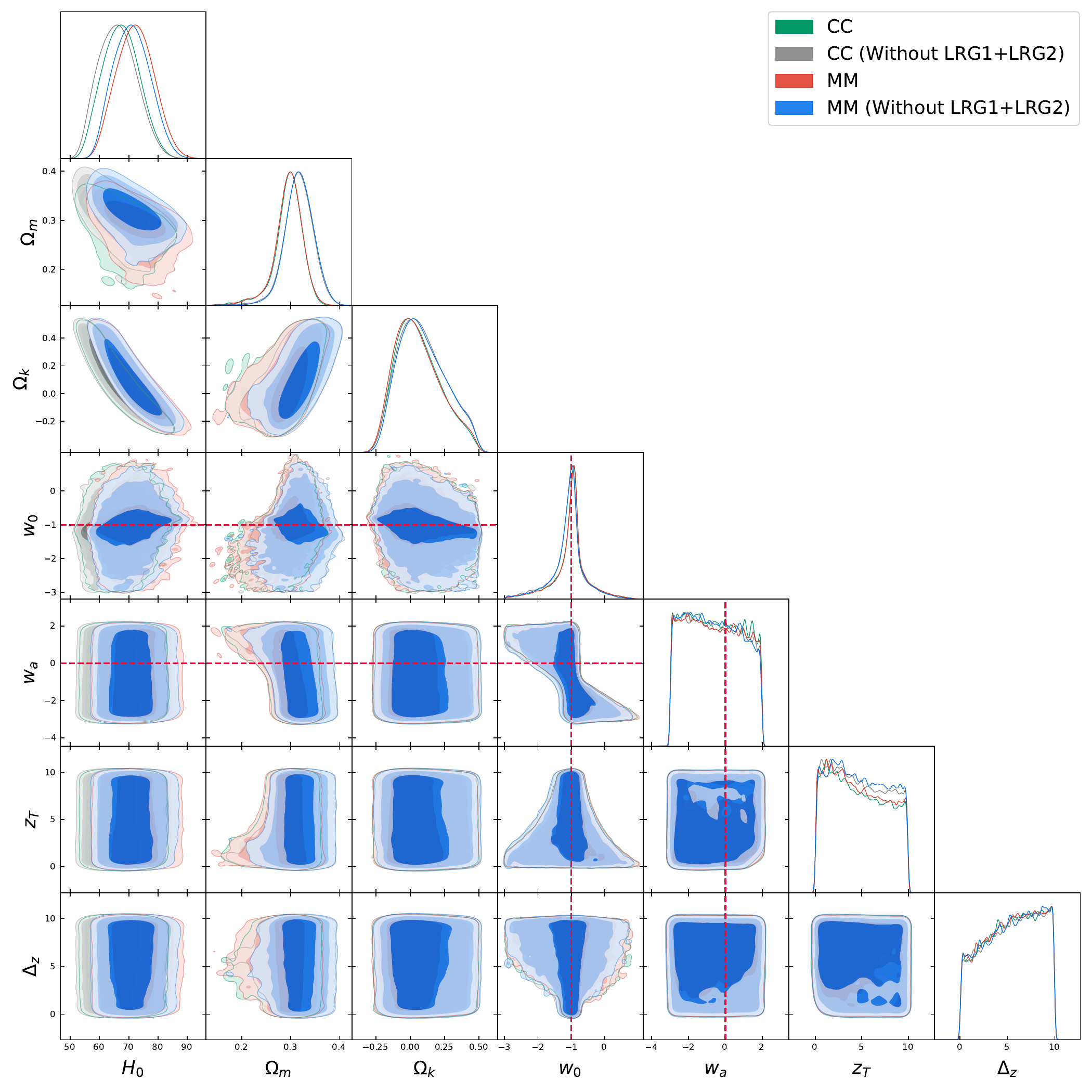}}
    \caption{Constraints on the TDE model (Eqn.~\ref{eq9}) with uniform priors on $M$ and $r_d$. The left panel corresponds to the flat scenario, while the right panel represents the non-flat scenario. The contours denote the 68\%, 95\%, and 99\% credible intervals. Red dotted lines indicate the standard $\Lambda$CDM values of $w_0 = -1$ and $w_a = 0$.}
    \label{fig11}
\end{figure}

\begin{figure}
    \centering
    \subfloat[Flat TDE]{\includegraphics[width=0.5\textwidth]{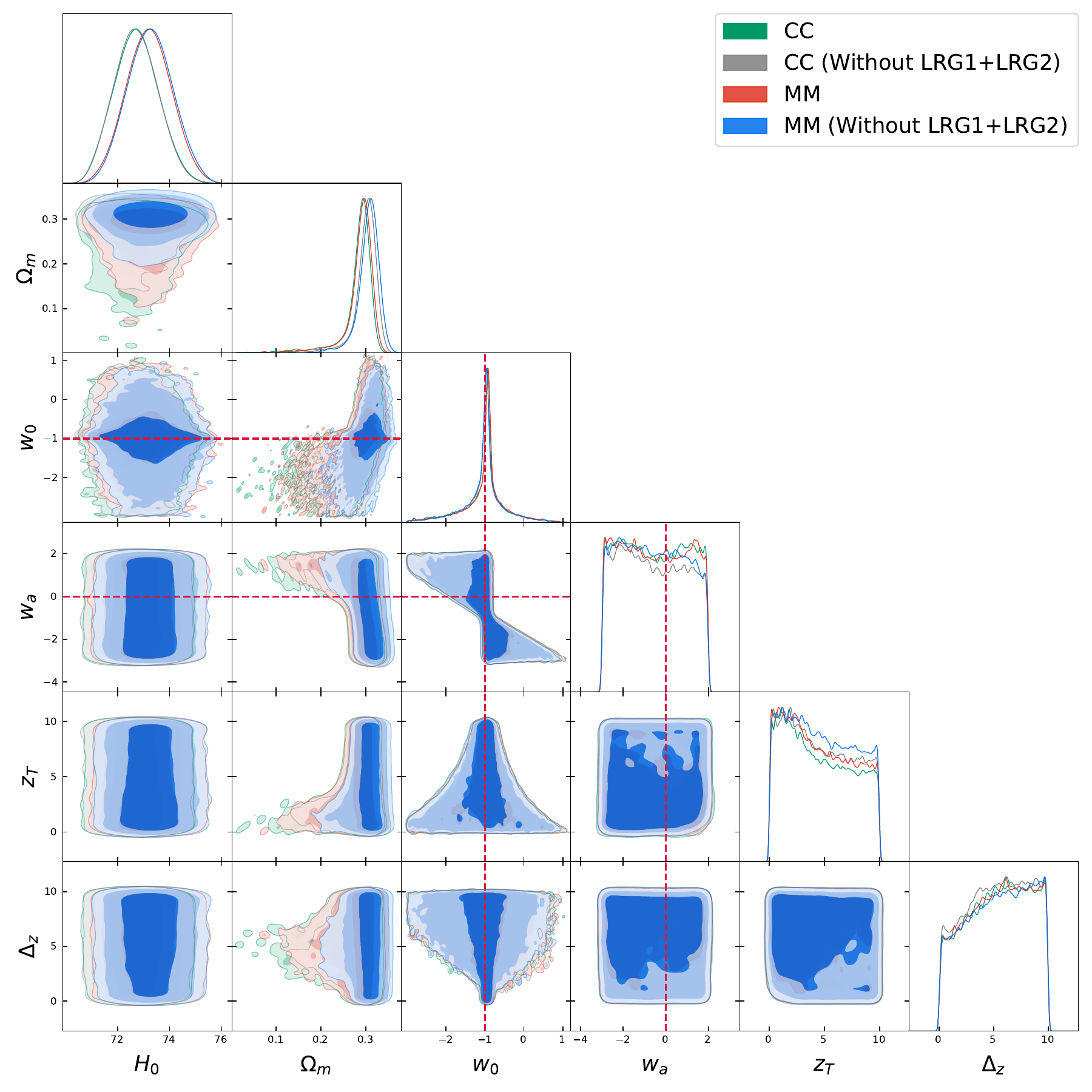}}
    \subfloat[Non-Flat TDE]{\includegraphics[width=0.5\textwidth]{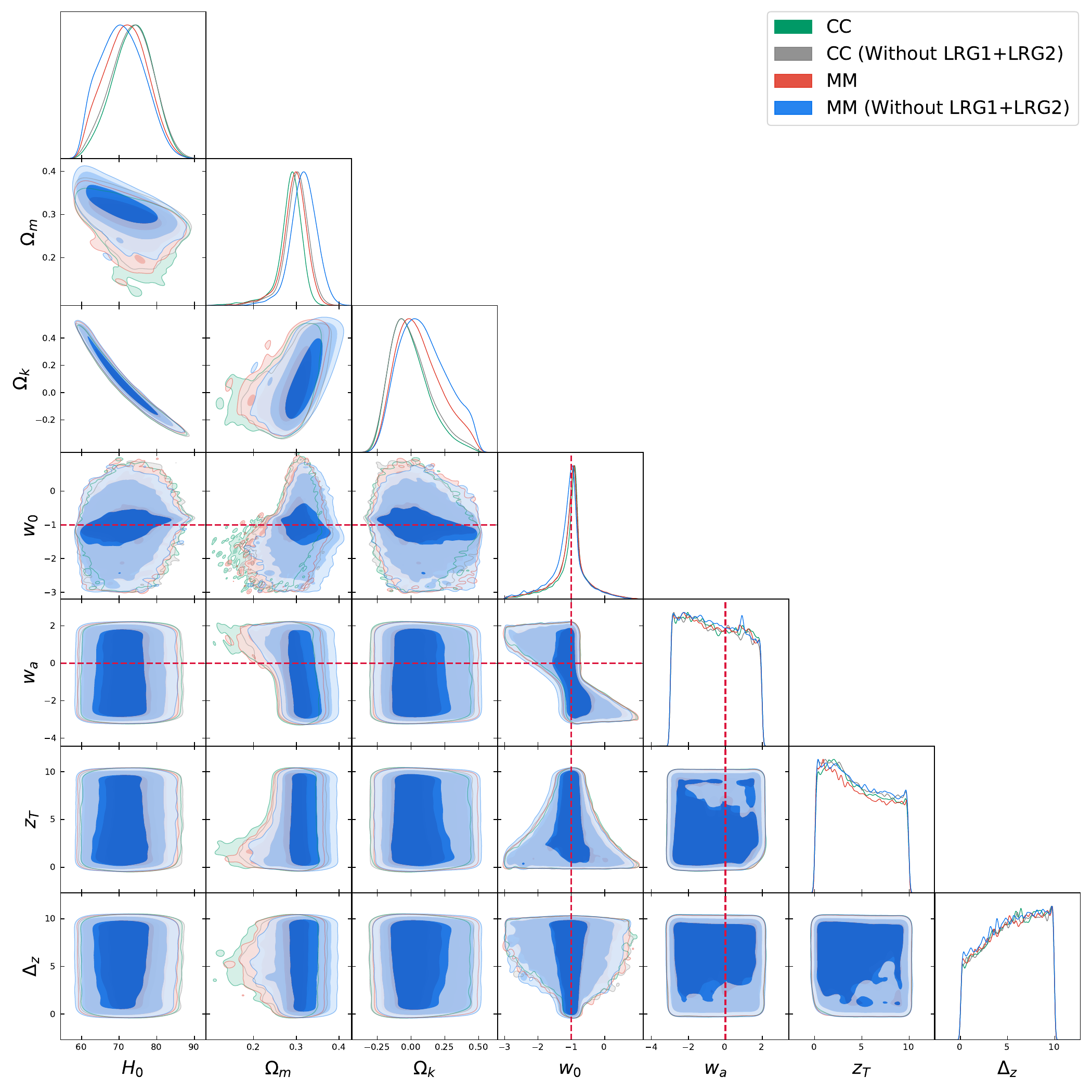}}
    \caption{Constraints on the TDE model (Eqn.~\ref{eq9}) with Gaussian prior on $M$ and uniform prior on $r_d$. The left panel corresponds to the flat scenario, while the right panel represents the non-flat scenario. The contours denote the 68\%, 95\%, and 99\% credible regions. The red dotted lines indicate the standard $\Lambda$CDM values of $w_0 = -1$ and $w_a = 0$.}
    \label{fig12}
\end{figure}
Refer to Figs.~\ref{fig11} and \ref{fig12}.
\begin{itemize}
    \item TDE parameterization shows no preference for flat or the non-flat scenarios for all dataset and prior combinations.
    \item For Base$+$CC dataset, the preference towards $\Lambda$CDM is positive in both flat and non-flat cases. This is different from all the other parameterizations (CPL, BA, JBP and EXP). This is also the case for the Base$+$MM dataset combination. However, when applying uniform priors on both $M$ and $r_d$ (in the flat scenario), we get a strong preference towards $\Lambda$CDM.
    \item The posteriors for $w_a$, $z_T$, and $\Delta_z$ are broad and not properly constrained (Figs.~\ref{fig11} and \ref{fig12}). Additionally, the error bars on $H_0$ and $w_0$ are large in comparison with other parameterizations. It can, however, be seen that the $w_0$ and $w_a$ values are consistent  with  the standard $\Lambda$CDM  value of $(-1, 0)$ to within $1\sigma$.
    \item For all dataset combinations and priors, it can be seen that in the non-flat TDE parameterization case, $\Omega_k$ and $H_0$ show a strong correlation. This is not the case for other non-flat parameterizations.
   \item In the case of non-flat TDE parameterization, irrespective of priors and CC or MM datasets, the error bars on $H_0$ are larger than the corresponding cases. This is due to the additional degrees of freedom ($z_T$ and $\Delta_z$) introduced in the TDE parameterization.
\end{itemize}

\subsection{Comparison with other works}
\label{comparison}

In this subsection, we compare our results with similar studies in the literature, albeit based on different combinations of datasets. We compare our results with those of \cite{Giare_2024} (Planck$+$DESI$+$PantheonPlus), \cite{zheng_2024} (DESI$+$SN$+$QSO) and \cite{desi_2024} (DESI$+$PantheonPlus$+$CMB and DESI$+$CMB). 
Note that our analysis does not include Planck CMB data; instead, we rely solely on geometric and expansion-based probes.

\subsubsection{CPL Parameterization}
\label{comparison_CPL}

\begin{itemize}
    \item As discussed in \cite{desi_2024} and \cite{Giare_2024}, CMB $+$ DESI gives a value of $w_0 = -0.44^{+0.34}_{-0.21}$ and $w_a = -1.79^{+0.48}_{-1.0}$. Including SNe Ia data, namely PantheonPlus compilation, decreases $w_0$ to $-0.827\pm0.063$ \cite{desi_2024}. These results are also reflected in our study even though we do not use CMB data. However, $w_a$ values are considerably larger ($\gtrsim 1.2\sigma$) when comparing with CMB$+$DESI dataset. The values are consistent within $1\sigma$ when including Pantheon$+$ dataset and also with those found in \cite{zheng_2024}.
    \item The value of $\Omega_k$ for DESI$+$CMB$+$PantheonPlus has been found  to be $0.0024\pm0.0016$  using DESI BAO$+$CMB in ~\cite{desi_2024}. We obtain similar results where we do not find any significant deviation from a flat Universe. 
    \item Similar to \cite{zheng_2024}, we find that the $w_0$ values increase very slightly with the inclusion of LRG1 and LRG2 data points and hence deviating from $\Lambda$CDM. However, as discussed in Sect.~\ref{CPL}, the deviation is very small.  
\end{itemize}

\subsubsection{BA Parameterization}
\label{comparison_BA}

\begin{itemize}
    \item With respect to \cite{Giare_2024}, the $w_0$ and $w_a$ values are very much consistent with their Planck$+$DESI$+$PantheonPlus results, even though we do not use Planck CMB data.
    \item \citet{zheng_2024} found that removing LRG1 and LRG2 data points makes the BA parameterization more consistent with the $\Lambda$CDM model. We find similar results following the trend of values of $w_0$ and $w_a$. 
\end{itemize}

\subsubsection{JBP Parameterization}
\label{comparison_JBP}

\begin{itemize}
    \item Similar to the CPL and the BA parameterization, our results agree with the $w_0$ value estimated in~\cite{Giare_2024} ($w_0 = -0.767\pm0.086$)  within $1\sigma$ from Planck$+$DESI$+$PantheonPlus dataset combination.
    \item Similar to BA parameterization, the central value of $w_a$ is consistent within $1\sigma$ with out results.
    \item Comparing with \cite{zheng_2024}, BAO$+$SNe Ia$+$QSO in their work provides a constraint of $w_a = -0.94^{+0.83}_{-0.87}$ for flat JBP and $-0.39^{+0.91}_{-1.18}$ for a non-flat cosmology. 
    In our case, we find that $w_0$ and $w_a$ values for all the prior and dataset combinations agree within $1\sigma$.
\end{itemize}

\subsubsection{EXP Parameterization}
\label{comparison_EXP}

\begin{itemize}
    \item Our $w_0$ values are consistent with those estimated in ~\cite{Giare_2024} to within $\sim1\sigma$. On the other hand $w_a$ values are discrepant with a significance $ \geq 2\sigma$.
\end{itemize}

\begin{table}[htbp!]
\caption{Gaussian Prior on $M$ for Flat Cosmologies(With LRG1$+$LRG2) using Base$+$CC data}
\label{table5}
\centering
    \begin{tabular}{|c|c|c|c|c|c|c|}
        \hline
        \thead{Model} & \thead{\boldmath $H_0$ (km/s/Mpc)} & \thead{\boldmath $\Omega_m$} & \thead{\boldmath $w_0$} & \thead{\boldmath $w_a$} & \thead{\boldmath $z_T$} & \thead{\boldmath $\Delta_z$}\\
        \hline
        $\Lambda$CDM & $73.02\pm0.84$ & $0.305\pm0.011$          & -                         & -                       & -           & -                       \\
        CPL          & $72.63\pm0.87$ & $0.29^{+0.044}_{-0.002}$ & $-0.87^{+0.066}_{-0.082}$ & $-0.34^{+0.58}_{-0.79}$ & -           & -                         \\
        BA           & $72.60\pm0.89$ & $0.29^{+0.032}_{-0.008}$ & $-0.88^{+0.055}_{-0.066}$ & $-0.25\pm0.33$          & -           & -                    \\
        JBP          & $72.59\pm0.87$ & $0.30^{+0.019}_{-0.016}$ & $-0.87\pm0.094$           & $-0.61\pm0.86$          & -           & -                         \\
        EXP          & $72.62\pm0.89$ & $0.29^{+0.036}_{-0.007}$ & $-0.95^{+0.062}_{-0.11}$  & $-0.16^{+0.25}_{-0.31}$ & -           & -                         \\
        TDE          & $72.72\pm0.84$ & $0.28^{+0.032}_{-0.008}$ & $-1.02^{+0.39}_{-0.46}$   & $-0.5\pm1.5$            & $4.5^{+2.3}_{-4.4}$ & $5.4^{+4.4}_{-2.1}$ \\
        \hline
    \end{tabular}
\end{table}

\begin{table}[htbp!]
\caption{Uniform Priors on $M$ and $r_d$ for Flat cosmologies(With LRG1$+$LRG2) using Base$+$CC data}
\label{table6}
\centering
    \begin{tabular}{|c|c|c|c|c|c|c|}
        \hline
        \thead{Model} & \thead{\boldmath $H_0$ (km/s/Mpc)} & \thead{\boldmath $\Omega_m$} & \thead{\boldmath $w_0$} & \thead{\boldmath $w_a$}  & \thead{\boldmath $z_T$} & \thead{\boldmath $\Delta_z$} \\
        \hline
        $\Lambda$CDM & $69.4\pm2.40$  & $0.31\pm0.01$            & -                       & -                      & - & - \\
        CPL          & $69.00\pm2.20$ & $0.29^{+0.05}_{-0.001}$  & $-0.87^{+0.06}_{-0.08}$ & $-0.25^{+1.2}_{-0.74}$  & - & - \\
        BA           & $69.00\pm2.40$ & $0.29^{+0.04}_{-0.01}$   & $-0.88^{+0.05}_{-0.07}$ & $-0.2^{+0.31}_{-0.36}$ & - & - \\
        JBP          & $69.20\pm2.40$ & $0.30\pm0.02$            & $-0.87\pm0.09$          & $-0.52\pm0.87$         & - & - \\
        EXP          & $69\pm2.4$     & $0.29^{+0.043}_{-0.004}$ & $-0.935^{+0.06}_{-0.13}$ & $-0.12^{+0.26}_{-0.32}$ & - & - \\
        TDE          & $69\pm2.3$     & $0.29^{+0.026}_{-0.012}$ & $-0.99\pm0.54$  & $-0.6\pm1.4$ & $4.7^{+2.6}_{-4.5}$ & $5.4^{+4.5}_{-1.6}$ \\ 
        \hline
    \end{tabular}
\end{table}

\begin{table}[htbp!]
\caption{Gaussian Priors on $M$ for Non-flat cosmologies(With LRG1$+$LRG2) using Base$+$CC data}
\label{table7}
\centering
    \begin{tabular}{|c|c|c|c|c|c|c|c|c|c|}
        \hline
        \thead{Model} & \thead{\boldmath $H_0$ (km/s/Mpc)} & \thead{\boldmath $\Omega_m$} & \thead{\boldmath $\Omega_k$} & \thead{\boldmath $w_0$} & \thead{\boldmath $w_a$}  & \thead{\boldmath $z_T$} & \thead{\boldmath $\Delta_z$}\\
        \hline
        $\Lambda$CDM & $72.82\pm0.87$ & $0.28\pm0.025$           & $0.074\pm0.075$ & -                         & -                      & - & - \\
        CPL          & $72.64\pm0.89$ & $0.29^{+0.063}_{-0.029}$ & $-0.04\pm0.15$ & $-0.87^{+0.10}_{-0.069}$  & $-0.33^{+0.62}_{-0.72}$ & - & - \\
        BA           & $72.63\pm0.88$ & $0.30^{+0.051}_{-0.033}$ & $-0.03\pm0.15$  & $-0.88^{+0.094}_{-0.061}$ & $-0.25^{+0.34}_{-0.30}$ & - & - \\
        JBP          & $72.62\pm0.88$ & $0.31\pm0.037$           & $-0.05\pm0.15$  & $-0.86^{+0.12}_{-0.084}$  & $-0.57^{+0.88}_{-0.76}$ & - & - \\
        EXP          & $72.63\pm0.89$ & $0.30^{+0.056}_{-0.031}$ & $-0.04\pm0.15$  & $-0.94\pm0.15$            & $-0.15\pm0.3$          & - & - \\
        TDE          & $73.5^{+5.8}_{-5.0}$ & $0.28^{+0.03}_{-0.017}$ & $-0.01^{+0.097}_{-0.17}$ & $-0.98\pm0.55$ & $-0.6^{+2.0}_{-2.3}$ & $4.6^{+1.9}_{-4.5}$ & $5.4^{+4.4}_{-2.2}$ \\
        \hline
    \end{tabular}
\end{table}

\begin{table}[htbp!]
\caption{Uniform Priors on $M$ and $r_d$ for Non-flat cosmologies (With LRG1$+$LRG2) using Base$+$CC data}
\label{table8}
\centering
    \begin{tabular}{|c|c|c|c|c|c|c|c|}
        \hline
        \thead{Model} & \thead{\boldmath $H_0$ (km/s/Mpc)} & \thead{\boldmath $\Omega_m$} & \thead{\boldmath $\Omega_k$} & \thead{\boldmath $w_0$} & \thead{\boldmath $w_a$} & \thead{\boldmath $z_T$} & \thead{\boldmath $\Delta_z$}\\
        \hline
        $\Lambda$CDM & $68.7\pm2.50$  & $0.28\pm0.03$          & $0.1\pm0.08$  & -                       & -    & - & -                   \\
        CPL          & $68.8\pm2.50$  & $0.28^{+0.07}_{-0.03}$ & $0.03\pm0.16$ & $-0.90\pm0.1$           & $-0.27^{+1.1}_{-0.84}$ & - & - \\
        BA           & $68.9\pm2.50$  & $0.29^{+0.06}_{-0.03}$ & $0.04\pm0.16$ & $-0.91^{+0.12}_{-0.07}$ & $-0.24\pm0.4$         & - & - \\
        JBP          & $68.9\pm2.50$  & $0.3\pm0.04$           & $0.02\pm0.16$ & $-0.89^{+0.15}_{-0.09}$ & $-0.54^{+0.97}_{-0.84}$ & - & - \\
        EXP          & $68.8\pm2.5$   & $0.28^{+0.064}_{-0.032}$ & $0.03\pm0.16$ & $-0.97\pm0.18$        & $-0.13\pm0.35$         & - & - \\
        TDE          & $67.4\pm6.2$   & $0.295\pm0.03$       & $0.06^{+0.12}_{-0.2}$ & $-1.03^{+0.37}_{-0.41}$ & $-0.6^{+2.1}_{-2.3}$ & $4.6^{+2.0}_{-4.5}$ & $5.4^{+4.4}_{-1.7}$ \\
        \hline
    \end{tabular}
\end{table}

\begin{table}[htbp!]
\caption{Gaussian Priors on $M$ for Flat Cosmologies (Without LRG1$+$LRG2) using Base$+$CC data}
\label{table9}
\centering
    \begin{tabular}{|c|c|c|c|c|c|c|}
        \hline
        \thead{Model} & \thead{\boldmath $H_0$ (km/s/Mpc)} & \thead{\boldmath $\Omega_m$} & \thead{\boldmath $w_0$} & \thead{\boldmath $w_a$} & \thead{\boldmath $z_T$} & \thead{\boldmath $\Delta_z$}\\
        \hline
        $\Lambda$CDM & $72.88\pm0.87$ & $0.31\pm0.012$           & -                         & -               & - & -         \\
        CPL          & $72.62\pm0.88$ & $0.30^{+0.045}_{-0.003}$ & $-0.89^{+0.076}_{-0.095}$ & $-0.44^{+1.4}_{-0.84}$ & - & - \\
        BA           & $72.63\pm0.88$ & $0.31^{+0.039}_{-0.061}$ & $-0.91^{+0.060}_{-0.078}$ & $-0.27^{+0.35}_{-0.42}$ & - & - \\
        JBP          & $72.57\pm0.89$ & $0.32\pm0.018$           & $-0.88\pm0.1$             & $-0.85\pm0.94$          & - & - \\
        EXP          & $72.62\pm0.88$ & $0.31^{+0.036}_{-0.008}$ & $-0.99^{+0.07}_{-0.13}$   & $-0.21^{+0.28}_{-0.34}$ & - & - \\
        TDE          & $72.71\pm0.84$ & $0.30^{+0.026}_{-0.012}$ & $-1.02\pm0.55$            & $-0.6^{+2.0}_{-2.3}$ & $4.6^{+2.2}_{-4.5}$ & $5.3^{+4.5}_{-1.8}$ \\
        \hline
    \end{tabular}
\end{table}

\begin{table}[htbp!]
\caption{Uniform Priors on $M$ and $r_d$ for Flat cosmologies (Without LRG1$+$LRG2) using Base$+$CC data}
\label{table10}
\centering
    \begin{tabular}{|c|c|c|c|c|c|c|}
        \hline
        \thead{Model} & \thead{\boldmath $H_0$ (km/s/Mpc)} & \thead{\boldmath $\Omega_m$} & \thead{\boldmath $w_0$} & \thead{\boldmath $w_a$}  & \thead{\boldmath $z_T$} & \thead{\boldmath $\Delta_z$} \\
        \hline
        $\Lambda$CDM & $68.8\pm2.40$  & $0.32\pm0.01$             & -                       & -            & - & -            \\
        CPL          & $68.70\pm2.40$ & $0.30^{+0.051}_{-0.0027}$ & $-0.90^{+0.07}_{-0.09}$ & $-0.33^{+1.4}_{-0.82}$ & - & - \\
        BA           & $68.70\pm2.30$ & $0.30\pm0.04$             & $-0.92^{+0.06}_{-0.08}$ & $-0.19^{+0.68}_{-0.40}$ & - & - \\
        JBP          & $68.70\pm2.50$ & $0.32\pm0.02$             & $-0.89\pm0.1$           & $-0.73\pm0.97$          & - & - \\
        EXP          & $68.7\pm2.4$   & $0.31^{+0.043}_{-0.0064}$ & $-0.98^{+0.073}_{-0.14}$ & $-0.16^{+0.3}_{-0.37}$ & - & -\\
        TDE          & $68.7\pm2.4$   & $0.31^{+0.026}_{-0.014}$  & $-1.04^{+0.35}_{-0.41}$  & $-0.6^{+1.7}_{-2.4}$ & $4.7^{+3.5}_{-4.5}$ & $5.4^{+4.4}_{-1.6}$ \\
        \hline
    \end{tabular}
\end{table}

\begin{table}[htbp!]
\caption{Gaussian Priors on $M$ for Non-flat cosmologies (Without LRG1$+$LRG2) using Base$+$CC data}
\label{table11}
\centering
    \begin{tabular}{|c|c|c|c|c|c|c|c|}
        \hline
        \thead{Model} & \thead{\boldmath $H_0$ (km/s/Mpc)} & \thead{\boldmath $\Omega_m$} & \thead{\boldmath $\Omega_k$} & \thead{\boldmath $w_0$} & \thead{\boldmath $w_a$} & \thead{\boldmath $z_T$} & \thead{\boldmath $\Delta_z$}\\
        \hline
        $\Lambda$CDM & $72.83\pm0.86$ & $0.31\pm0.027$           & $0.025\pm0.077$ & -                       & -           & - & -           \\
        CPL          & $72.65\pm0.89$ & $0.31^{+0.062}_{-0.03}$  & $-0.03\pm0.16$ & $-0.89^{+0.12}_{-0.078}$ & $-0.45^{+0.72}_{-0.82}$ & - & - \\
        BA           & $72.63\pm0.9$  & $0.31^{+0.056}_{-0.033}$ & $-0.02\pm0.16$ & $-0.92^{+0.11}_{-0.07}$  & $-0.29^{+0.44}_{-0.37}$ & - & - \\
        JBP          & $72.61\pm0.9$  & $0.32\pm0.037$           & $-0.05\pm0.15$ & $-0.88^{+0.14}_{-0.09}$  & $-0.79^{+0.97}_{-0.87}$ & - & - \\
        EXP          & $72.63\pm0.89$ & $0.31^{+0.058}_{-0.032}$ & $-0.02\pm0.16$ & $-1.0^{+0.015}_{-0.13}$ & $-0.21\pm0.36$         & - & -\\
        TDE          & $73.4^{+6.0}_{-5.3}$ & $0.3^{+0.029}_{-0.022}$ & $0.0^{+0.1}_{-0.18}$ & $-1.03^{+0.37}_{-0.41}$ & $-0.6^{+1.9}_{-2.3}$ & $4.7^{+5.0}_{-4.5}$ & $5.4^{+4.4}_{-1.9}$ \\ 
        \hline
    \end{tabular}
\end{table}

\begin{table}[htbp!]
\caption{Uniform Priors on $M$ and $r_d$ for Non-flat cosmologies (Without LRG1$+$LRG2) using Base$+$CC data}
\label{table12}
\centering
    \begin{tabular}{|c|c|c|c|c|c|c|c|}
        \hline
        \thead{Model} & \thead{\boldmath $H_0$ (km/s/Mpc)} & \thead{\boldmath $\Omega_m$} & \thead{\boldmath $\Omega_k$} & \thead{\boldmath $w_0$} & \thead{\boldmath $w_a$} & \thead{\boldmath $z_T$} & \thead{\boldmath $\Delta_z$}\\
        \hline
        $\Lambda$CDM & $68.6\pm2.50$  & $0.31\pm0.03$            & $0.05\pm0.08$ & -                        & -      & - & -                \\
        CPL          & $68.4\pm2.50$  & $0.28^{+0.073}_{-0.032}$ & $0.07\pm0.17$ & $-0.96^{+0.17}_{-0.09}$  & $-0.39^{+1.4}_{-0.95}$ & - & - \\
        BA           & $68.93\pm2.60$ & $0.29^{+0.06}_{-0.04}$   & $0.08\pm0.17$ & $-0.98^{+0.16}_{-0.08}$  & $-0.31^{+0.6}_{-0.45}$ & - & - \\
        JBP          & $68.5\pm2.50$  & $0.31\pm0.04$            & $0.04\pm0.16$ & $-0.94^{+0.18}_{-0.098}$ & $-0.7\pm1.0$          & - & - \\
        EXP          & $68.3\pm2.6$   & $0.29^{+0.065}_{-0.035}$ & $0.07\pm0.17$ & $-1.07^{+0.22}_{-0.17}$  & $-0.22\pm0.4$         & - & - \\
        TDE          & $66.2^{+5.9}_{-7.2}$   & $0.32\pm0.031$ & $0.09^{+0.13}_{-0.22}$ & $-1.09^{+0.4}_{-0.36}$ & $-0.6^{+1.6}_{-2.3}$ & $4.7\pm0.29$ & $5.4^{+4.4}_{-2.1}$ \\
        \hline
    \end{tabular}
\end{table}

\begin{table}[htbp!]
\caption{Gaussian Priors on $M$ for Flat Cosmologies (With LRG1$+$LRG2) using Base$+$MM data}
\label{table13}
\centering
    \begin{tabular}{|c|c|c|c|c|c|c|c|}
        \hline
        \thead{Model} & \thead{\boldmath $H_0$ (km/s/Mpc)} & \thead{\boldmath $\Omega_m$} & \thead{\boldmath $w_0$} & \thead{\boldmath $w_a$}  & \thead{\boldmath $z_T$} & \thead{\boldmath $\Delta_z$}\\
        \hline
        $\Lambda$CDM & $73.57\pm0.87$ & $0.31\pm0.012$           & -                         & -           & - & -            \\
        CPL          & $73.12\pm0.90$ & $0.29^{+0.046}_{-0.002}$ & $-0.87^{+0.065}_{-0.085}$ & $-0.37^{+1.2}_{-0.77}$ & - & - \\
        BA           & $73.07\pm0.88$ & $0.3^{+0.032}_{-0.010}$  & $-0.88^{+0.056}_{-0.065}$ & $-0.24\pm0.33$          & - & - \\
        JBP          & $73.08\pm0.91$ & $0.3\pm0.018$            & $-0.87\pm0.096$           & $-0.55^{+0.92}_{-0.84}$ & - & - \\
        EXP          & $73.1\pm0.9$   & $0.29^{+0.038}_{-0.007}$ & $-0.944^{+0.064}_{-0.12}$ & $-0.14^{+0.26}_{-0.32}$ & - & - \\
        TDE          & $73.2\pm0.87$  & $0.29^{+0.03}_{-0.01}$   & $-1^{+0.38}_{-0.43}$   & $-0.5\pm1.5$ & $4.5^{+2.9}_{-4.5}$ & $5.4^{+4.5}_{-1.7}$ \\
        \hline
    \end{tabular}
\end{table}

\begin{table}[htbp!]
\caption{Uniform Priors on $M$ and $r_d$ for Flat cosmologies(With LRG1$+$LRG2) using Base$+$MM data}
\label{table14}
\centering
    \begin{tabular}{|c|c|c|c|c|c|c|c|}
        \hline
        \thead{Model} & \thead{\boldmath $H_0$ (km/s/Mpc)} & \thead{\boldmath $\Omega_m$} & \thead{\boldmath $w_0$} & \thead{\boldmath $w_a$} & \thead{\boldmath $z_T$} & \thead{\boldmath $\Delta_z$} \\
        \hline
        $\Lambda$CDM & $73.6^{+2.8}_{-3.1}$ & $0.31\pm0.012$           & -                         & -                 & - & -      \\
        CPL          & $73.6^{+2.7}_{-3.2}$ & $0.29^{+0.041}_{-0.006}$ & $-0.87^{+0.066}_{-0.081}$ & $-0.33^{+0.61}_{-0.76}$ & - & - \\
        BA           & $73.7^{+2.7}_{-3.1}$ & $0.3^{+0.032}_{-0.01}$   & $-0.88^{+0.056}_{-0.065}$ & $-0.24\pm0.33$          & - & - \\
        JBP          & $73.6^{+2.7}_{-3.1}$ & $0.3\pm0.02$             & $-0.87\pm0.096$           & $-0.56^{+0.94}_{-0.84}$ & - & - \\
        EXP          & $73.7^{+2.8}_{-3.1}$ & $0.29^{+0.04}_{-0.006}$  & $-0.94^{+0.06}_{-0.12}$   & $-0.15^{+0.26}_{-0.32}$ & - & - \\
        TDE          & $73.6\pm2.8$ & $0.29^{+0.026}_{-0.011}$  & $-0.98\pm0.54$  & $-0.6\pm1.4$ & $4.7^{+2.1}_{-4.5}$ & $5.4^{+4.4}_{-1.7}$ \\
        \hline
    \end{tabular}
\end{table}

\begin{table}[htbp!]
\caption{Gaussian Priors on $M$ for Non-flat cosmologies (With LRG1$+$LRG2) using Base$+$MM data}
\label{table15}
\centering
    \begin{tabular}{|c|c|c|c|c|c|c|c|c|}
        \hline
        \thead{Model} & \thead{\boldmath $H_0$ (km/s/Mpc)} & \thead{\boldmath $\Omega_m$} & \thead{\boldmath $\Omega_k$} & \thead{\boldmath $w_0$} & \thead{\boldmath $w_a$} & \thead{\boldmath $z_T$} & \thead{\boldmath $\Delta_z$}\\
        \hline
        $\Lambda$CDM & $73.27\pm0.85$ & $0.28\pm0.025$           & $0.096\pm 0.078$ & -                         & -         & -  & -          \\
        CPL          & $73.12\pm0.87$ & $0.28^{+0.068}_{-0.032}$ & $0.03\pm0.17$    & $-0.89^{+0.130}_{-0.072}$ & $-0.31^{+0.69}_{-0.83}$ & - & - \\
        BA           & $73.1\pm0.89$  & $0.29^{+0.052}_{-0.038}$ & $0.03\pm0.16$    & $-0.91^{+0.12}_{-0.061}$ & $-0.27\pm0.36$         & - & - \\
        JBP          & $73.09\pm0.9$  & $0.30\pm0.039$           & $0.01\pm0.16$    & $-0.89^{+0.15}_{-0.091}$  & $-0.55\pm0.91$        & - & - \\
        EXP          & $73.09\pm0.89$ & $0.29^{+0.056}_{-0.036}$ & $0.03\pm0.17$    & $-0.98^{+0.015}_{-0.012}$ & $-0.16\pm0.33$        & - & - \\
        TDE          & $71.8\pm5.6$   & $0.30^{+0.029}_{-0.021}$  & $0.06^{+0.12}_{-0.21}$ & $-1.02^{+0.37}_{-0.41}$ & $-0.6^{+1.6}_{-2.3}$ & $4.7^{+5.1}_{-4.6}$ & $5.4^{+4.5}_{-2.0}$ \\
        \hline
    \end{tabular}
\end{table}

\begin{table}[htbp!]
\caption{Uniform Priors on $M$ and $r_d$ for Non-flat cosmologies (With LRG1$+$LRG2) using Base$+$MM data}
\label{table16}
\centering
    \begin{tabular}{|c|c|c|c|c|c|c|c|}
        \hline
        \thead{Model} & \thead{\boldmath $H_0$ (km/s/Mpc)} & \thead{\boldmath $\Omega_m$} & \thead{\boldmath $\Omega_k$} & \thead{\boldmath $w_0$} & \thead{\boldmath $w_a$} & \thead{\boldmath $z_T$} & \thead{\boldmath $\Delta_z$}\\
        \hline
        $\Lambda$CDM & $73.7^{+2.7}_{-3.2}$ & $0.28\pm0.025$           & $0.096\pm0.08$ & -                        & -         & - & -              \\
        CPL          & $73.7^{+2.7}_{-3.2}$ & $0.28^{+0.07}_{-0.03}$   & $0.03\pm0.17$  & $-0.89^{+0.13}_{-0.073}$ & $-0.30^{+0.69}_{-0.86}$ & - & - \\
        BA           & $73.7^{+2.7}_{-3.2}$ & $0.29^{+0.053}_{-0.037}$ & $0.03\pm0.17$  & $-0.91^{+0.12}_{-0.06}$  & $-0.27^{+0.39}_{-0.34}$ & - & - \\
        JBP          & $73.7^{+2.7}_{-3.1}$ & $0.3\pm0.04$             & $0.01\pm0.16$  & $-0.89^{+0.15}_{-0.09}$  & $-0.57^{+0.96}_{-0.87}$ & - & - \\
        EXP          & $73.6^{+2.7}_{-3.2}$ & $0.29^{+0.06}_{0.037}$   & $0.03\pm0.17$  & $-0.97\pm0.17$           & $-0.15\pm0.32$        & - & - \\
        TDE          & $72.2\pm6.2$         & $0.3^{+0.029}_{-0.021}$  & $0.06^{+0.12}_{-0.2}$ & $-1.01\pm0.54$  & $-0.6^{+1.5}_{-2.3}$ & $4.7^{+2.2}_{-4.6}$ & $5.4^{+4.5}_{-1.9}$ \\
        \hline
    \end{tabular}
\end{table}

\begin{table}[htbp!]
\caption{Gaussian Prior on $M$ for Flat Cosmologies (Without LRG1$+$LRG2) using Base$+$MM data}
\label{table17}
\centering
    \begin{tabular}{|c|c|c|c|c|c|c|}
        \hline
        \thead{Model} & \thead{\boldmath $H_0$ (km/s/Mpc)} & \thead{\boldmath $\Omega_m$} & \thead{\boldmath $w_0$} & \thead{\boldmath $w_a$} & \thead{\boldmath $z_T$} & \thead{\boldmath $\Delta_z$} \\
        \hline
        $\Lambda$CDM & $73.4\pm0.86$  & $0.32\pm0.013$           &  -                        & -              & - & -          \\
        CPL          & $73.16\pm0.9$  & $0.3^{+0.046}_{-0.005}$  & $-0.90^{+0.076}_{-0.094}$ & $-0.37^{+1.4}_{-0.84}$ & - & - \\
        BA           & $73.18\pm0.89$ & $0.31^{+0.037}_{-0.011}$ & $-0.92^{+0.065}_{-0.074}$ & $-0.24\pm0.4$           & - & - \\
        JBP          & $73.10\pm0.89$ & $0.32\pm0.019$           & $-0.89\pm0.11$            & $-0.74\pm0.97$         & - & - \\
        EXP          & $73.14\pm0.9$  & $0.31^{+0.041}_{-0.008}$ & $-0.98^{+0.075}_{-0.14}$  & $-0.17^{+0.31}_{-0.36}$ & - & - \\
        TDE          & $73.26\pm0.87$ & $0.31^{+0.025}_{-0.015}$ & $-1.03^{+0.34}_{-0.4}$ & $-0.6^{+1.2}_{-2.3}$ & $4.7^{+5.1}_{-4.6}$ & $5.4^{+4.4}_{-1.7}$ \\
        \hline
    \end{tabular}
\end{table}

\begin{table}[htbp!]
\caption{Uniform Priors on $M$ and $r_d$ for Flat cosmologies (Without LRG1$+$LRG2) using Base$+$MM data}
\label{table18}
\centering
    \begin{tabular}{|c|c|c|c|c|c|c|c|}
        \hline
        \thead{Model} & \thead{\boldmath $H_0$ (km/s/Mpc)} & \thead{\boldmath $\Omega_m$} & \thead{\boldmath $w_0$} & \thead{\boldmath $w_a$} & \thead{\boldmath $z_T$} & \thead{\boldmath $\Delta_z$} \\
        \hline
        $\Lambda$CDM & $73.6^{+2.7}_{-3.2}$ & $0.32\pm0.013$         & -                         & -              & - & -          \\
        CPL          & $73.6^{+2.7}_{-3.2}$ & $0.3\pm0.05$           & $-0.9^{+0.077}_{-0.093}$  & $-0.37^{+1.4}_{-0.85}$ & - & -  \\
        BA           & $73.7^{+2.7}_{-3.2}$ & $0.31^{+0.04}_{-0.08}$ & $-0.92^{+0.064}_{-0.077}$ & $-0.23^{+0.38}_{-0.43}$ & - & - \\
        JBP          & $73.6^{+2.8}_{-3.1}$ & $0.32\pm0.02$          & $-0.89\pm0.11$            & $-0.74\pm0.98$          & - & - \\
        EXP          & $73.7^{+2.8}_{-3.1}$ & $0.31^{+0.04}_{-0.01}$ & $-0.99^{+0.076}_{-0.14}$   & $-0.17\pm0.34$          & - & - \\
        TDE          & $73.7^{+2.7}_{-3.0}$ & $0.31^{+0.023}_{-0.015}$ & $-1.02^{+0.34}_{-0.39}$ & $-0.6^{+1.3}_{-2.3}$ & $4.7\pm2.9$ & $5.5^{+4.4}_{-1.8}$ \\
        \hline
    \end{tabular}
\end{table}

\begin{table}[htbp!]
\caption{Gaussian Prior on $M$ for Non-flat cosmologies (Without LRG1$+$LRG2) using Base$+$MM data}
\label{table19}
\centering
    \begin{tabular}{|c|c|c|c|c|c|c|c|}
        \hline
        \thead{Model} & \thead{\boldmath $H_0$ (km/s/Mpc)} & \thead{\boldmath $\Omega_m$} & \thead{\boldmath $\Omega_k$} & \thead{\boldmath $w_0$} & \thead{\boldmath $w_a$} & \thead{\boldmath $z_T$} & \thead{\boldmath $\Delta_z$} \\
        \hline
        $\Lambda$CDM & $73.28\pm0.88$ & $0.31\pm0.028$           & $0.046\pm0.081$ & -                        & -          & - & -              \\
        CPL          & $73.16\pm0.9$  & $0.29^{+0.069}_{-0.035}$ & $0.07\pm0.17$   & $-0.96^{+0.18}_{-0.08}$  & $-0.40^{+0.9}_{-1.0}$ & - & -   \\
        BA           & $73.14\pm0.88$ & $0.29^{+0.056}_{-0.041}$ & $0.08\pm0.18$   & $-0.98^{+0.17}_{-0.072}$ & $-0.33^{+0.56}_{-0.42}$ & - & - \\
        JBP          & $73.13\pm0.87$ & $0.31\pm0.04$            & $0.04\pm0.17$   & $-0.95^{+0.2}_{-0.1}$    & $-0.7\pm1$             & - & - \\
        EXP          & $73.15\pm0.91$ & $0.29^{+0.056}_{-0.041}$ & $0.08\pm0.18$   & $-1.09^{+0.24}_{-0.13}$  & $-0.24\pm0.47$          & - & - \\
        TDE          & $70.7^{+5.7}_{-6.5}$ & $0.32\pm0.031$ & $0.09^{+0.14}_{-0.22}$ & $-1.09\pm0.53$ & $-0.6^{+1.5}_{-2.3}$ & $4.7^{+5.1}_{-4.6}$ & $5.4^{+4.5}_{-2.3}$ \\
        \hline
    \end{tabular}
\end{table}

\begin{table}[htbp!]
\caption{Uniform Priors on $M$ and $r_d$ for Non-flat cosmologies (Without LRG1$+$LRG2) using Base$+$MM data}
\label{table20}
\centering
    \begin{tabular}{|c|c|c|c|c|c|c|c|}
        \hline
        \thead{Model} & \thead{\boldmath $H_0$ (km/s/Mpc)} & \thead{\boldmath $\Omega_m$} & \thead{\boldmath $\Omega_k$} & \thead{\boldmath $w_0$} & \thead{\boldmath $w_a$} & \thead{\boldmath $z_T$} & \thead{\boldmath $\Delta_z$}\\
        \hline
        $\Lambda$CDM & $73.7^{+2.7}_{-3.2}$ & $0.31\pm0.028$           & $0.047\pm0.08$ & -                        & -             & - & -          \\
        CPL          & $73.7^{+2.7}_{-3.1}$ & $0.29^{+0.067}_{-0.037}$ & $0.07\pm0.17$  & $-0.96^{+0.18}_{-0.08}$  & $-0.41\pm0.95$     & - & -     \\
        BA           & $73.7^{+2.7}_{-3.2}$ & $0.29^{+0.058}_{-0.04}$  & $0.08\pm0.18$  & $-0.98^{+0.17}_{-0.072}$ & $-0.33^{+0.58}_{-0.44}$ & - & - \\
        JBP          & $73.7^{+2.7}_{-3.2}$ & $0.31\pm0.041$           & $0.04\pm0.17$  & $-0.94^{+0.2}_{-0.1}$    & $-0.7\pm1.0$            & - & - \\
        EXP          & $73.7^{+2.7}_{-3.2}$ & $0.29^{+0.061}_{-0.04}$  & $0.07\pm0.18$  & $-1.08^{+0.23}_{-0.14}$  & $-0.23\pm0.48$         & - & - \\
        TDE          & $71.2\pm5.9$ & $0.32\pm0.028$ & $0.09^{+0.13}_{-0.21}$ & $-1.08^{+0.38}_{-0.34}$ & $-0.6^{+1.4}_{-2.2}$ & $4.8\pm2.9$ & $5.4^{+4.4}_{-1.7}$ \\
        \hline
    \end{tabular}
\end{table}

\begin{table}[htbp!]
    \caption{Bayes' factors for various DE parameterizations comparing flat ($\Omega_k = 0$) and non-flat ($\Omega_k = 0$) scenarios using the Base$+$CC dataset. The columns represent the  results with and without the inclusion of LRG1 and LRG2, further subdivided based on the choice of priors: uniform priors on $M$ and $r_d$, and a Gaussian prior on $M$. For each parameterization, the null hypothesis corresponds to its flat counterpart.}
    \label{table21}
    \centering
    \begin{tabular}{|c|c|c||c|c|}
         \hline
         \multirow{2}{*}{\thead{Model}} & \multicolumn{2}{c||}{\thead{With LRG1 and LRG2}} & \multicolumn{2}{c|}{\thead{Without LRG1 and LRG2}} \\
         \cline{2-5}
         & \thead{Uniform} & \thead{Gaussian} & \thead{Uniform} & \thead{Gaussian} \\
         \hline
         $\Lambda$CDM & 2.5 & 3.35  & 4.2 & 4.95 \\
         CPL          & 2.4 & 2.51  & 2.1 & 2.5 \\
         BA           & 2.4 & 2.53  & 2.1 & 2.44 \\
         JBP          & 2.5 & 2.5   & 2.3 & 2.5 \\
         EXP          & 2.4 & 2.48  & 2.12 & 2.41 \\
         TDE          & 2.51 & 2.61 & 2.3 & 2.53 \\   
         \hline
    \end{tabular}
\end{table}

\begin{table}[htbp!]
    \caption{Bayes' factors for various DE parameterizations comparing flat ($\Omega_k = 0$) and non-flat ($\Omega_k = 0$) scenarios using the Base$+$MM dataset. The columns present results with and without the inclusion of LRG1 and LRG2, further subdivided based on the choice of priors: uniform priors on $M$ and $r_d$, and a Gaussian prior on $M$. For each parameterization, the null hypothesis corresponds to its flat counterpart.}
    \label{table22}
    \centering
    \begin{tabular}{|c|c|c||c|c|}
         \hline
         \multirow{2}{*}{\thead{Model}} & \multicolumn{2}{c||}{\thead{With LRG1 and LRG2}} & \multicolumn{2}{c|}{\thead{Without LRG1 and LRG2}} \\
         \cline{2-5}
         & \thead{Uniform} & \thead{Gaussian} & \thead{Uniform} & \thead{Gaussian} \\
         \hline
         $\Lambda$CDM & 29.4 & 2.44 & 52.0 & 4.18 \\
         CPL          & 2.4  & 2.4  & 2.1  & 2.08  \\
         BA           & 2.5  & 2.38  & 2.0  & 2.03 \\
         JBP          & 2.4  & 2.5  & 2.3  & 2.16  \\
         EXP          & 2.4  & 2.4  & 2.05  & 2.08  \\
         TDE          & 2.53 & 2.44 & 2.1 & 2.1 \\
         \hline
    \end{tabular}
\end{table}

\begin{table}[htbp!]
    \caption{Bayes' factors for different cosmological models using the Base$+$CC dataset, with and without the inclusion of LRG1 and LRG2. The results are further categorized based on the choice of priors: uniform priors on $M$ and $r_d$, and a Gaussian prior on $M$. Additionally, separate analyses are conducted for both flat and non-flat cases, where the null hypothesis is either the flat $\Lambda$CDM or the non-flat $\Lambda$CDM model, respectively.}
    \label{table23}
    \centering
    \begin{tabular}{|c|c|c||c|c|}
         \hline
         \multirow{2}{*}{\thead{Model}} & \multicolumn{2}{c||}{\thead{With LRG1 and LRG2}} & \multicolumn{2}{c|}{\thead{Without LRG1 and LRG2}} \\
         \cline{2-5}
         & \thead{Uniform} & \thead{Gaussian} & \thead{Uniform} & \thead{Gaussian} \\
         \hline
         \multicolumn{5}{|c|}{\thead{Flat Universe}} \\
         \hline
         $\Lambda$CDM & 1      & 1      & 1      & 1    \\
         CPL          & 27.9   & 27.9   & 58.0   & 53.5 \\
         BA           & 54.1   & 52.5   & 111.1  & 103.5 \\
         JBP          & 22.9   & 23.1   & 44.3   & 40.9 \\
         EXP          & 68.0   & 68.0   & 135.7  & 129.0 \\
         TDE          & 9.2    & 9.3    & 22.4    & 22.4 \\
         \hline
         \multicolumn{5}{|c|}{\thead{Non-flat Universe}} \\
         \hline
         $\Lambda$CDM & 1      & 1      & 1      & 1    \\
         CPL          & 27.1   & 20.9   & 29.1   & 26.6  \\
         BA           & 51.9   & 39.6   & 55.1   & 51  \\
         JBP          & 22.9   & 16.9   & 23.8   & 20.3  \\
         EXP          & 65.4   & 50.4   & 68.0   & 62.8    \\
         TDE          & 9.3    & 7.2   & 12.2 & 11.5 \\
         \hline
    \end{tabular}
\end{table}

\begin{table}[htbp!]
    \caption{Bayes' factors for different cosmological models using the Base$+$MM dataset, with and without the inclusion of LRG1 and LRG2. The results are further categorized based on the choice of priors: uniform priors on $M$ and $r_d$, and a Gaussian prior on $M$. Additionally, separate analyses are conducted for both flat and non-flat cases, where the null hypothesis is either the flat $\Lambda$CDM or the non-flat $\Lambda$CDM model, respectively.}
    \label{table24}
    \centering
    \begin{tabular}{|c|c|c||c|c|}
         \hline
         \multirow{2}{*}{\thead{Model}} & \multicolumn{2}{c||}{\thead{With LRG1 and LRG2}} & \multicolumn{2}{c|}{\thead{Without LRG1 and LRG2}} \\
         \cline{2-5}
         & \thead{Uniform} & \thead{Gaussian} & \thead{Uniform} & \thead{Gaussian} \\
         \hline
         \multicolumn{5}{|c|}{\thead{Flat Universe}} \\
         \hline
         $\Lambda$CDM & 1      & 1      & 1      & 1      \\
         CPL          & 314.2  & 26     & 685.4  & 56.3  \\
         BA           & 595.9  & 48.9   & 1326   & 109.9  \\
         JBP          & 259.8  & 21.3   & 523.2  & 43  \\
         EXP          & 765.1  & 63.4   & 1636   & 133  \\
         TDE          & 107.8  & 8.7    & 284.3  & 22.9 \\
         \hline
         \multicolumn{5}{|c|}{\thead{Non-flat Universe}} \\
         \hline
         $\Lambda$CDM & 1      & 1      & 1      & 1      \\
         CPL          & 25.5   & 25.5   & 27.8   & 27.9   \\
         BA           & 48.4   & 47.9   & 52.4   & 53.5   \\
         JBP          & 21.3   & 21.5   & 22.4   & 22.2   \\
         EXP          & 61.6   & 62.8   & 64.7   & 66.0   \\
         TDE          & 9.3    & 8.7    & 11.5   & 11.6  \\
         \hline
    \end{tabular}
\end{table}

\section{Comparison of $w(z)$ with DESI results}
\label{sec:behaviour}

\begin{figure}
    \centering
    \subfloat[With LRG points]{\includegraphics[width=0.5\textwidth]{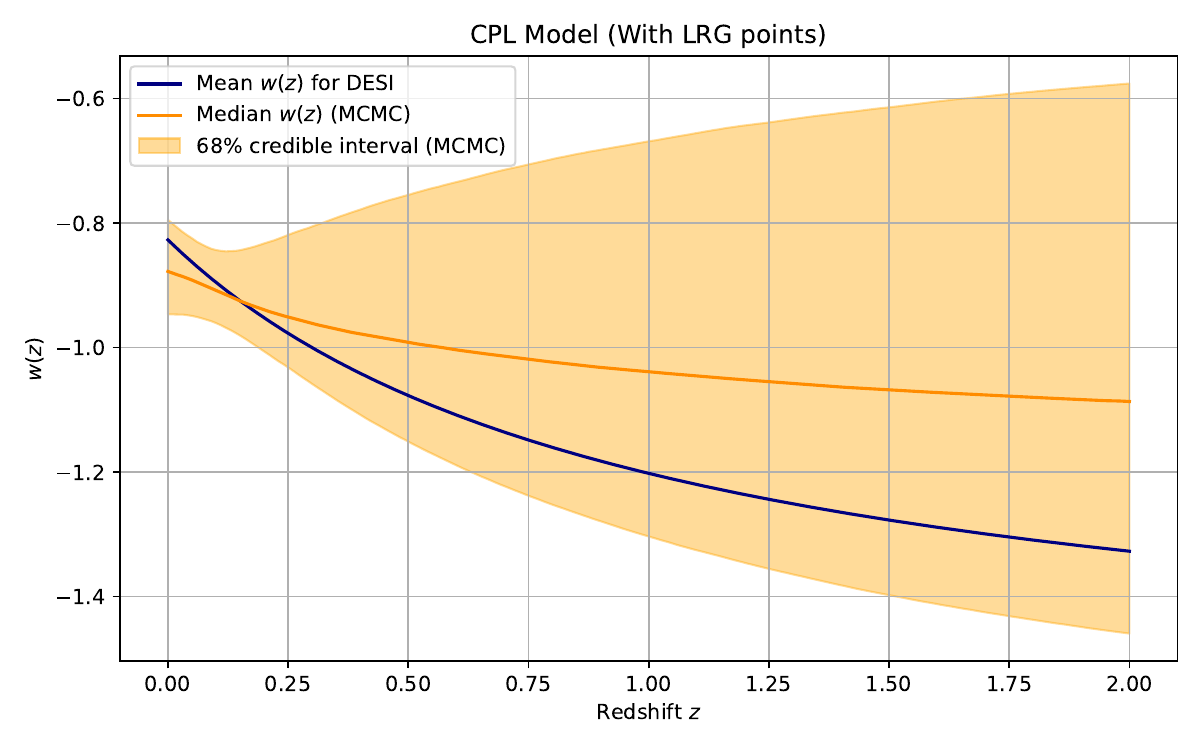}}
    \subfloat[Without LRG points]{\includegraphics[width=0.5\textwidth]{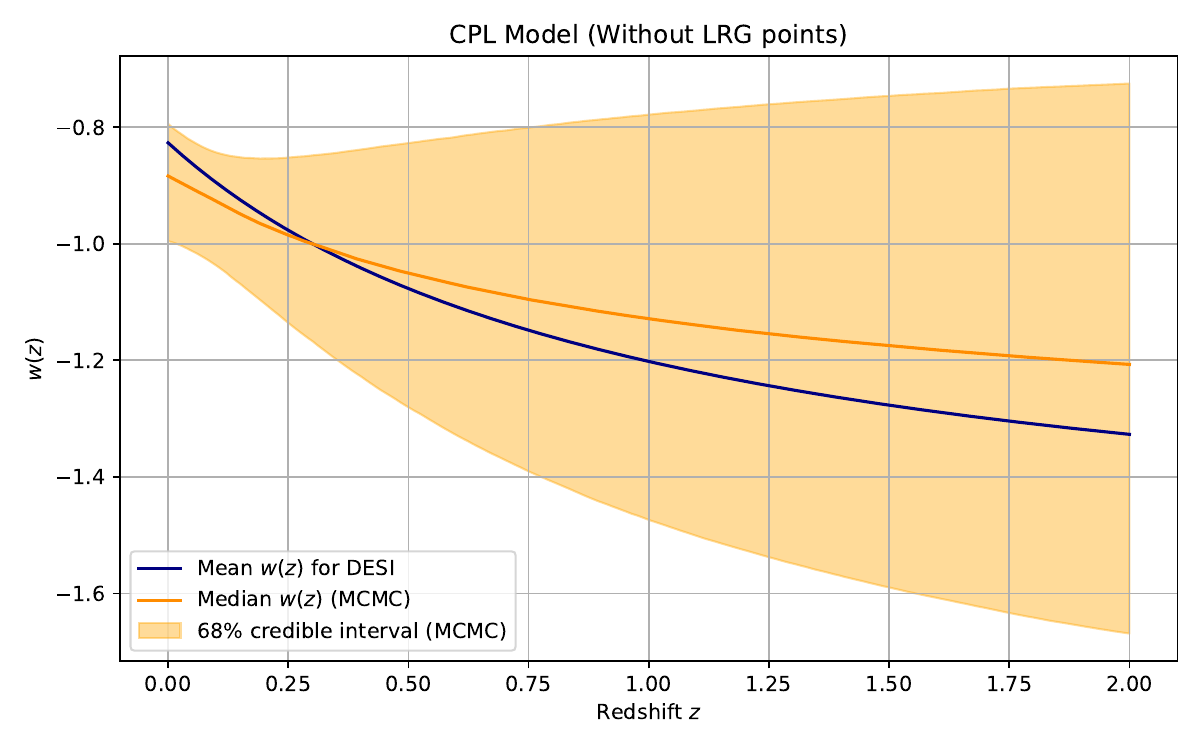}}
    \caption{Behaviour of $w(z)$ for CPL model (Eqn.~\ref{eq5}). The navy blue curve denotes the mean $w(z)$ value from DESI DR1. The dark orange curve is the median $w(z)$ value found in this work while the orange shaded part represents the 68\% credible interval of $w(z)$.}
    \label{fig13}
\end{figure}

\begin{figure}
    \centering
    \subfloat[With LRG points]{\includegraphics[width=0.5\textwidth]{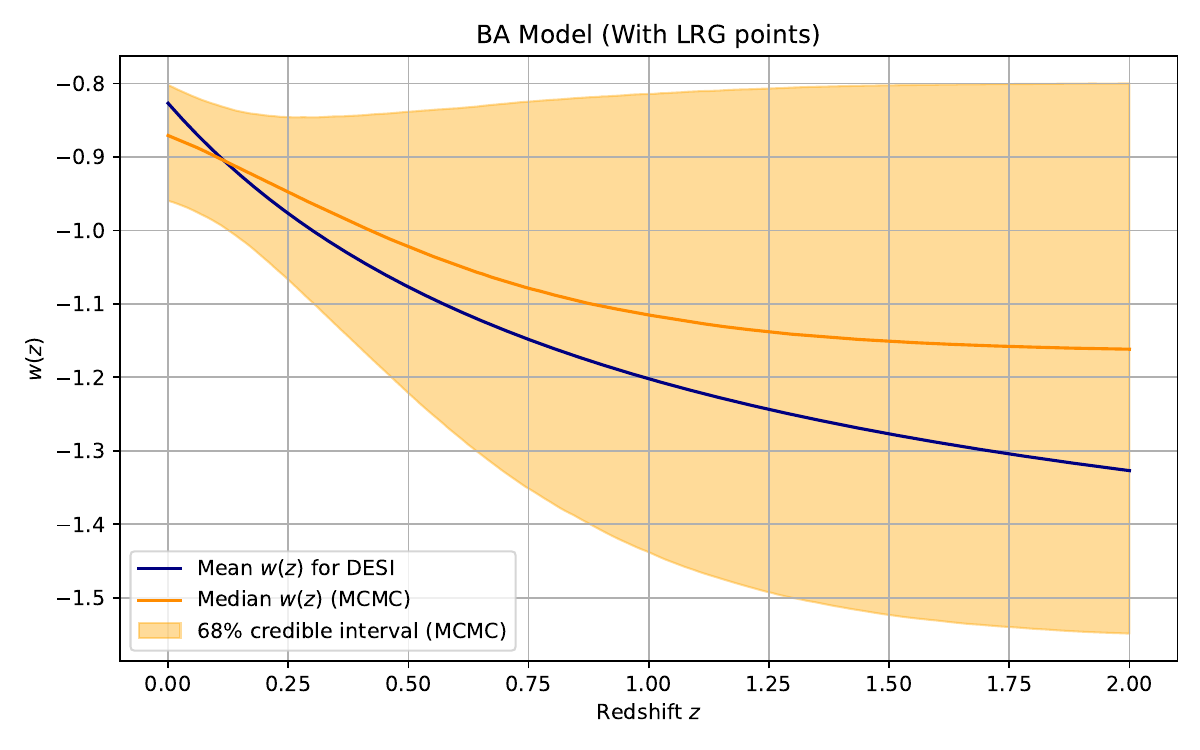}}
    \subfloat[Without LRG points]{\includegraphics[width=0.5\textwidth]{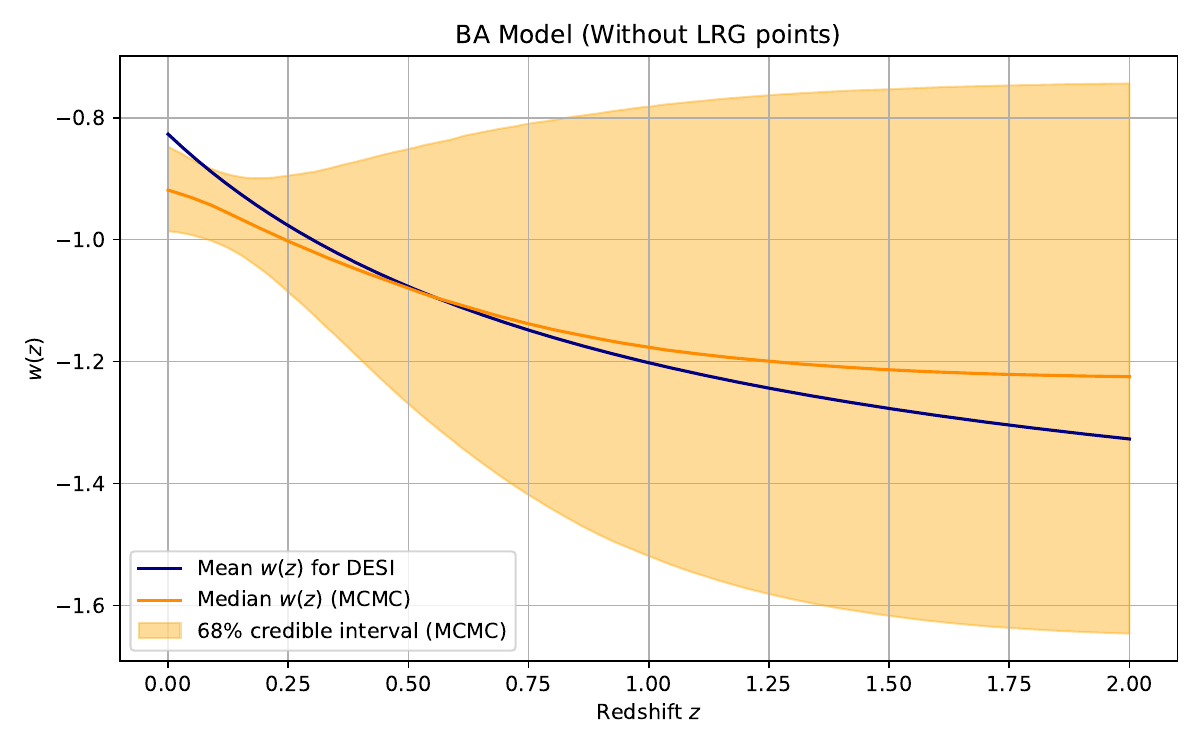}}
    \caption{Behaviour of $w(z)$ for BA model (Eqn.~\ref{eq6}). The navy blue curve denotes the mean $w(z)$ value from DESI DR1. The dark orange curve is the median $w(z)$ value found in this work while the orange shaded part represents the 68\% credible interval of $w(z)$.}
    \label{fig14}
\end{figure}

\begin{figure}
    \centering
    \subfloat[With LRG points]{\includegraphics[width=0.5\textwidth]{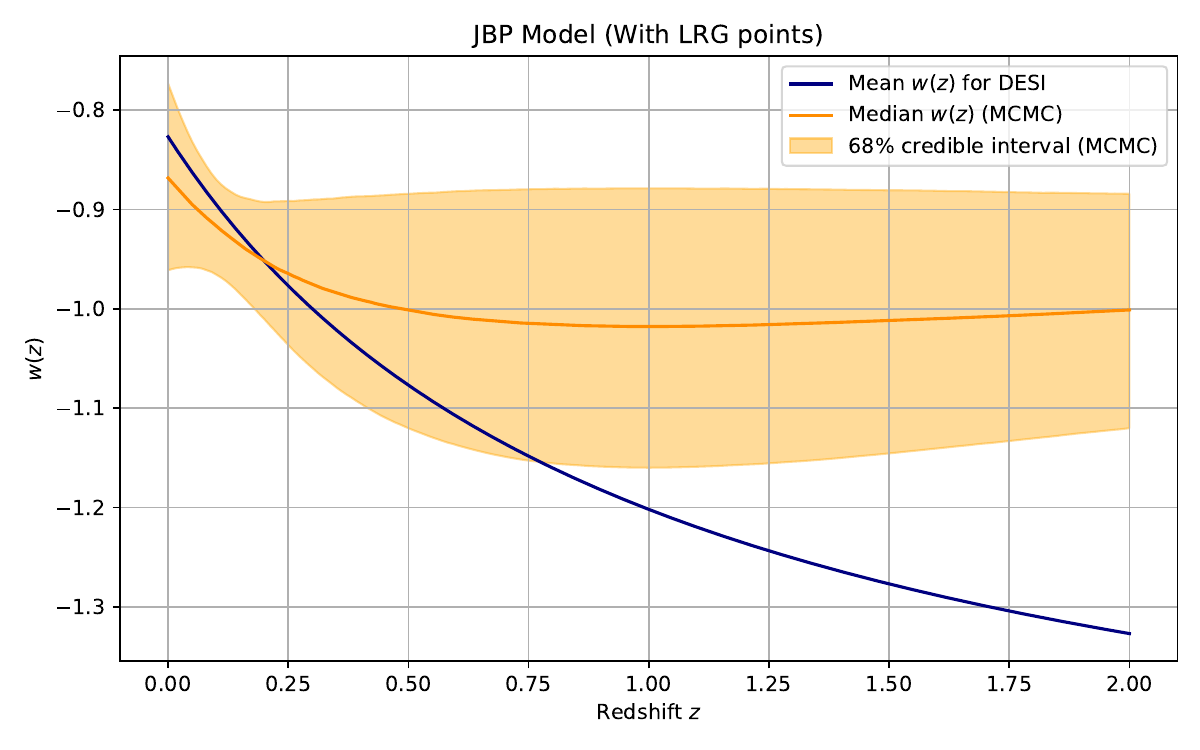}}
    \subfloat[Without LRG points]{\includegraphics[width=0.5\textwidth]{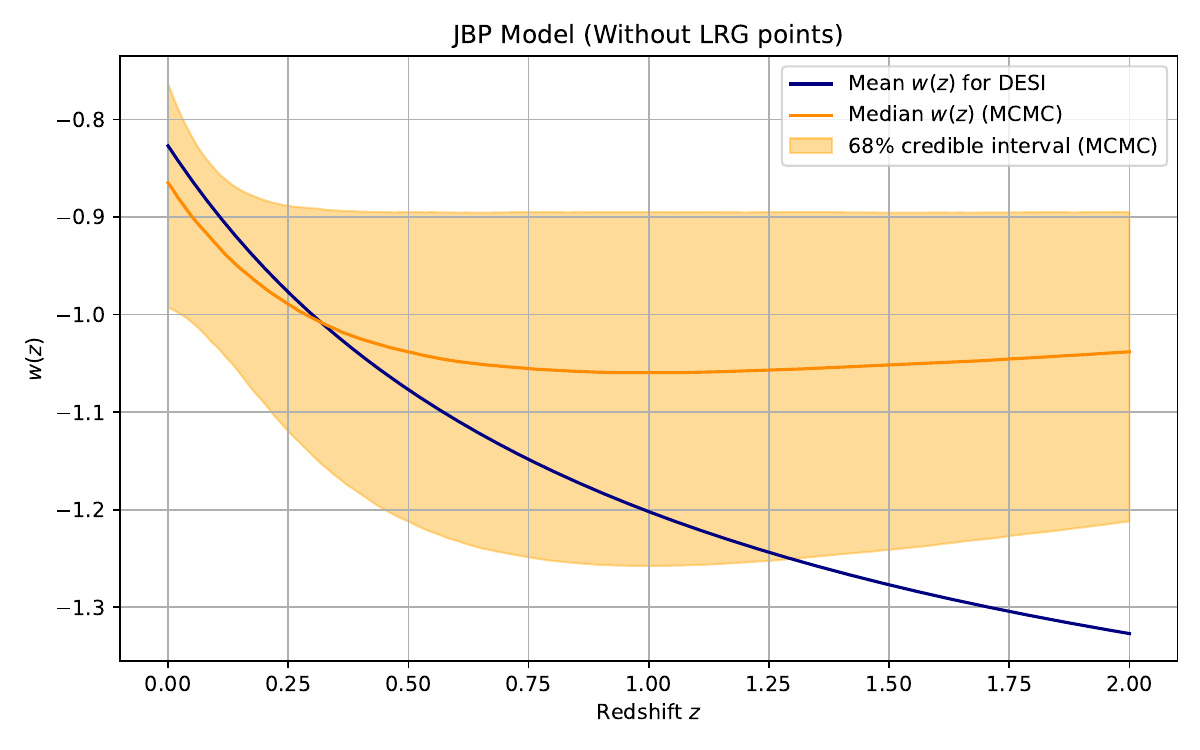}}
    \caption{Behaviour of $w(z)$ for JBP model (Eqn.~\ref{eq7}). The navy blue curve denotes the mean $w(z)$ value from DESI DR1. The dark orange curve is the median $w(z)$ value found in this work while the orange shaded part represents the 68\% credible interval of $w(z)$.}
    \label{fig15}
\end{figure}

\begin{figure}
    \centering
    \subfloat[With LRG points]{\includegraphics[width=0.5\textwidth]{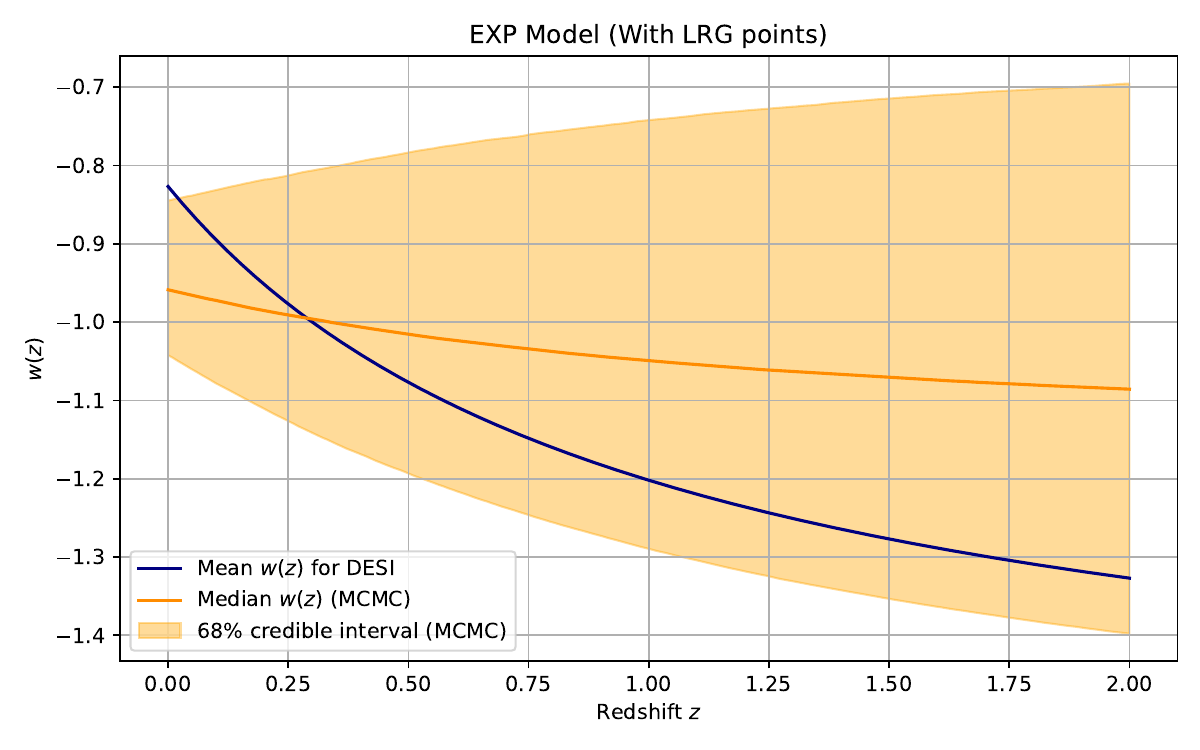}}
    \subfloat[Without LRG points]{\includegraphics[width=0.5\textwidth]{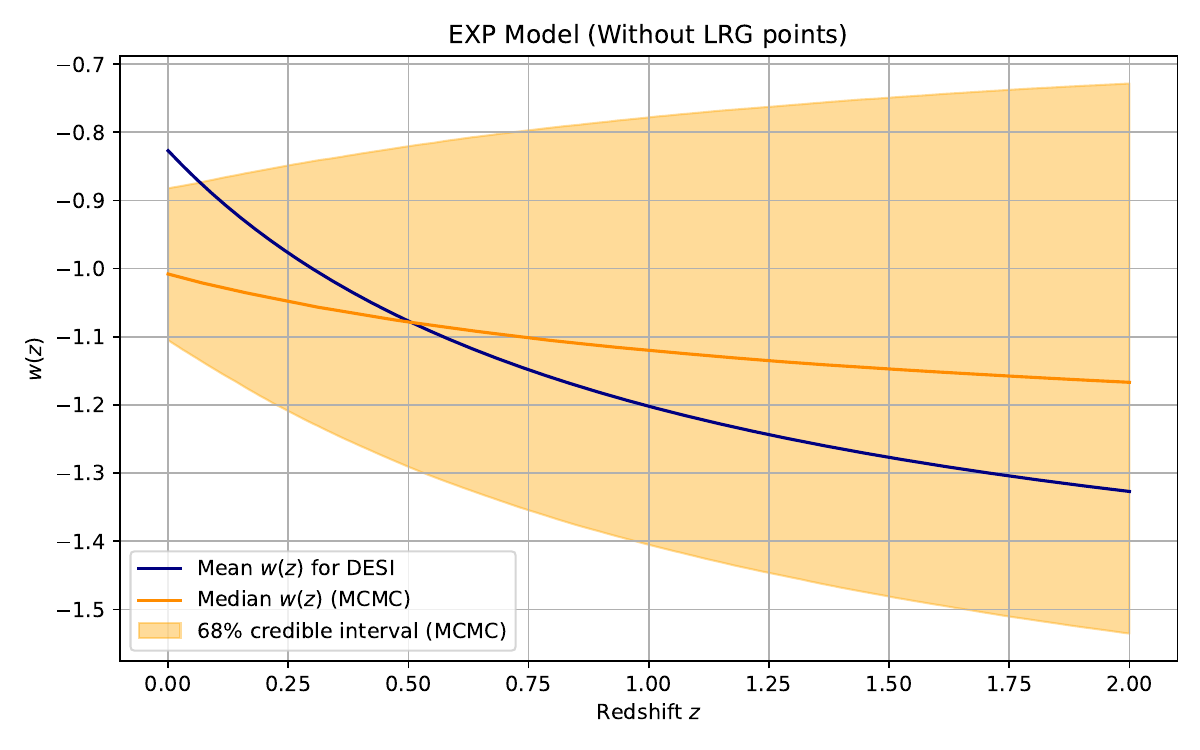}}
    \caption{Behaviour of $w(z)$ for EXP model (Eqn.~\ref{eq8}). The navy blue curve denotes the mean $w(z)$ value from DESI DR1. The dark orange curve is the median $w(z)$ value found in this work while the orange shaded part represents the 68\% credible interval of $w(z)$.}
    \label{fig16}
\end{figure}

\begin{figure}
    \centering
    \subfloat[With LRG points]{\includegraphics[width=0.5\textwidth]{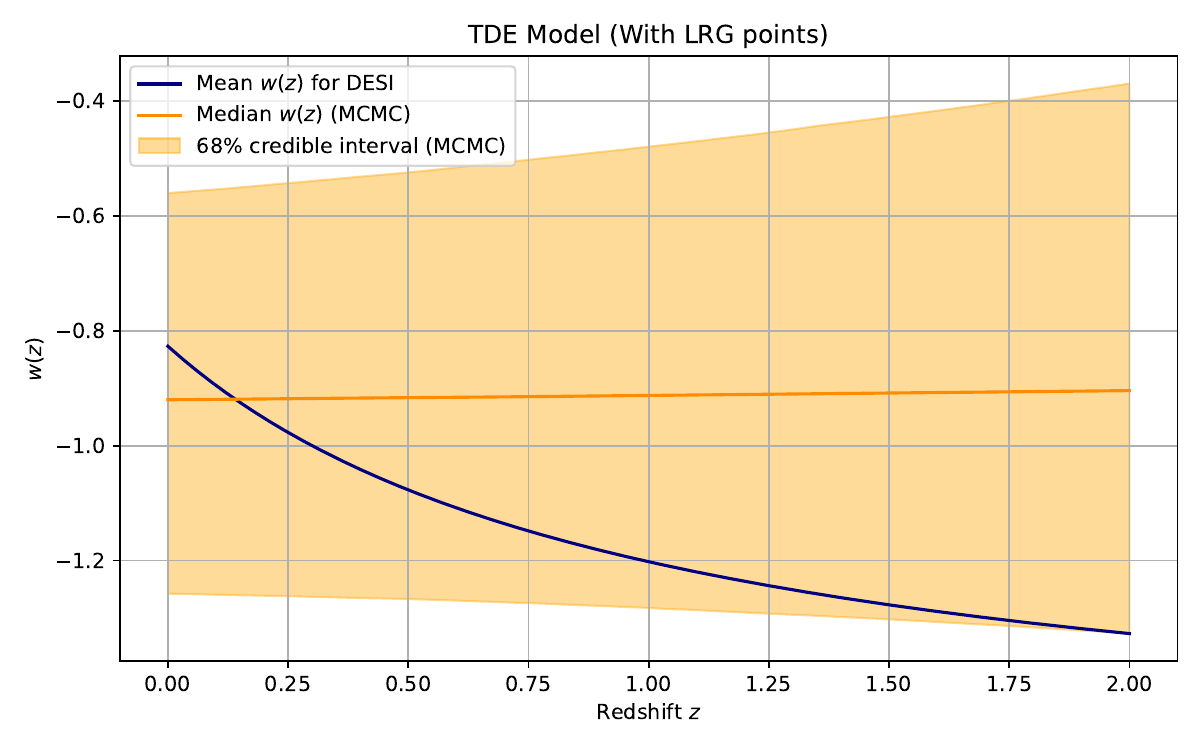}}
    \subfloat[Without LRG points]{\includegraphics[width=0.5\textwidth]{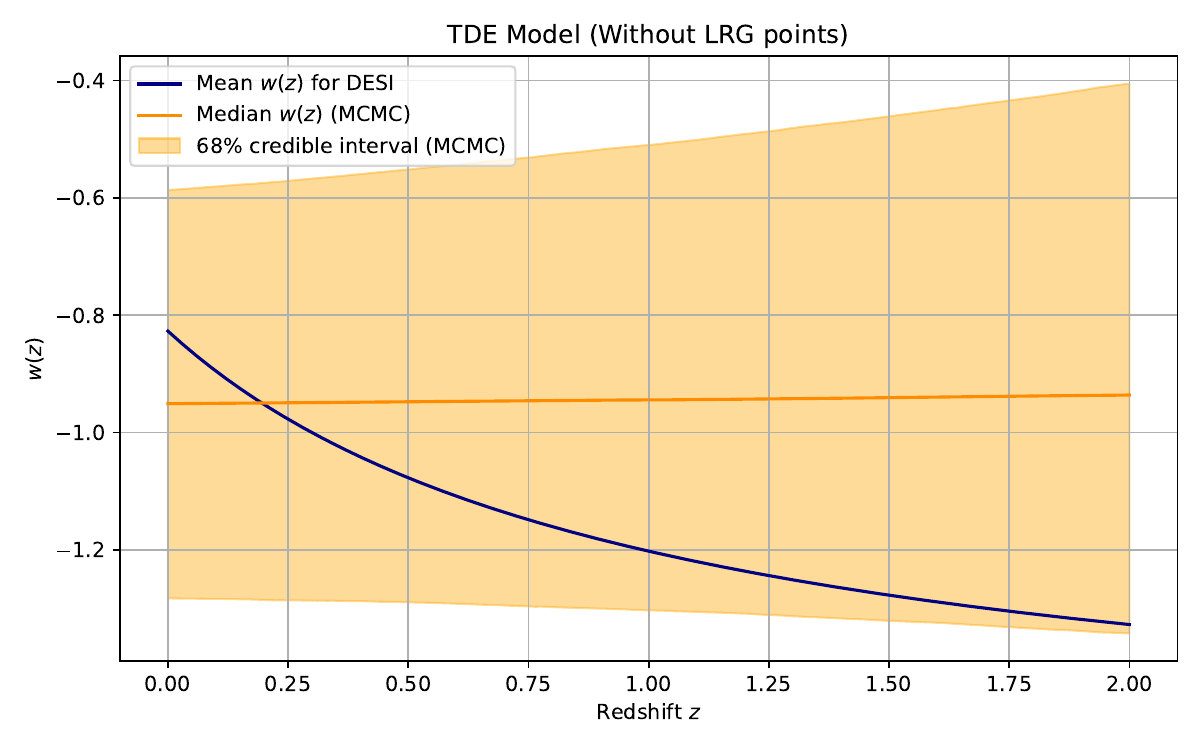}}
    \caption{Behaviour of $w(z)$ for TDE model (Eqn.~\ref{eq9}). The navy blue curve denotes the mean $w(z)$ value from DESI DR1. The dark orange curve is the median $w(z)$ value found in this work while the orange shaded part represents the 68\% credible interval of $w(z)$.}
    \label{fig17}
\end{figure}

\rthis{In this section, we investigate the behavior of the dark energy equation of state, $w(z)$, across various parameterizations (CPL, BA, JBP, EXP, and TDE), using the constraints presented in Tables~\ref{table5}-\ref{table20}. As discussed in Sec.~\ref{sec:results}, the constraints on these parameterizations are robust, regardless of the priors imposed or the dataset combinations employed. Furthermore, we observe that the spatial curvature parameter $\Omega_k$ does not significantly influence the determination of $w_0$ and $w_a$ (see Sec.~\ref{gen_results}). However, the inclusion of LRG1 and LRG2 data points appear to play a critical role in driving the preference for dynamical dark energy in several parameterizations—particularly CPL, BA, and JBP. Therefore, in this section, we categorize the $w(z)$ behaviour based on the presence or absence of LRG points and select the most stringent constraints within each category for analysis. Figures~\ref{fig13}-\ref{fig17} illustrate the evolution of $w(z)$ for each parameterization, distinguishing between datasets that include LRG points and those that do not. In each figure, the dark orange curve represents the median $w(z)$ from our analysis, with the shaded orange region indicating the 68\% credible interval. For comparison, the navy blue curve shows the DESI DR1 constraint on $w(z)$ for the CPL parameterization using the CMB$+$BAO$+$PantheonPlus \cite{desi_2024} dataset. We superimpose this curve on all parameterizations to highlight differences and similarities in their trends.}

\rthis{Due to the parametric forms and the usage of low-redshift data, each of the $w(z)$ curves show large error bands with increase in redshift. The large error bars on $w_a$ for all the parameterizations also play a role in increasing the credible intervals because $w_a$ controls the evolution of $w(z)$.} 

\rthis{In the CPL parameterization (Figure~\ref{fig13}), we notice a prominent phantom crossing behavior ($\sim2\sigma$) where $w(z)<-1$, for the case where the LRG points are included. This is less prominent in the case when LRG points are excluded ($\sim1\sigma$).}

\rthis{In the BA parameterization (Figure~\ref{fig14}), we again notice a prominent phantom crossing behavior ($\sim2\sigma$) for the case with LRG points included. Similar to the CPL case, this becomes less prominent in the case when LRG points are excluded($\sim1\sigma$).}

\rthis{In the JBP parameterization (Figure~\ref{fig15}), we again notice a prominent phantom crossing behavior ($\sim2\sigma$) for the case with LRG points included. This is less pronounced in the case when the LRG points are excluded($\sim1\sigma$).} 

\rthis{In the all the three cases (CPL, BA and JBP), the phantom crossing can be noticed even in the case without LRG points but it is not statistically significant($\sim1\sigma$). Another point to note is that $w(z)$ for JBP deviates significantly from the $w(z)$ for CPL (determined from DESI results) at redshifts close to 1.} 

\rthis{For the EXP parameterization (Figure~\ref{fig16}), we notice no phantom crossing behavior which is expected since this parameterization is more consistent with the $\Lambda$CDM model as discussed. The same applies to the TDE parameterization (Figure~\ref{fig17}) as well. The large credible interval in the TDE parameterization $w(z)$ is due to posteriors being poorly constrained. In the TDE case, it should also be noted that due to the non-linearity of the parameterization, $w(z=0)\neq w_0$.}

\rthis{Therefore, the trends of $w(z)$ which we obtain as a function of redshift are consistent with the DESI best-fit values for all the parameterizations used, once we incorporate the uncertainties in the best-fit parameters. }

\section{Conclusions}
\label{sec:conclusion}
In this work, we have used geometric probes in the late Universe, viz. DESI BAO, PantheonPlus SNe Ia, QSO, and either the Cosmic Chronometer or Megamasers dataset—to constrain $\Lambda$CDM,  as well as five distinct parameterizations of dark energy, viz. CPL, BA, JBP, EXP, and TDE. Our analysis is carried out in multiple parts, considering cases where we include as well  exclude LRG1 and LRG2 datapoints from DESI DRI. We apply both uniform priors on $M$ and $r_d$, as well as a Gaussian prior on $M$ along with a uniform prior on $r_d$ (cf. Figures~\ref{fig1}-\ref{fig12} and Tables~\ref{table5}-\ref{table20}). The models are compared using the Bayes' Factor (Tables~\ref{table21}, \ref{table22}, \ref{table23}, \ref{table24}), and we contrast our findings with other recent studies (Sect.~\ref{comparison}).

We find that (flat/non-flat) $\Lambda$CDM model shows a strong preference over other parameterizations (Tables~\ref{table23} and \ref{table24}) in all cases. We also find that late Universe probes provide inconclusive evidence for distinguishing between flat and non-flat scenarios for all parameterizations (Tables~\ref{table21} and \ref{table22}).

In CPL parameterization (Eqn.~\ref{eq5}), $w_0$ shows a slight deviation from its $\Lambda$CDM value of -1, $(1.5 - 1.8)\sigma$, in certain cases (see Sect.~\ref{CPL}). However, a more general version of the CPL parameterization, the exponential parameterization (Eqn.~\ref{eq8}) wherein we include terms upto the second power of $z$, seems to nullify this deviation (Sect.~\ref{exp}). The $w_0$ values of the CPL and EXP parameterizations shift towards -1, when excluding LRG1 and LRG2 data points. From this we find hints that LRG1 and LRG2 indeed drive the deviation from $\Lambda$CDM model, as discussed in \cite{zheng_2024} and \cite{Wang_2024}. The same discussion also applies to BA and JBP parameterizations. However, as mentioned above and in Sect.~\ref{exp}, the $w_0$ and $w_a$ values of the EXP parameterization show very small deviations from $(-1, 0)$. This is due to the inclusion of higher orders of $z$ \cite{nesseris_2025}. Although the DE parameters exhibit some variations depending on the choice of priors, no clear pattern is discerned.
In this work, we have also examined the TDE parameterization. We find that the posteriors for this are not well constrained. Our estimated values for this parameterization indicate potential consistency with $\Lambda$CDM. 


From our analysis, it is also clear that the Base$+$CC dataset show slightly higher discrepancy in $w_0$ values for CPL parameterization than the Base$+$MM dataset (Please refer to Sect.~\ref{sec:results}). The BA and JBP parameterizations show similar discrepancies in both datasets. However, both Base$+$CC and Base$+$MM show that the discrepancy is driven primarily by LRG1 and LRG2 points.

\rthis{We also noticed that phantom crossing exists in CPL, BA and JBP parameterizations (Sect.~\ref{sec:behaviour}) for the dataset combinations used in this work. It is prominent in the case of dataset which includes LRG1 and LRG2 BAO data points but not statistically significant when excluding these points.}

To conclude, our analysis provides indications of dynamical dark energy in the case of CPL, BA and JBP parameterizations (supported by the discrepancies in $w_0$ and $w_a$ values compared to the $\Lambda$CDM values for the same) reinforcing the need for further investigation. However, the current late universe data provide strong evidence for the $\Lambda$CDM model as found using Bayes' factor values. EXP and TDE parameterizations show good consistency with $\Lambda$CDM values of $(w_0 = -1, w_a = 0)$.
\rthis{We note that this is not in contradiction with the DESI DRI results, since on its own DESI DRI is consistent with flat $\Lambda$CDM. Only when the DESI DRI data is combined with CMB and different combinations of Type Ia SNe, one obtains evidence for dynamical dark energy between $2.5-4\sigma$. Here, we have only considered low redshift probes and not included CMB \cite{park_2024, park_2025}.}


\vspace{10pt}
\noindent\textbf{Note added:} During the final stages of our work, the DESI DR2 \cite{desi_2025} results were published. However, we do not think that the results in our work will show any major changes due to usage of DESI DR1. We would also like to mention a few works  similar to ours \cite{wu_2025, wolf_2025, Shlivko_2025, Silva_2025}, which appeared on arXiv, while this manuscript was under preparation. However, our dataset combinations are different and we also use the Bayes' factor for model comparison, instead of information theory techniques used in these works. Furthermore, we also examine the TDE parameterization, which has not been discussed in the other works.

\begin{acknowledgments}
SB would like to extend his gratitude to the University Grants Commission (UGC), Govt. of India for their continuous support through the Junior Research Fellowship, which has played a crucial role in the successful completion of our research.\rthis{ We also thank the anonymous referee for very constructive and useful comments on our manuscript.}
Computational work was supported by the National Supercomputing Mission (NSM), Government of India, through access to the ``PARAM SEVA'' facility at IIT Hyderabad. The NSM is implemented by the Centre for Development of Advanced Computing (C-DAC) with funding from the Ministry of Electronics and Information Technology (MeitY) and the Department of Science and Technology (DST). We also acknowledge the use of  IUCAA HPC Computing facilities.
\end{acknowledgments}

\bibliography{References}

\end{document}